\documentclass[pra,showpacs,groupedaddress,amssymb,twocolumn,notitlepage,nofootinbib,longbibliography,floatfix,superscriptaddress]{revtex4-1}

\usepackage{graphicx,amsmath,dsfont}
\usepackage[dvipsnames]{xcolor}
\usepackage{subcaption}
\usepackage{animate}
\usepackage{tabularx}
\usepackage{enumerate}
\usepackage{float}
\usepackage{upgreek}
\usepackage{soul}
\usepackage[]{svg}

\usepackage[capitalize]{cleveref}

\newcounter{protocol}

\usepackage{color}
\usepackage{amsthm}
\usepackage{amsmath,amssymb,amsbsy}

\usepackage{xcolor}
\pagecolor{white}
\usepackage[english]{babel}

\usepackage{calrsfs}

\usepackage{xparse}   

\NewDocumentCommand{\RInv}{m}{\underset{\leftarrow}{#1}\phantom{}}

\newcommand{\inv}{\text{-1}}

\newcommand{\ket}[1]{\left| #1 \right\rangle}

\newcommand{\expect}[1]{\left\langle#1\right\rangle}
\newcommand{\eea}{\end{eqnarray}}
\newcommand{\bea}{\begin{eqnarray}}
\newcommand{\ee}{\end{align}}
\newcommand{\be}{\begin{align}}

    \newcommand{\id}{\mathds{1}}

\begin{document}

\title{Towards the Geometry of Quantum Optical Measurements:\\ Kraus Trajectories in the Instrumental Weyl-Heisenberg Group Manifold}

\author{Tzula B. Propp}
\email{propp@physics.leidenuniv.nl}
\affiliation{Leiden Institute of Physics, Leiden University, 2333 CA Leiden, The Netherlands}

\author{Chris S. Jackson}
\email{omgphysics@gmail.com}
\affiliation{Ginzton Lab, Stanford University, 348 Via Pueblo Mall, Stanford, CA 94305, United States}
\affiliation{Perimeter Institute for Theoretical Physics, 31 Caroline Street North, Waterloo, ON N2L 2Y5, Canada}

\begin{abstract}

Since ancient Greek times, geometry has been a method by which humans can make inferences, by the use of similar triangles, about the world far beyond their immediate senses.
Similarity plays an equally important role in quantum measurement, except that instead of between triangles on wax tablets and in the sky, the similarity is between the infinitesimal generators of transformation (i.e. rotation, affine canonical transformation) on the page and within the atom.
For many quantum optical measurements of a single mode, these generators produce Kraus operators that are contained within a twelve-dimensional Lie group: the Instrumental Weyl-Heisenberg group manifold  ($\mathrm{IWH}^{\textbf{12}}$).
In particular, $\mathrm{IWH}^{\textbf{12}}$ is general enough to include all diffusively generated 'dyne measurements. In this paper, $\mathrm{IWH}^{\textbf{12}}$ will be used to develop two more general classes of 'dyne measurement: one that interpolates between diffusive homodyne and diffusive heterodyne (``slomodyne'') and one that adds to 'dyne measurements adaptive squeezing and displacement.
Our analysis uses an approach known as the Instrument Manifold Program (IMP).
In particular, a Harish-Chandra-like decomposition of $\mathrm{IWH}^{\textbf{12}}$ is used to calculate the vielbein transformation from a right-invariant basis to the corresponding coordinate basis, allowing us to universally and autonomously navigate $\mathrm{IWH}^{\textbf{12}}$ and locate the Kraus-operator of whatever selective process is being considered as it evolves in time according to the measurement record.

\end{abstract} 

\maketitle
\pagestyle{plain}

\section{Introduction}

Measurements are properly described using state-independent objects: elements of a positive-operator valued measure (POVM) and Kraus operators \cite{Davies1976,Holevo1982,Kraus1983}.\footnotemark[1] Such a description can make manifest the geometry of the measurement process itself.
This is especially true in quantum optics, where the measurements of interest are those of light---in the simplest case---in a single mode.

This paper is an application of the so-called Instrument Manifold Program (IMP): an ongoing effort to systematically unify each measuring apparatus in quantum science with its native manifold, the so-called \emph{instrumental group (IG) manifold}.
The IMP consists of two basic principles, which guide our study here of linear diffusive quantum optical measurements and the associated twelve-dimensional Instrumental Weyl-Heisenberg group manifold ($\mathrm{IWH}^{\textbf{12}}$):

\emph{\textbf{The Principle of Instrument Universality} |
While a Hilbert space of a quantum system neccessarily carries a representation of the observables measured, the measuring instrument must itself be presented in a way that is manifestly independent of the particular Hilbert space representation.}

The concatenated actions of a diffusive instrument generate a group, the IG manifold, the existence of which is referred to as universal
\cite{Poincare83,Poincare99,Weyl13,Bourbaki71,Jackson2023}. 
Geometrically, it is helpful to picture (at least heuristically) the IG manifold as a stack of generalized phase spaces; for spin systems these are nested spheres \cite{JacksonCaves2023Quantum}, and for Weyl-Heisenberg these are stacked phase planes \cite{Jackson2023SPQM}.\footnotemark[2] 

\emph{\textbf{The Principle of Instrument Autonomy} |
An instrumental group manifold is a type of automaton, where any coordinate can serve as the memory for generating adaptive behaviors.}

Here memory refers precisely to the usual stack or tape that plays a key role in the theory of automata, one of the original ideas of modern information theory used by the pioneers of the field \cite{Rosenblueth1943,vonNeumann1945,Wiener1948,Turing1950,vonNeumann1966,Jackson2025}.

By applying these principles to continuous measurements of a quantum optical mode, the IMP enables a state-independent, dynamic description of POVM elements and Kraus operators evolving on the IG manifold i.e. studying continuous measurement in measurement's own terms.
This framework of Kraus operators diffusing within an IG manifold has already resulted in the discovery of a previously unknown measurement | the so-called isotropic spin measurement (ISM) \cite{Shojaee2018,JacksonCaves2023Quantum} which realizes the spin-coherent-state POVM, a.k.a. the measurement of a spin's direction \cite{d2002quantum} | and, in this paper, we add more new measurements to the list. What allows the IMP to embody these principles of universality and autonomy is essentially the application of two mathematical theories: the theory of Lie's transformation groups and the theory of Cartan's differential forms.

Before the IMP came together, the importance of applying Lie group theory to the problem of diffusively generated POVMs had already been recognized in optics by Warszawski et al. \cite{Warszawski2020}; the $\mathrm{IWH}^{\textbf{12}}$ can be considered as a special case of their general Lie group for an arbitrary number of modes.
However, particularly because their technique is based in thermo-entangled states instead of Kraus operators directly, the extent to which the methods of \cite{Warszawski2020} are able to see that the IG manifold obeys the principles of universality and autonomy is unclear and their techniques therefore do not generalize in an obvious way to spin systems, for example.

As for diffusive measurements of spin systems, there is another significant body of work that preceded the IMP, based on the work of Chantasri and Atalaya et al \cite{chantasri2018simultaneous}.
However, particularly because they assume the spin being measured is always a spin-half or ``qubit'' representation, the diffusive instrument implicit in \cite{chantasri2018simultaneous} is not universal, meaning it does not make sense for spins other than a half.
Part of the surprise of ISM is that by simply adding the orthogonal spin component $J_y$ to their simultaneous measurement of $J_z$ and $J_x$, a  universal, finite-dimensional IG manifold emerges \cite{JacksonCaves2023Quantum}.

Returning to the mathematics that make the IMP possible, it is worth taking a moment to reflect on the emergence of these two theories and how they accompanied some of the biggest theoretical revolutions of the 18th and 19th centuries.
Lie's theory of transformation groups was a synthesis of Jacobi's theory of canonical transformations and Galois' theory of algebraic groups \cite{Jacobi1837,Galois1846,Edwards1984,Jacobi1866,Lie1880,Ackerman1975,Hawkins1991,Hawkins2000,Borel2001,Fraser2023}.
With historical hindsight, 
we can see there was a shift from a Physics of pure analysis
\cite{Newton1687,Lagrange1788} 
towards a Physics based in transformation \cite{Tacquet1651,Alexander2014}
, a shift that enabled both thermodynamics and the theory of parallel transport to make sense of path-dependent processes, resulting in the discovery of entropy, phase space, general relativity, and quantum mechanics \cite{Clausius1850,Clausius1865,Boltzmann1872,nolte2010tangled,ricci1895dei,cassidy1992uncertainty,jammer1966conceptual}.
An ability to directly handle these ubiquitous path-dependent structures was exactly what the advances of the Lie bracket and differential forms \cite{Cartan1899,Cartan1922,CartanDelphenich1922} (which are equivalent) so elegantly accomplished.
Meanwhile in the 20th century, these early advances that were fundamentally based in the structure of path-dependent processes would further develop into the canonical commutation relation, the theory of phase space symbols, quasiprobabilities, and coherent states \cite{Weyl1927,Wigner1932,Husimi1940,Moyal1949,Glauber1963,
Weyl1928,Wigner1931,Wigner1939,Perelomov1972}, all of which form the foundation for our here-presented work

To navigate our way through the manifold $\mathrm{IWH}^{\textbf{12}}$, the easiest approach within the IMP---also serving as a pedagogical introduction---is by means of a Harish-Chandra-like group decomposition \cite{Knapp2002} of $\mathrm{IWH}^{\textbf{12}}$ i.e. a quantum optician's normal-ordering.
However, Harish-Chandra decompositions, though algebraically powerful and easy to work with, are not the best for understanding the geometry of the POVM.
For the advanced reader, we additionally introduce a new half-algebraic half-geometric Iwasawa-like decomposition, which geometrically separates the POVM from the post-measurement unitary while preserving some of the algebraic convenience of the Harish-Chandra-like decomposition. This immediately reveals the fiber bundle structure underlying the geometry of quantum optical measurements i.e. the POV manifold, which we preview in this paper but leave a proper introduction to in a later publication.
Furthermore, Iwasawa decompositions enable a new geometric measure of POVM purity even for non-trace-class measurements such as homodyne and diffusive homodyne (where we can now elucidate the speedup attained by modern diffusive homodyne over Barchielli's 1990 scheme \cite{Barchielli1990}).
We also introduce here a new method to leverage the ostensibility of the Kraus-opertor distribution to remotely (state-independently) sample Kraus operators far from the origin of the phase plane.\footnotetext[1]{For the reader unfamiliar with Kraus operators, they primarily appear in quantum theory when updating quantum states conditioned on a particular measurement outcome $k$ e.g. $\rho_{|k}= \Omega_k \rho_0 \Omega_k^\dagger / |\Omega_k \rho_0 \Omega_k^\dagger|$. For an unconditional channel, they allow the construction of a completely positive and trace-preserving map $\rho_0\rightarrow \sum_k \Omega_k \rho_0 \Omega_k^\dagger$. They also appear in the Born rule where POVM elements, defined $\Pi_k = \Omega_k^\dagger \Omega_k$, are used to calculate the probability of observing an outcome conditioned on the presence of a quantum state ${\rm Pr}(k|\rho_0) = {\rm Tr} \,\Pi_k \rho_0$. By conservation of probability, $\sum_k{\rm Pr}(k|\rho_0)=1\rightarrow \sum_k \Pi_k  = \id$, so that the POVM elements are positive operator-valued partitions of unity i.e. a measure. Note that the POVM elements may be coarse grained i.e. not projecting onto pure states. In the case that there is only a single Kraus operator, it must be unitary, so that the first two equations collapse to a single case which is the integrated form of the von Neumann equation \cite{vonNeumann1927,vonNeumann1932}, the POVM element is the identity, and the third equation simply restates the normalization condition for the density matrix. For a thorough review see Ref. \cite{Preskill1998}.}\footnotetext[2]{This sweeps under the rug the distinction between the instrumental group and the POV manifold; in effect, for $\mathrm{IWH}^{\textbf{12}}$ the post-measurement unitary produces a second stack of phase planes.}\setcounter{footnote}{2}\footnote{In particular, the phase plane associated with the POVM unitary; there is a second phase plane associated with the post-measurement unitary. It is these, together, that stack up in the direction of the complexified ruler to form the geometric structure of $\mathrm{IWH}^{\textbf{12}}$.}


Generically, each increment of a diffusive measurement contributes a differential stochastic Kraus element
\bea\label{dKraus}
d\Omega_{d\vec{W}_t} = \sqrt{d\mu(d\vec{W}_t)} \;e^{-\frac{\vec{L}^\dagger\cdot \vec{L} + \vec{L}\cdot\vec{L}}{2}\kappa dt + \sqrt{\kappa} \vec{L}\cdot d\vec{W}_t},
\eea where $\vec{L}$ is the vector of Lindblad operators \cite{Lindblad1976,GKS1976,Jackson2025} describing system-meter interactions (here considered to have uniform strength $\kappa$), $d\vec{W}_t$ is a vector of Wiener increments, and $d\mu(d\vec{W}_t)$ is the zero-mean Gaussian measure with variance $dt$.
When concatenated, these generate time-dependent trajectories for the Kraus operators upon integration of the time series $d\vec{W}_{[0,T)}$ \cite{Benoist2019,Benoist2021}. If the integrated Kraus elements are parametrized with coordinates $x^\mu$ i.e. $\Omega_{x^\mu_t}$, then these trajectories---as one ``flies through'' the measurement record---are isomorphic to trajectories $x^\mu_t$ within a manifold which is $\mathrm{IWH}^{\textbf{12}}$ for linear diffusive optical measurements (i.e. those where the Lindblad operators in $\vec{L}$ are linear in the field quadratures) with arbitrary squeezing and rotating. 

These are precisely the Kraus trajectories of interest to us in this paper, as they give a visualizable sense of the geometry of such optical measurement processes within $\mathrm{IWH}^{\textbf{12}}$ independent of any particular choice of quantum state. Each measurement process is defined by its Lindblad operators $\vec{L}$ and any unitary feedforward. As we detail in this paper, these uniquely specify for each measurement process a right-invariant forward generator $\RInv{\updelta_t}$ on $\mathrm{IWH}^{\textbf{12}}$. To calculate the time-evolving coordinates $x^\mu_t$ of the Kraus trajectory, we must evaluate $\boldsymbol{d}x^\mu (\RInv{\updelta_t})$ which will define our stochastic, time-evolving parameters---that is, we must project a right-invariant forward generator $\RInv{\updelta_t}$ onto a coordinate basis via the one forms $\boldsymbol{d}x^\mu$. We detail this process in the sections to come.

For measurements of the electromagnetic field, diffusive processes include measurement schemes for single field quadratures (or ``homodynes'') and two non-commuting field quadratures (or ``heterodynes''). They also provide natural unravelings of Lindblad decoherence channels in terms of continuous phase space quadratures as alternatives to e.g. the method of quantum jumps \cite{Dalibard1992,Carmichael1993}; emphatically,  Lindblad operators are not jump operators for diffusive processes. In this way, the IMP provides a stepping stone towards quantitative, state-independent descriptions of open quantum systems.

\begin{figure}[t!]
\centering
\includegraphics[width=\linewidth]{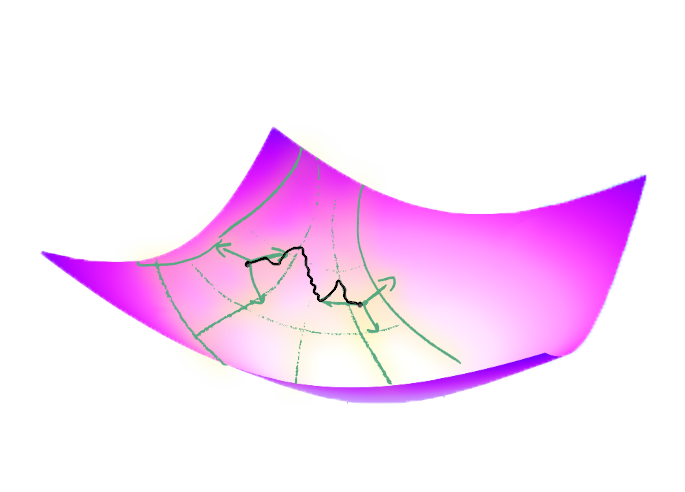}
\caption{\label{fig:manifoldwithtrajectory} Each Kraus operator trajectory corresponds to a path through $\mathrm{IWH}^{\textbf{12}}$, which is in some dimensions deterministic and in others stochastic, depending on the measurement. For all measurements, the ruler coordinate evolves deterministically with time. The right-invariant frame that co-propagates with the trajectory as the Kraus operator autonomously navigates $\mathrm{IWH}^{\textbf{12}}$ co-evolves in the coordinate frame during the trajectory due to the non-commutative group structure (geometrically, the curved manifold).  Ignoring the post-measurement unitary (for now), the structure of $\mathrm{IWH}^{\textbf{12}}$ is a stack of generalized phase spaces, with the ruler coordinate determining the layer of the stack. Collapsing the stack results in the trajectory in the coordinates of $\mathrm{IWH}^{\textbf{12}}$. The traditional phase space of quantum optics lives on the top of this stack. Including the post-measurement unitary doubles the generalized phase spaces, so that two exist in parallel within each layer of the stack. This doubled-structure is part of what motivates the reconceptualization of $\mathrm{IWH}^{\textbf{12}}$ as a fiber bundle and the subsequent discussion of the POV Manifold in Section IV.}
\end{figure}

When a diffusive measurement admits a description in terms of an IG manifold, Eq. \ref{dKraus} can be recast in terms of group elements, a Kraus operator density (KOD), and an unraveling in terms of a diffusing ensemble of Kraus operator trajectories on a manifold (Fig. \ref{fig:manifoldwithtrajectory}). For single- and double-quadrature measurements of the electromagnetic field, including (possibly adaptive feedforward) displacement and squeezing, this group is $\mathrm{IWH}^{\textbf{12}}$. Notably, this group includes measurement operators that are non-trace-class, yet can still be seen to purify to rank-$1$ projectors in the quantum representation.  This appears to be another innovation produced by the geometric methods of this paper: a notion of Hilbert-space-independent purification which could be amenable to the non-trace-class issues that appear in the theory of von Neumann algebras \cite{Witten2018}.\\

This paper is organized as follows. In section II, we introduce the reader to the Instrumental Weyl-Heisenberg group manifold, its decomposition in Harish-Chandra-like coordinates, and its native universal action: the right-invariant derivative.
In section IIC, advanced readers are invited to appreciate the other coordinate systems (Cartan-Berger and our here-introduced Iwasawa), and we rigorously define both Hilbert space purity and a generalized, geometric purity.
In section IIIA-D, we present the right-invariant forward generators for the various 'dyne measurements of quantum optics, including our new slomodyne and three adaptive 'dyne measurements (``adaptidynes'') approximating the phase POVM, analyzing their behavior and performance.
In section IIIE, advanced readers are introduced to our trick to manipulate the ostensible distribution to sample regions of phase space far from the origin.
In section IV, we briefly preview an important geometric results: the POV base manifold, over which the post-measurement unitaries form the fiber.
In section V we briefly discuss the decomposition of the Born rule into a dynamical, state-independent component, and a static, geometric, and state-dependent component, and the implications of this result.
We conclude with section VI.



\section{The Theory of Kraus Trajectories}

\subsection{First Approach: Coordinatizing $\mathrm{IWH}^{\textbf{12}}$ with a Harish-Chandra-like decomposition}

In Eq. \ref{dKraus}, multiple paths through the measurement record can result in the same Kraus operator (albeit, with different weights). This motivated the development by Wiseman of the ostensible distribution \cite{Wiseman1996} $D_t(g)$, here, the KOD.\footnote{Formally, the KOD lives in an involutive Banach space. In this way, the IMP provides an involutive Banach algebra which can be followed all the way to Hilbert space \cite{dixmier1977calgebras}.} Given a group decomposition $G$, a diffusively evolving Kraus element can be rewritten

\bea\label{dGroup}
d\Omega_g(t) = \sqrt{d_{\mathrm L} g D_t(g)}K_g
\eea where $K_g$ is the representation of group element $g\in G$ and $d_{\mathrm L} g$ the left-invariant (because Kraus operators pile up right to left) Haar measure over the group elements.

One of the ways to study the KOD $D_t(g)$ is through its unraveling in terms of Kraus operator trajectories.\footnote{The KOD is formally defined through its Wiener path integral: $d_{\mathrm L} g D_t(g) = \int_{dW_{[0,t) \in \gamma^{-1} (g)}} \mathcal{D}_\mu [dW_{[0,t)}]$. Here $\gamma(dW_{[0,t)})  = g$ is the time-ordered exponential from the space of Wiener paths to the group elements. In many cases the KOD also admits elegant solutions via diffusion equations, studied elsewhere \cite{Jackson2023,Jackson2025}.} For any functional of group elements $f(g)$ (e.g. $f(g)={\rm Tr}[K_g^\dagger K_g \rho]$ i.e. proportional to the Born rule), we can estimate

\bea\label{KrausEstimate}
\int d_{\mathrm L} g D_t (g) f(g) \sim \frac{1}{N} \sum_{i=1}^N f(g_t^{(i)})
\eea with $g_t^{(i)}\in G$ a time-evolved group element corresponding to a (simulated or real) sample path $dW_{[0,t)}$.

Having motivated and defined our Kraus trajectories, we narrow our focus explicitly to a class of measurements of the electromagnetic field that are diffusive (i.e. measurements resulting from iterated stochastic processes) and linear (i.e. with Lindblad operators $\vec{L}$ linear in the field quadrature operators) of a single mode of the electromagnetic field. For these, the group of Kraus elements closes and is denoted $\mathrm{IWH}^{\textbf{12}}$ \cite{Warszawski2020}. Intuitively, this closure is because the commutator of two operators of degree $d_1$ and $d_2$ will have degree $d_1 + d_2 - 2$, so for $d_{1/2} \leq 2,$ the Lie algebra is guaranteed to close. This enables a universal covering group, as shown in Ref.~\cite{Jackson2023} via the Harish-Chandra-like decomposition of a group element:

\bea\label{HCDef}
g = e^{1z}e^{a^\dagger \nu} e^{-A^\dagger \omega}e^{-H_0 \rho} e^{-A\zeta^*}e^{a\mu^*}
\eea where the single mode creation and annihilation operators satisfy the canonical bosonic commutation relation $[a,a^\dagger]=1$, $A=\frac{a^2}{2}$, $H_0 = a^\dagger a + \frac{1}{2}$, and $\{z,\nu,\omega,\rho,\zeta,\mu\}$ are $6$ complex coordinates covering a $12$ (real) dimensional manifold. As emphasized by Knapp \cite{Knapp2002}, the HC decomposition is natural due to its algebraic structure, which is reflected in the (relative) simplicity of the stochastic differential equations (SDEs) associated with the evolution of HC coordinates (e.g. Eqs.~\ref{FHetSDEs:zeta}-\ref{FHetSDEs:z}); $a$ and $A$ are both nilpotent. However, the non-unitarity of the operators on the left of the ruler term ($e^{-H_0 \rho})$ obscures the form of the POVM element's group representation $\pi(g) = g ^\dagger g$ (Cartan conjugation) where  we note $K^\dagger_g K_g = K_{g^\dagger g} = K_{\pi(g)}$ by construction (that is, $K_g$ is a dagger representation of the instrumental group \cite{Jackson2023,JacksonCaves2023Quantum}). This obfuscation is removed by the move to Iwasawa or Cartan-Berger coordinates, as detailed in the next section.

For all coordinates, the natural universal group action is the right invariant derivative (RID), defined for a function $f$ evaluated at group element $g$:

\bea\label{RIDDef}
\RInv{X}(f)[g] \equiv \lim_{\epsilon\rightarrow 0} \frac{f(e^{\epsilon X}g) - f(g)}{\epsilon}
\eea where $X$ is understood to be a generator of the Lie group.

As each Kraus operator is left-multipled by the differential stochastic Kraus operator in Eq. \ref{dKraus}, the associated group element $g$ is multiplied by $e^{\updelta_t}$ where we define the SDE infinitesimal forward generator $\updelta_t = X_\gamma \epsilon_t^\gamma$ (note the Einstein summation notation). Here $X_\gamma$ are the generators of the group (e.g. $1,a^\pm,A^\pm,H_0$ for HC) and $\epsilon_t^\gamma$ is an infinitesimal vector of coefficients, some proportional to $dW_t$ (stochastic), some proportional to $dt$ (ballistic).

We now perform a conceptual leap. We identify for every element of the Lie algebra $X\in\mathfrak{g}$ a vector field in the tangent bundle of the Lie group: the right-invariant vector field that moves the group element according to $\epsilon_t^\gamma$. Thus, we may rewrite the exponentiated infinitesimal forward generator as a right-invariant vector field

\bea\label{RInvDelta}
e^{\RInv{\updelta_t}} = e^{\RInv{X_\gamma} \epsilon_t^\gamma}. \eea This object, which uniquely defines each incremental stochastic measurement process, naturally generates right-invariant motion within the group. However, this motion must be coordinatized in order to be amenable to analysis. Since the chain rule is basis invariant, we may write the gradient of a function $df = dx^\mu \partial_\mu [f] = \theta_R^\gamma \RInv{X}_\gamma [f]$, with $\theta_R^\gamma$ the right-invariant one form. To generate SDEs, all that remains to be done is to expand the coordinate one forms in terms of the right-invariant one forms $\boldsymbol{d}x^\mu = W^\mu_\gamma \theta_R^\gamma$ (with $W^\mu_\gamma$ the vielbein\footnote{We note that, while the contemporary mathematics literature refers to the transformation from right-invariant derivatives to coordinate derivatives as a vielbein, this is in direct conflict with Cartan's notion of a triad from which the term is derived; at the level of right-invariant derivatives to coordinate derivatives alone, there is no notion of parallel transport. A proper vielbein will appear in our discussion of the POV manifold in a later publication.}), and evaluate these on the right-invariant infinitesimal forward generator. These define stochastic variables

\bea\label{StochasticEqGenerator}
dx_t^\mu\equiv\boldsymbol{d}x^\mu (\RInv{\updelta_t}) 
\eea which obey SDEs. This process for using the IMP to generate SDEs is detailed elsewhere \cite{Jackson2023,JacksonCaves2023Quantum}, and is worked out for our measurements of interest in section III. The end result is that, once a measurement characterized by $\RInv{\updelta_t}$ is specified and a coordinate choice is made (either Harish-Chandra-like as in Eq. \ref{HCDef}, Cartan-Berger as in Eq. \ref{CartanCoords}, or Iwasawa as in Eq. \ref{IwasawaDef}), we obtain associated SDEs that completely describes the measurement process.

Once these SDEs are derived, they may be solved to sample Kraus trajectories, with some of the coordinates evolving purely ballistically depending on only $dt$ (deterministic, as in the ruler coordinate), and other depending also on the Wiener increments $dW_t$ (stochastic). Since the Kraus operators are defined completely---though not uniquely\footnote{We can use a different ostensible distributions than the KOD to sample other regions of phase space, as we detail in the next section.}---by their representation and the KOD, we can use the behavior of their stochastically-evolved representations to study the Kraus operators defined in Eq. \ref{dGroup} themselves through their action on an unspecified quantum state i.e. their behavior as the conditional channel $d\Omega_g(t) \odot d\Omega_g(t)^\dagger$.\footnote{For those unfamiliar with this notation, the $\odot$ indicates where a density matrix goes upon evaluation \cite{CavesNotes2017}.} We note that such a channel is not normalization-preserving as it is \emph{selective}, corresponding to a specific path through a measurement record as illustrated in Fig. \ref{fig:manifoldwithtrajectory}.

We can approximate a channel---selective or unconditional---using Eq. \ref{KrausEstimate} with $f(g)\rightarrow   K_{g}\odot K_g^\dagger$, such that

\bea\label{dGroupEstimate}
\int d\Omega_g(t) \odot d\Omega_g(t)^\dagger \sim \frac{1}{N} \sum_{i=1}^N K_{g^{(i)}_t} \odot K_{g^{(i)}_t}^\dagger
\eea Thus, each of our simulated Kraus trajectories $K_{g^{(i)}_t}$ where the group element $g^{(i)}_t$ coincides with our group element of interest $g$ informs us about the possible measurement records that led to that outcome at time $t$.

We emphasize that while we use the language of estimators, the right hand side of Eq. \ref{dGroupEstimate} is ostensible; it depends on how we sample phase space. Simply put, an instrument is not a distribution, but the capacity to produce distributions. Thus, care must be taken in interpreting the RHS of Eq. \ref{dGroupEstimate} as a true estimate of the LHS, as it only estimates the channel on a small region of phase space, as we discuss more in section IIIE.

\subsection{Advanced Approach: Cartan-Berger and Iwasawa Coordinates, POVM Purities}

Harish-Chandra-like coordinates are natural due to their algebraic simplicity, and produce the simplest SDEs to solve. However, the obfuscation of the POVM structure is inconvenient and inelegant from a geometric perspective. This is resolved by moving to Cartan-Berger or Iwasawa coordinates.
Whereas the Harish-Chandra-like coordinates (which we use throughout this paper to generate our trajectories) correspond to normal ordering of Kraus operators ($K\propto\bar{N}P N$, with $P$ positive and $N$ nilpotent, and $\bar{N}$ conjugate nilpotent\footnote{Technically, infinite nilpotency is ill defined, but we sweep this technicality under the rug. The point is, the algebra of the Harish-Chandra-like decomposition is nice because the algebra of $a$ and $a^\dagger$ is nice.}), the Cartan-Berger coordinates have as a representation the singular value decomposition ($K\propto U \Sigma V^\dagger$). Iwasawa coordinates interpolate between the two: the square root of the POVM element $\pi$ is normal ordered, but the leftmost post-measurement unitary is left explicit i.e. $K=U':\sqrt{\pi}:\propto U' P N$.\footnote{For those unfamiliar with the notation, $:A:$ indicates normal ordering of an operator $A$ \cite{Glauber1963coherence}.}

Maximum elegance is achieved with the fully geometric Cartan-Berger decomposition \cite{Helgason1978} of the group element, where both pre-measurement and post-measurement unitaries are manifest:

\bea\label{CartanCoords}
g = e^{1z_{CB}} D_\beta S_\eta e^{-iH_0 \psi'} S_{\xi} e^{-H_0 a_{CB}'} S_{\xi}^{-1} D_{\alpha}^{-1}.
\eea Here the prime marks the Kraus-level ruler; we write $a_{CB}\equiv 2a_{CB}'$ for the ruler of the POVM element itself, $\pi(g)\propto e^{-H_0 a_{CB}}$. Importantly, for a trace-class group element, the POVM purity defined in Ref. \cite{vanEnk2017} is easiest calculated in Cartan coordinates:

\begin{figure}[t]
\centering
\includegraphics[width=\linewidth]{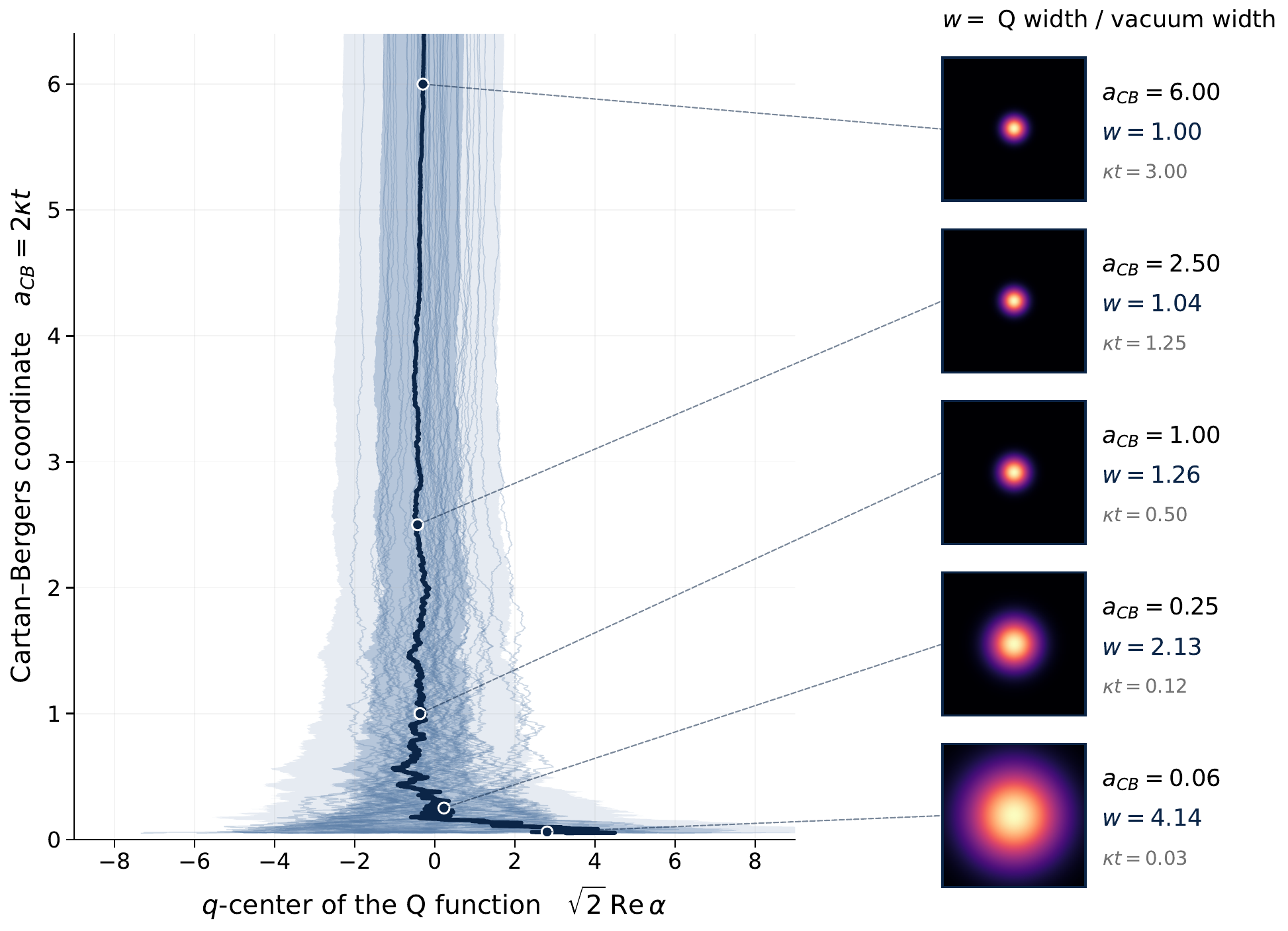}
\caption{\label{fig:aCBforHeterodyne} The Kraus trajectories for L\"{u}ders Heterodyne with associated POVM Q-functions, which are constructed analogously to Q-functions for states \cite{Husimi1940}. The width of the Q-function relative to the minimum uncertainty coherent state is $w=1/\sqrt{1-e^{-2a_{CB}'}}=1/\sqrt{1-e^{-a_{CB}}}$ with associated POVM purity ${\rm tanh}\, a_{CB}'$. As the measurement purifies, its stochastic motion disappears. We only plot $q$-quadrature behavior but the Physics is cylindrically symmetric. Although we only calculate using L\"{u}ders, the behavior of other Heterodynes are identical in terms of purity and displacement. The shaded regions correspond to the width (and twice the width) of the central, bolded trajectory whose behavior we follow as Q-functions. For the advanced reader: note that in the absence of POVM squeezing, all three coordinate systems have the same ruler, so that our geometric POVM purity defined in Eq. \ref{GeometricPOVMPurity} agrees identically with the Hilbert space POVM purity in Eq. \ref{POVMPurity} for heterodyne, since it contains no squeeze.}
\end{figure}

\begin{figure}[t]
\centering
\includegraphics[width=\linewidth]{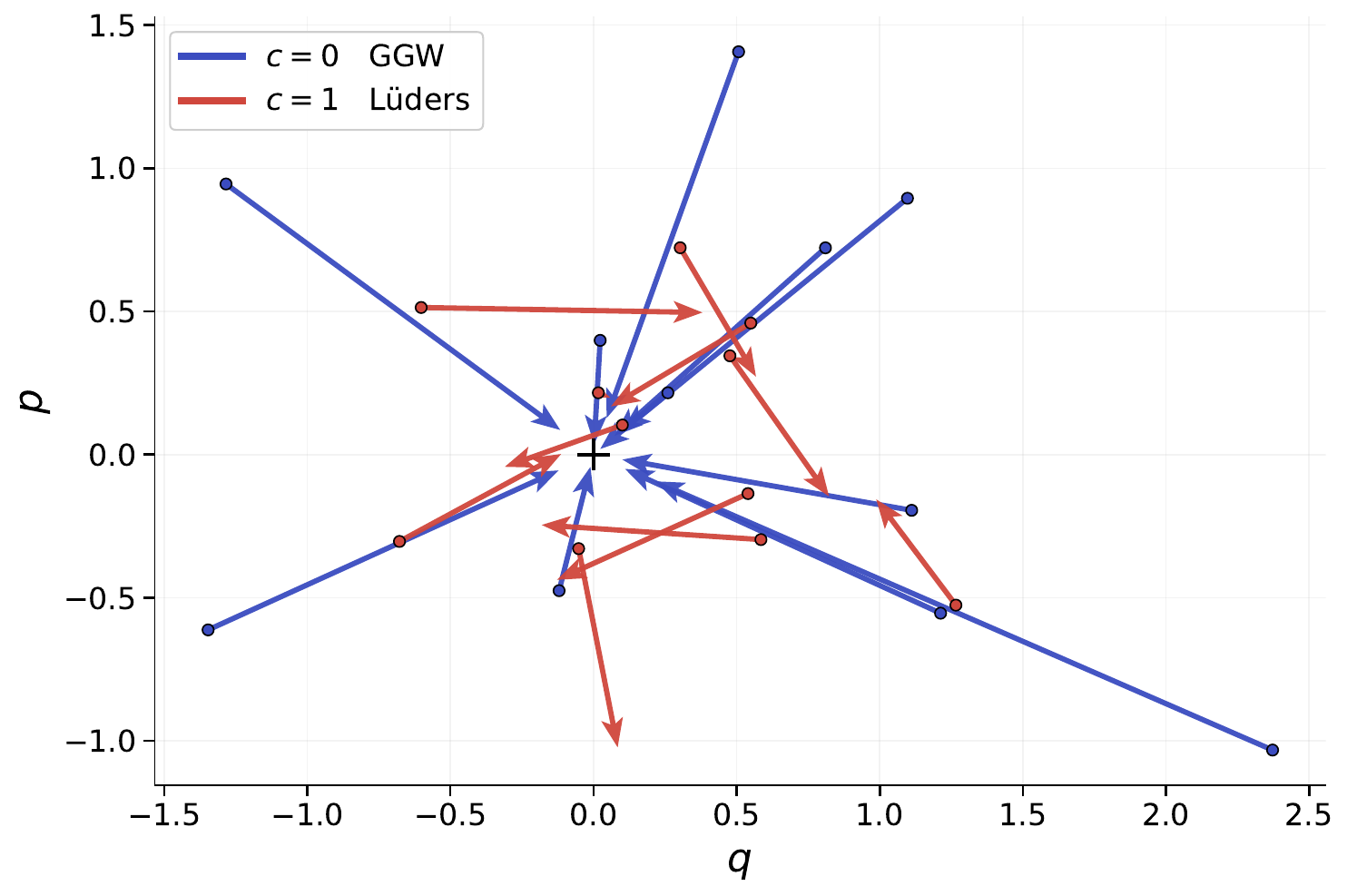}
\caption{\label{fig:kickforHeterodyne} Here the strength of the post-measurement unitary---here, just a displacement---is displayed for several trajectories, with standard phase space coordinates $p$ and $q$: each arrow runs from the center of the operator $K_g^\dagger K_g$\footnote{We note that $K_g^\dagger K_g$ is not a POVM element proper, but proportional to it; it is missing the Haar measure and KOD which, along with $K_g$, go into the full Kraus operator $\Omega_g$. Thus, $K_g^\dagger K_g$ is a representation of POVM group elements $\Omega_g^\dagger \Omega_g$.} ($\alpha$ in the Cartan-Berger coordinates of Eq. \ref{CartanCoords}) to the center of the operator $K_gK_g^\dagger$ ($\beta$), for the same 11 measurement records at $\kappa t=1.2$, computed from the SDEs of Eqs. \ref{FHetSDEs:zeta}-\ref{FHetSDEs:z}. For GGW ($c=0$) the post-measurement unitary pulls every state radially towards the origin, $\beta=e^{-\rho}\alpha$. For heterodyne individual L\"{u}ders step ($c=1$) carries no post-measurement unitary, but the accumulated Kraus operator does: its kicks are random, and of the same order as $|\alpha|$.}
\end{figure}

\bea\label{POVMPurity}
{\rm Pur}(\pi(g)) \equiv \frac{{\rm Tr} \,\pi(g)^2}{({\rm Tr} \,\pi(g))^2}= {\rm tanh} a_{CB}'.
\eea However, in this present work not all group elements of interest are trace class. Here, the Cartan-Berger ruler degenerates ($a_{CB}=0$ and ${\rm Tr}\,\pi(g)=\infty$), and Eq. \ref{POVMPurity} is ill-defined ($\infty/\infty$).

Thus, it is most preferable to work in a new coordinate system that interpolates between Harish-Chandra-like and Cartan-Berger, inspired by the work of Iwasawa \cite{Iwasawa1949,Knapp2002}:

\bea\label{IwasawaDef}
g = e^{1z_I} D_\gamma S_\omega e^{-H_0 \rho_{I}} e^{-A\chi^*}e^{a\lambda^*}
\eea where $D_\alpha = e^{\alpha a^\dagger - \alpha^* a}$, and the unitary squeeze operator written in stereographic coordinates $S_\xi =e^{-A^\dagger \eta + A \eta^*}$ where $|\xi| = {\rm tanh} \,|\eta|$ and ${\rm arg} \,\xi={\rm arg} \,\eta$ \footnote{Or explicitly, $S_\xi= e^{-A^\dagger \xi} e^{-H_0 {\rm ln} \frac{1}{\sqrt{1-|\xi|^2}}} e^{A\xi^*}$ such that $S_\eta S_\xi = S_{\frac{\eta +\xi}{1+\eta^*\xi}} e^{\frac{iH_0}{2} {\rm arg}\, \frac{1+\eta \xi^*}{1+\eta^* \xi}} $ i.e. the holonomy of the hyperboloid.}. To recover Harish-Chandra-like coordinates from Iwasawa coordinates, one normal orders the left unitary and commutes through to the positive term. Coordinate transformations between Harish-Chandra-like, Iwasawa, and Cartan-Bergers coordinates are presented briefly in Appendix \ref{app:coordtrans}.

Examining Eq. \ref{IwasawaDef}, we can clearly see the six-dimensional structure of the manifold of POVM elements (the POV manifold); in Iwasawa coordinates, the left unitary represents the post-measurement unitary, which lives in the fiber over this manifold, and the POVM elements are written simply

\bea\label{IwasawaPOVM}
\pi(g) = e^{2 {\rm Re}\,z_I} e^{a^\dagger\lambda}e^{-A^\dagger \chi}e^{-2H_0 {\rm Re}\,\rho_{I}} e^{-A\chi^*}e^{a\lambda^*}.
\eea

Thus, the real part of the Iwasawa ruler coordinate ${\rm Re}\,\rho_I$ will function as a measure of a generalized, geometric POVM purity, defined as

\bea\label{GeometricPOVMPurity}
{\rm GPur}(\pi(g)) \equiv {\rm tanh}\, {\rm Re}\,[\rho_I].
\eea For a POVM element without squeeze ($\chi=0$), Eq. \ref{IwasawaPOVM} is a displaced thermal operator and Eq. \ref{GeometricPOVMPurity} reduces exactly to the Hilbert space purity of Eq. \ref{POVMPurity}. We use this to describe geometrically the purification of homodyne in the next section, despite the associated Kraus operators forming a type-III von Neumann algebra i.e. not having finite trace.

\section{Nonadaptive 'Dynes}

Along with photon counting,\footnote{Photon counting is a non-diffusive measurement and treated from the view of the IMP elsewhere \cite{Jackson2022photodetector}. For photon counting, fundamental limits to performance are enabled by a state-independent POVM-level analysis \cite{ProppNet,Propp2019,ProppPOVM}.} 'dyne measurements form the backbone of quantum optics experiments. Homodyne and heterodyne inherit their names from the classical radio-engineering techniques; by mixing incoming light with either a single frequency local oscillator (homo), or two local oscillators with distinct frequencies (hetero), one is able to extract phase and amplitude information about the incoming electromagnetic field by measuring variations in the intensity at the output. In the quantum optical case there are multiple instantiations of both homodyne and heterodyne, as well as their variations: general-dyne (heterodyne with unbalanced transmission to both detectors) and the here-introduced slomodyne (effectively, homodyne with a variable-rate rotation). In this section, we analyze the 'dyne measurements in terms of their associated Kraus trajectories.

Having established the connection between evolving Kraus operator representations and the Kraus operators themselves in the prior section, we now focus on the stochastic trajectories traced out in the measurement record, each corresponding to a representation of a time-evolving Kraus operator. While it is tempting to consider the operators as matrices, we emphasize that for some of our measurements of interest this leads to pathological issues; we instead emphasize the abstract picture on generalized phase space: the representations of Kraus operators are elements of a finite Lie group i.e. the $12$-dimensional Instrumental Weyl-Heisenberg group.

\subsection{Heterodynes}

We begin with a treatment of heterodyne. Despite requiring two detectors physically, it is mathematically the simplest measurement to analyze, and nicely serves as a warm up to illustrate the techniques that we use in evaluating all our here-presented measurement processes.

Heterodyne is a quantum optical measurement involving two local oscillators performed to project onto coherent states. There are several, nonequivalent ways to do this (e.g. simultaneously measuring two field quadratures \cite{ArthursKelly1965}, measuring the annihilation operator \cite{WisemanMilburn2010}), depending on the system-meter interaction, giving rise to L\"{u}ders \cite{Luders1951} and non-L\"{u}ders infinitesimal instruments; that is, Kraus operators without a post-measurement unitary. For a integrated measurement to be L\"{u}ders, the full Kraus operators should be the square roots of POVM elements and similarly for the associated representations $g=\sqrt{\pi}$. As we will now show, L\"{u}ders and non-L\"{u}ders heterodyne are nearly identical in performance and neither measurement when integrated over time remain L\"{u}ders. Interestingly, none of this will be the case for homodyne, as explore in IIIB.


As detailed in Ref. \cite{Jackson2023}, the simultaneous measurement of two non-commuting field quadratures at equal rates $\kappa$ gives rise to a forward generator

\bea\label{updeltaHeterodyneL}
\updelta_t^{\text{Lud-Het}}
&= -2H_0 \kappa dt + Q \sqrt{\kappa} dW^q_t + P\sqrt{\kappa}dW^p_t\\
&= -2H_0 \kappa dt + a \sqrt{\kappa}dZ_t^* + a^\dagger \sqrt{\kappa}dZ_t
\eea whose associated infinitesimal Kraus elements are L\"{u}ders. Similarly, the heterodyne that is performed in practice requires measuring the annihilation operator, as developed by Goetsch, Graham, and Wiseman (GGW) \cite{WisemanMilburn1993,GoetschGraham1994,WisemanMilburn2010}. This gives a forward generator

\bea\label{updeltaHeterodyneGGW}
\updelta_t^{\text{GGW}}
&= -2 a^\dagger a \kappa dt + 2a\sqrt{\kappa} dZ_t^*\\
&= -2 H_0 \kappa dt + \kappa dt + 2a\sqrt{\kappa} dZ_t^* .
\eea

Observe that Eqs. \ref{updeltaHeterodyneL} and \ref{updeltaHeterodyneGGW} differ by an overall rescaling and a unitary displacement. Thus we can construct a flexible heterodyne

\bea\label{updeltaHeterodyneF}
\updelta_t^{\text{f-het}}
&= \updelta_t^{\text{GGW}} + c  (a^\dagger \sqrt{\kappa} dZ_t - a \sqrt{\kappa} dZ_t^* - \kappa dt),
\eea
corresponding to the two Lindblad operators
\bea
    L_q &= Q + iP (1-c)\\
    &= \sqrt2 a - iP c
\eea
and
\bea
    L_p &= P - iQ (1-c)\\
    &= -i\sqrt2 a + iQ c
\eea
which recover GGW for $c=0$ and L\"{u}ders for $c=1$.
We can construct SDEs for the flexible heterodyne in Eq. \ref{updeltaHeterodyneF} by following a standard procedure of the IMP, which we go through in detail here.

First, we promote $\updelta_t^{\text{f-het}}$ to the right-invariant forward generator

\begin{widetext}
\bea\label{RIDupdeltaHeterodyneF}
\RInv{\updelta_t}^{\text{f-het}}
= \kappa dt\left(-2 \RInv{H_0} + (1-c)\RInv{1}\right)
+ (2-c) \sqrt{\kappa} dZ_t^* \RInv{a}
+ c \sqrt{\kappa} dZ_t \RInv{\,a^\dagger}.
\eea
\end{widetext} We emphasize the inclusion of the scalar direction $\RInv{1}$ corresponding to changes in the real part of the Harish-Chandra-like coordinate $z$.

In principle, all that now remains is to use Eq. \ref{StochasticEqGenerator} to calculate the SDEs for each coordinate. However, doing this requires first rewriting the coordinate derivatives $\partial_\mu$ and one forms $\mathbf{d}x^\mu$ in terms of right-invariant derivatives $\RInv{X}_\gamma$ and $\boldsymbol{\theta}_R^\gamma$. This is done by taking right invariant derivatives as defined in Eq. \ref{RIDDef} of the group elements defined in Eq. \ref{HCDef}, and solving the resulting system of equations for the complexified right-invariant one-forms. We thus calculate 

\bea \label{WierbienHC:1}
\boldsymbol{\theta}_\mathrm{R}^1 + i \boldsymbol{\theta}_\mathrm{R}^{i1}  &=& \mathbf{d}z - \nu e^{\rho}\,\mathbf{d}\mu^* - \tfrac{1}{2}\nu^2 e^{2\rho}\,\mathbf{d}\zeta^* \\ \label{WierbienHC:adagger}
\boldsymbol{\theta}_\mathrm{R}^{a^\dagger} + i\boldsymbol{\theta}_\mathrm{R}^{ia^\dagger} &=& \mathbf{d}\nu + \nu\,\mathbf{d}\rho + \omega e^{\rho}\,\mathbf{d}\mu^* + \nu\omega e^{2\rho}\,\mathbf{d}\zeta^* \,\,\\ \label{WierbienHC:a}
\boldsymbol{\theta}_\mathrm{R}^{a} - i\boldsymbol{\theta}_\mathrm{R}^{-ia} &=& e^{\rho}\,\mathbf{d}\mu^* + \nu e^{2\rho}\,\mathbf{d}\zeta^* \\ \label{WierbienHC:ADagger}
\boldsymbol{\theta}_\mathrm{R}^{A^\dagger} + i\boldsymbol{\theta}_\mathrm{R}^{iA^\dagger} &=& -\mathbf{d}\omega - 2\omega\,\mathbf{d}\rho - \omega^2 e^{2\rho}\,\mathbf{d}\zeta^* \\ \label{WierbienHC:H0}
\boldsymbol{\theta}_\mathrm{R}^{H_o} + i\boldsymbol{\theta}_\mathrm{R}^{iH_o} &=& -\mathbf{d}\rho - \omega e^{2\rho}\,\mathbf{d}\zeta^* \\ \label{WierbienHC:A}
\boldsymbol{\theta}_\mathrm{R}^{A} - i\boldsymbol{\theta}_\mathrm{R}^{-iA} &=& -e^{2\rho}\,\mathbf{d}\zeta^*.
\eea

Thus, we identify for Harish-Chandra-like coordinates the complex vielbein\footnote{Technically, the IG is a principal bundle over the 6-dimensional POV base manifold and it is in the base that there is a 6-bein.  The remaining 6 dimensions along the principal fibers solder the right-invariant data onto the POV manifold. But for the purposes of this paper, it is sufficient to treat Eqs. \ref{WierbienHC:1}-\ref{WierbienHC:A} as a stand-alone $12$-bein.} coefficients ${T^\gamma}_\mu$ relating the complex right-invariant one-forms to the complex coordinate one-forms
\bea
    \boldsymbol{\theta}_\mathrm{R}^\gamma = {T^\gamma}_\mu \mathbf{d}x^\mu,
\eea
with the tilde defining complexified one-forms respecting the sign conventions in Eqs. \ref{WierbienHC:a} and \ref{WierbienHC:A}, i.e.
\bea
    \tilde{\boldsymbol{\theta}}_\mathrm{R}^{A} &\equiv \boldsymbol{\theta}_\mathrm{R}^{A} - i\boldsymbol{\theta}_\mathrm{R}^{-iA},\\
    \tilde{\boldsymbol{\theta}}_\mathrm{R}^{A^\dagger} &\equiv \boldsymbol{\theta}_\mathrm{R}^{A^\dagger} + i\boldsymbol{\theta}_\mathrm{R}^{iA^\dagger},
\eea
and similarly for $a$ and $a^\dagger$.\footnote{Note that we were not guaranteed complex analyticity \emph{ab initio}, hence treating the real and imaginary parts separately until this point.}

With the complexified right-invariant one-forms, the coordinate one-forms can be solved in terms of them to give the inverse expressions
\bea \label{WierbienHCInv:zeta}
\mathbf{d}\zeta^* &=& -e^{-2\rho} \tilde{\boldsymbol{\theta}}_\mathrm{R}^A
 \\ \label{WierbienHCInv:rho}
\mathbf{d}\rho &=& -\tilde{\boldsymbol{\theta}}_\mathrm{R}^{H_0} + \omega \tilde{\boldsymbol{\theta}}_\mathrm{R}^A \\ \label{WierbienHCInv:omega}
\mathbf{d}\omega &=& -\tilde{\boldsymbol{\theta}}_\mathrm{R}^{A^\dagger} + 2\omega \tilde{\boldsymbol{\theta}}_\mathrm{R}^{H_0} - \omega^2 \tilde{\boldsymbol{\theta}}_\mathrm{R}^{A}  \\ \label{WierbienHCInv:mu}
\mathbf{d}\mu^* &=& e^{-\rho} (\tilde{\boldsymbol{\theta}}_\mathrm{R}^a + \nu \tilde{\boldsymbol{\theta}}_\mathrm{R}^A) \\ \label{WierbienHCInv:nu}
\mathbf{d}\nu &=& \tilde{\boldsymbol{\theta}}_\mathrm{R}^{a^\dagger} + \nu\tilde{\boldsymbol{\theta}}_\mathrm{R}^{H_0} - \omega \tilde{\boldsymbol{\theta}}_\mathrm{R}^{a} - \nu \omega \tilde{\boldsymbol{\theta}}_\mathrm{R}^{A}  \\ \label{WierbienHCInv:z}
\mathbf{d}z &=& \tilde{\boldsymbol{\theta}}_\mathrm{R}^{1} + \nu\tilde{\boldsymbol{\theta}}_\mathrm{R}^{a} + \frac{1}{2}\nu^2\tilde{\boldsymbol{\theta}}_\mathrm{R}^{A}. \\ \nonumber
\eea

Emphatically, Eqs. \ref{WierbienHC:1}- \ref{WierbienHC:A} and \ref{WierbienHCInv:zeta}-\ref{WierbienHCInv:z} will hold for every measurement we consider in this paper.

This allows us to rewrite coordinate one-forms in terms of right-invariant one forms multiplied by functions of the coordinates, which will in the next step be reinterpreted as stochastic parameters to derive our SDEs.

With the hard work done, it is now straightforward to generate SDEs for our measurements of interest, here the flexible heterodyne, using the definition of our time-evolving stochastic differential in terms of the one form given in Eq. \ref{StochasticEqGenerator}. Since $\RInv{\updelta_t}$ is a stochastic forward generator, application of it to a stochastic vector requires care; when $\mathbf{d}x^\mu$ is rewritten in terms of right-invariant one forms and functions of coordinates using Eqs. \ref{WierbienHCInv:zeta}-\ref{WierbienHCInv:z}, a Stratonovich product must be inserted between the coordinates and the stochastic terms in $\RInv{\updelta_t}$ \cite{Gardiner2009,Chirikjian2009,Jackson2026Notes}. This is defined
\bea\label{stratonovich}
dx_t^\mu \circledS f(x_t^\nu)\equiv dx_t^\mu f(x_t^\nu) + \frac{1}{2} dx_t^\mu df(x_t^\nu).
\eea\\
We note that Eq. \ref{stratonovich} is symmetric i.e. $dx_t^\mu \circledS f(x^\nu)=f(x_t^\nu)\circledS  dx_t^\mu $.

Thus, we can calculate SDEs for the flexible heterodyne using the defining relationship between right invariant one forms $\boldsymbol{\theta}_\mathrm{R}^\gamma$ and derivatives $\RInv{X}_\gamma$ from differential geometry i.e. that they are orthogonal $\boldsymbol{\theta}_\mathrm{R}^\gamma (\RInv{X}_\nu )= \delta_\nu^\gamma$. In this way, each right-invariant one form in $\mathbf{d}x^\mu$ will pick out the (stochastic) coefficient of a right-invariant one form in the forward generator $\RInv{\updelta_t}$.

Proceeding in this manner, we derive the full set of SDEs for flexible heterodyne in Harish-Chandra-like coordinates:

\begin{widetext}
\begin{align}\label{FHetSDEs:zeta}
   d\zeta_t^* &= 0 \\\label{FHetSDEs:rho}
    d\rho_t &= 2\kappa dt \\\label{FHetSDEs:omega}
    d\omega_t &= -4\omega_t\kappa dt \\\label{FHetSDEs:mu}
   d\mu_t^* &= e^{-\rho_t} \circledS (2-c) \sqrt{\kappa} dZ^*_t
   &= e^{-\rho_t} (2-c) \sqrt{\kappa} dZ^*_t\\\label{FHetSDEs:nu}
    d\nu_t &= c \sqrt{\kappa} dZ_t -2\nu_t\kappa dt - \omega_t\circledS(2-c) \sqrt{\kappa}\, dZ_t^*
    &=  c \sqrt{\kappa} dZ_t -2\nu_t\kappa dt -\omega_t(2-c) \sqrt{\kappa}\, dZ_t^*\\\label{FHetSDEs:z}
    dz_t &= (1-c) \kappa dt + \nu_t \circledS (2-c) \sqrt{\kappa} dZ_t^*
    &=  \Big(1-\tfrac{1}{2}c^2\Big) \kappa dt + \nu_t (2-c) \sqrt{\kappa} dZ_t^*
\end{align}
\end{widetext} where in the last line we have used the Stratonovich product and evaluated $dZ_t\,dZ_t^*=dt$ and $dZ_t^2=0$. Observe that both $\mu_t$ and $\nu_t$ evolve stochastically in the complex plane, while the ruler $\rho_t$ is purely real and ballistic, admitting the simple analytic solution $\rho_t=2\kappa t$. Similarly, $\omega_t$ has the simple solution that it is identically zero; this is because it, like all other parameters, starts at zero as an initial condition. Since $\zeta_t$ is also identically zero, we observe there is no squeeze in heterodyne. Observing that $\nu_t$ evolves stochastically, care must be taken in Eq. \ref{FHetSDEs:z} when evaluating the Stratonovich product. 

Starting from the identity (all coordinates zero at $t=0$), these SDEs integrate to the filter functions, i.e. the coordinates as explicit functionals of the measurement record $dZ_{[0,t)}$:
\begin{widetext}
\bea
\zeta^*_t=\omega_t &=& 0 \\ \label{FHetFilt:rho}
\rho_t &=& 2\kappa t \\\label{FHetFilt:mu}
\mu_t &=& (2-c)\sqrt{\kappa}\int_0^t e^{-2\kappa s}\,dZ_s \\ \label{FHetFilt:nu}
\nu_t &=& c\sqrt{\kappa}\int_0^t e^{-2\kappa(t-s)}\,dZ_s \\ \label{FHetFilt:z}
z_t &=& \Bigl(1-\tfrac12 c^2\Bigr)\kappa t + c(2-c)\,\kappa\int_0^t\!\!\int_0^{s} e^{-2\kappa(s-s')}\,dZ_{s'}\,dZ^*_s .
\eea
\end{widetext}
Eqs. \ref{FHetFilt:mu}-\ref{FHetFilt:nu} make explicit that the POVM coordinate $\mu^*_t$ weights the \emph{earliest} part of the record most heavily, and the post-measurement coordinate $\nu_t$ the \emph{latest}. These filter functions are a special case of a general result that holds for every forward generator in this paper.

 Note that the ruler coordinate $\rho_t$ evolves deterministically at the same speed regardless of choice of heterodyne scheme i.e. independently of $c$. This manifests in Fig. \ref{fig:aCBforHeterodyne}, where we show the behavior of the Kraus trajectories, as all exhibiting the same purification. Indeed, the only physical difference between schemes is the rate at which they diffuse at very short times (Eq. \ref{FHetSDEs:mu}), and the strength of the post-measurement unitary (Eq. \ref{FHetSDEs:nu}). This can be seen in Fig. \ref{fig:kickforHeterodyne}. Notably, the post-measurement unitary does \emph{not} vanish for L\"{u}ders heterodyne: each L\"{u}ders step is Hermitian, but the steps do not commute ($[H_0,a]\neq0$), so the accumulated Kraus operator carries a post-measurement displacement (in Cartan-Berger coordinates) $\beta-\alpha=(\nu-\mu)/(1+e^{-\rho})$, with $\mu_t$ weighting early and $\nu_t$ late measurement noise. In this way, we can see that maintaining the L\"{u}dersianity of a measurement, in general, requires adaptivity;\footnote{In particular, choosing $c=1-\tanh\kappa t$ in Eq.~\ref{RIDupdeltaHeterodyneF} and adding a term to the forward generator $\frac{2\kappa}{1+e^{-2\kappa t}}\bigl(\nu_t\RInv{a}^\dagger-\nu_t^*\RInv{a}\bigr)dt$ results in a heterodyne that is truly L\"{u}dersian i.e. having no post-measurement unitary after piling up of the Kraus operators. Adaptive 'dyne measurements are treated in Section IIID.} a locally L\"{u}dersian diffusive measurement may iterate to a non-L\"{u}ders final measurement, in analogy to local and global non-Euclideanity (i.e. to be torsion-free). 
 
\begin{figure*}[t]
\centering
\includegraphics[width=\linewidth]{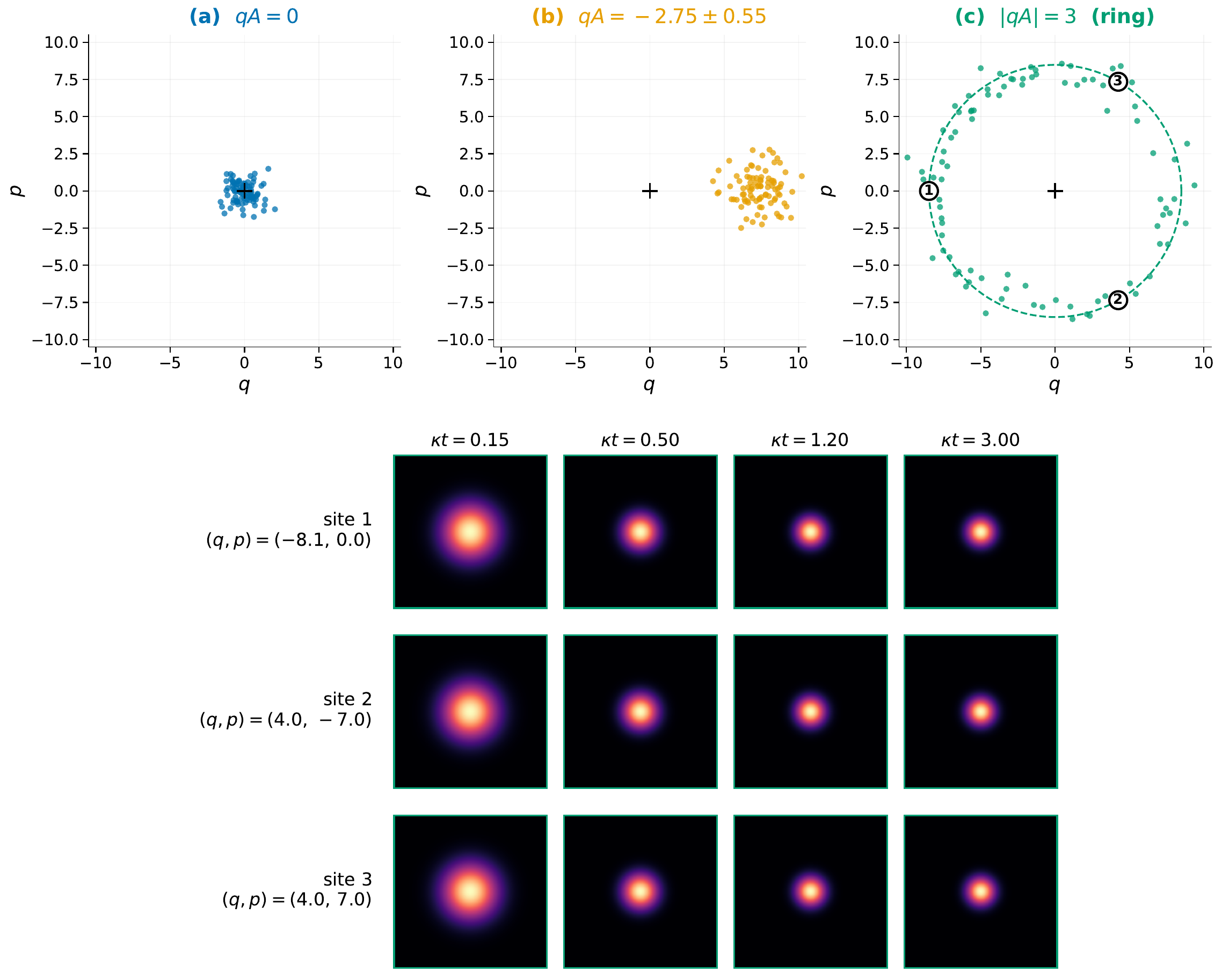}
\caption{\label{fig:samplingregionsforHetero} Here we illustrate our ability to sample various regions of phase space, as denoted in the three panels (top): (a) the origin, (b) a smeared-out Gaussian (here with width $0.55$) using a cluster of shift values, and (c) a ring. The behavior of the POVM Q-functions with time (bottom), shown for three sites on the ring of (c), is identical across location due to the symmetry of heterodyne; thermal states that cool to coherent states.}
\end{figure*}

 The rescaling of Kraus operators also differs as determined by $dz_t$, but at the level of individual trajectories this is not physically meaningful. Indeed, $dz_t$ also depends on a choice of gauge, that we have here implicitly chosen so as to  sample Kraus trajectories that originate near the origin. Mathematically, this is due to the Gaussian weighting inherent in the Gaussian measure $d\mu(dW_t)$. But the KOD is only one such ostensible distribution, and there is no reason \emph{a priori} to expect measurement outcomes near the origin in phase space, nor can we expect such outcomes to be representative of the whole possible set of measurement records (although for heterodyne, this is completely the case). To sample far from the origin in phase space requires a trick, which we detail in section IIIE. For heterodyne, the symmetry of the problem simplifies the sampling; not only is there a cylindrical symmetry, but the behavior of the Kraus trajectories are uniform (thermal states approaching coherent states) across the phase plane as seen in Fig. \ref{fig:samplingregionsforHetero}.

\begin{figure}[t!]
\centering
\includegraphics[width=\linewidth]{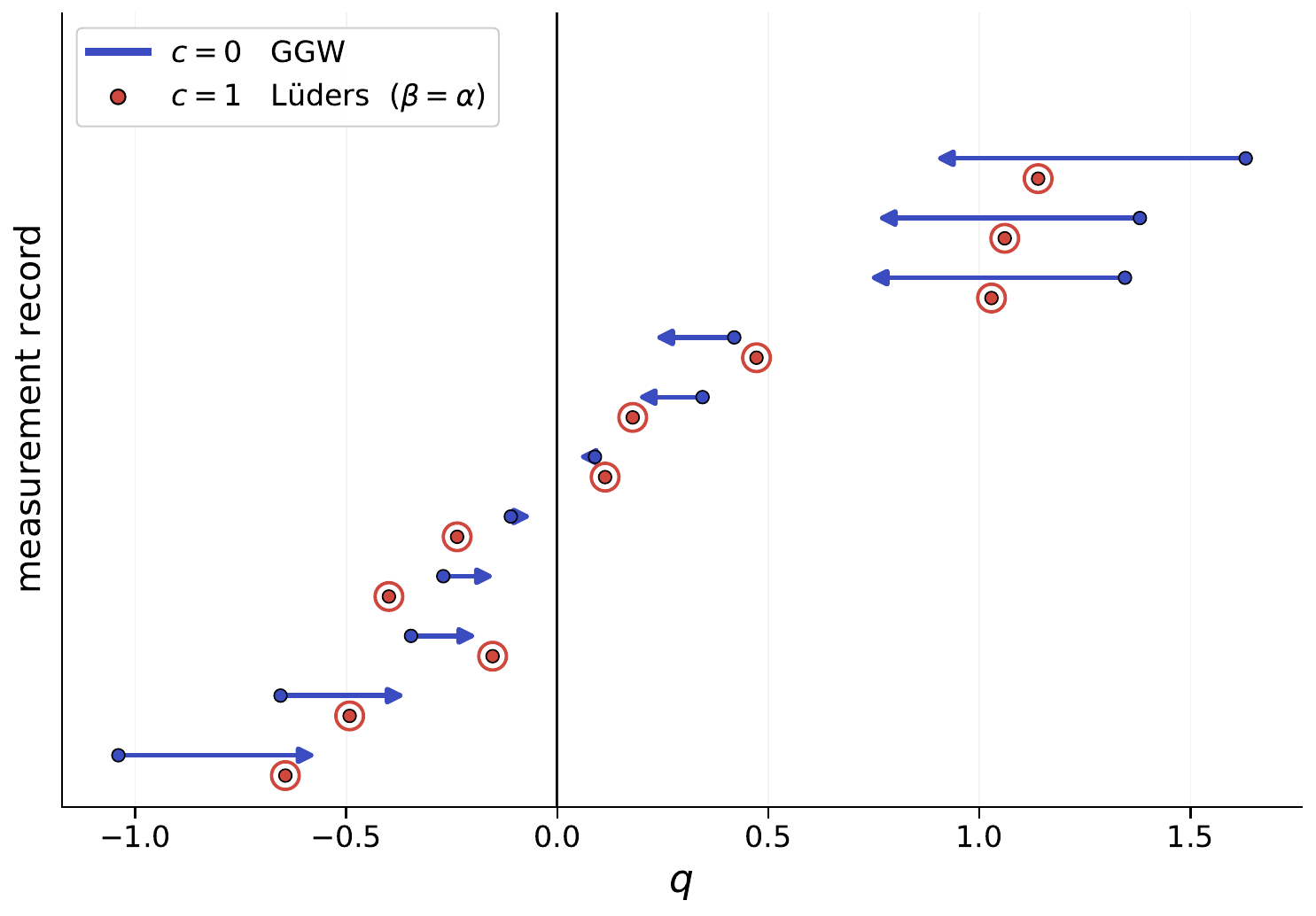}
\caption{\label{fig:kickforHomodyne} The homodyne counterpart of Fig. \ref{fig:kickforHeterodyne}. Homodyne POVM elements are ridges, flat along $p$, so only the $q$-coordinate of their center is defined; each row is one of 11 measurement records at $\kappa t=1.2$ (the height carries no meaning), and each arrow runs from the Cartan-Berger POVM coodrinate $\alpha_q$ (the center of $K_g^\dagger K_g$) to the Cartan-Berger post-measurement unitary coordinate  $\beta_q$ (the center of $K_gK_g^\dagger$). For GGW homodyne ($c=0$) the post-measurement unitary pulls every ridge towards the $p$ axis, $\beta_q=e^{-\rho_t}\alpha_q$, and also squeezes it (not shown). For L\"{u}ders homodyne ($c=1$, circled) all steps are in the same, commuting, right invariant directions $Q$ and $Q^2$, so $K_g=K_g^\dagger=K_{g^\dagger}$ and the post-measurement unitary vanishes identically, $\beta_q=\alpha_q$.}
\end{figure}

\subsection{Homodynes}

Having completed our analysis of heterodyne, we move onto homodyne. Homodyne, like heterodyne, has two well-known implementations; the first is a L\"{u}ders homodyne attained by directly coupling a meter to a system's field quadrature. The second is the scheme that is used in practice \cite{WisemanMilburn1993,GoetschGraham1994,WisemanMilburn2010}, involving again coupling to an annihilation operator and applying a deterministic unitary squeeze. We interpolate between these using a flexible homodyne with Lindblad operator
\bea
    L &= \Big(Q + iP (1-c)\Big)/\sqrt2\\
    &= a - iP c/\sqrt2
\eea
and consequent forward generator (which we immediately write as a stochastic IG vector field in the RID basis)

\begin{widetext}
\bea\label{RIDupdeltaHomoF}
\RInv{\updelta_t}^{\text{f-hom}}
= \kappa dt\left(-\frac12\RInv{H_0} + \frac{1-c}{4}\RInv{1}
- \frac{2-c}{2}\RInv{A}  - \frac{c}{2} \RInv{A}^\dagger\right)
+ \sqrt{\kappa}dW_t\left(\frac{2-c}{2}  \RInv{a} + \frac{c}{2} \RInv{a}^\dagger\right)
\eea
\end{widetext}
which yields L\"{u}ders homodyne for $c = 1$, and GGW homodyne for $c = 0$. We proceed as in the prior section, using Eqs. \ref{WierbienHCInv:zeta}-\ref{WierbienHCInv:z} to evaluate Eq. \ref{StochasticEqGenerator} for our forward generator in Eq. \ref{RIDupdeltaHomoF}. This yields a set of SDEs

\begin{figure*}[t]
\centering
\includegraphics[width=\linewidth]{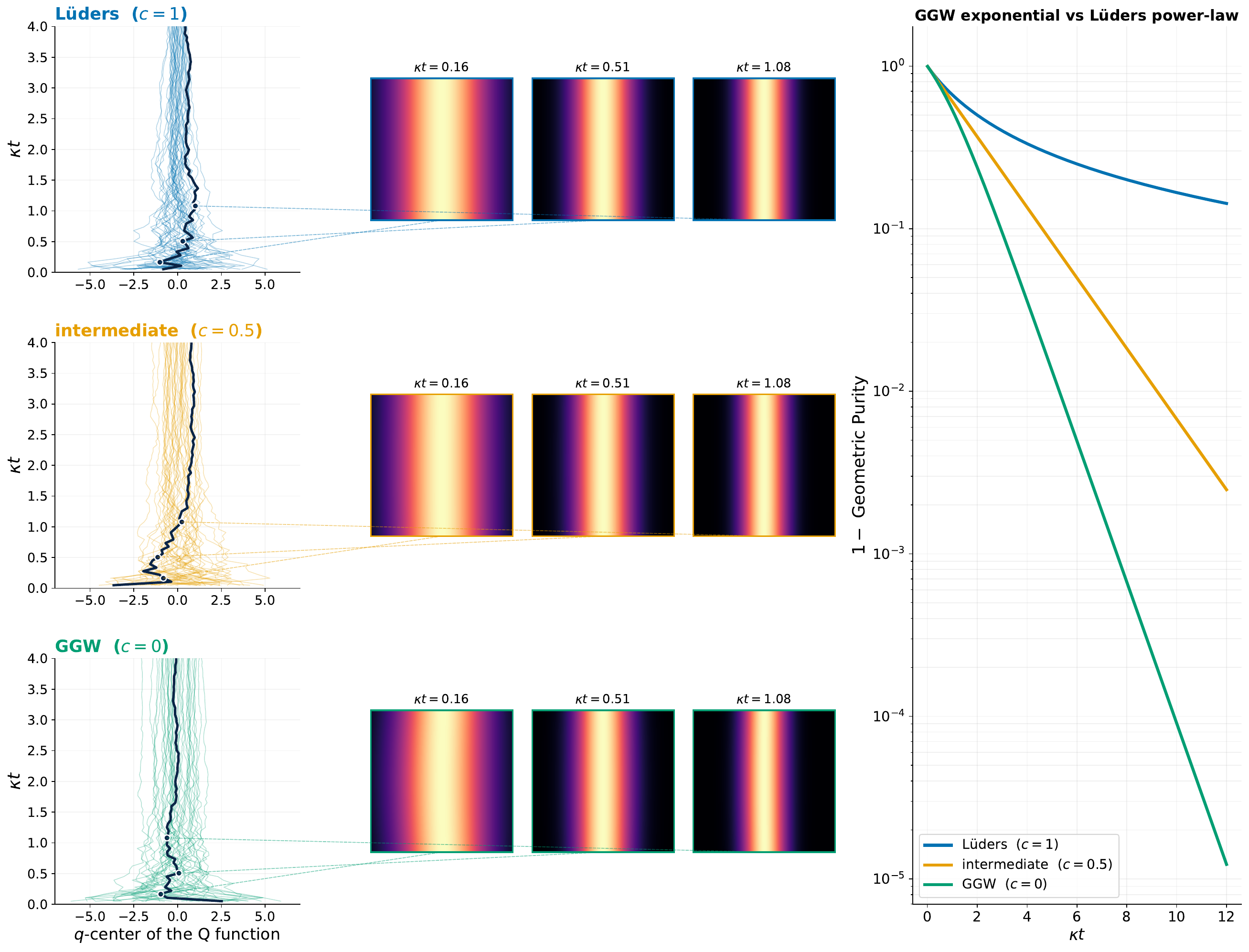}
\caption{\label{fig:homodyne} The Kraus trajectories for the various homodynes with associated POVM Q-functions. In comparison to heterodyne, we observe many differences between the schemes. As the measurement purifies, its stochastic motion slows down in all cases. This is coupled with a different scaling of the purification: the geometric purity (defined in Eq. \ref{GeometricPOVMPurity}) of L\"{u}ders approaches one as a power law, ${\rm tanh} [\tfrac{1}{2}{\rm ln}(1+\kappa t)]=\kappa t/(2+\kappa t)$, whereas GGW approaches it exponentially, ${\rm tanh} [\kappa t/2]$.}
\end{figure*}

\begin{widetext}
\bea\label{FHomoSDEs:zeta}
d\zeta_t^* &=& \frac{2-c}{2} e^{-2\rho_t}  \kappa dt \\ \label{FHomoSDEs:rho}
d\rho_t &=& \left(\frac{1}{2}  - \frac{2-c}{2}\omega_t\right)\kappa dt \\ \label{FHomoSDEs:omega}
d\omega_t &=& \left(\frac{c}{2} -\omega_t + \frac{2-c}{2}\omega_t^2 \right)\kappa dt \\  \label{FHomoSDEs:mu}
d\mu_t^* &=& \frac{2-c}{2}e^{-\rho_t} \Big(\sqrt{\kappa} dW_t - \nu_t\kappa dt\Big)\\\label{FHomoSDEs:nu}
d\nu_t &=& - \nu_t\left( \frac{1}{2} - \frac{2-c}{2}\omega_t \right)\kappa dt + \Big( \frac{c}{2} - \frac{2-c}{2}\omega_t \Big) \sqrt{\kappa} dW_t \\ \label{FHomoSDEs:z}
dz_t &=& \left(\frac{2-c^2}{8} - \frac{(2-c)^2}{8}\omega_t\right)\kappa dt + \frac{2-c}{2}\nu_t \sqrt{\kappa} dW_t - \frac{2-c}{4} \nu_t^2 \kappa dt.
\eea
\end{widetext} We can immediately observe that, in comparison to heterodyne, the squeeze sector is non-zero; the coordinate $\zeta$ has a deterministically increasing squeeze that turns off as the ruler coordinate $\rho$ increases, and the post-measurement unitary squeeze coordinate $\omega$ in Eq. \ref{FHomoSDEs:omega} has a fixed point structure, which will become significant in our treatment of our new slomodyne measurement in the next section. Most importantly, we see that the ruler coordinate in Eq. \ref{FHomoSDEs:rho} is driven by the squeeze; this results in a difference in the rate of purification between homodynes, as shown in Fig. \ref{fig:homodyne}. Here, we cannot use the Hilbert-space dependent definition of POVM purity given in Eq. \ref{POVMPurity}, as at all times between $t=0$ (when the Kraus operators are the identity) and $t=\infty$ (when the Kraus operators project onto quadrature eigenstates) the Kraus operators are non-trace class. Instead, we utilize the geometric purity defined in Eq. \ref{GeometricPOVMPurity} to show the speedup attained by GGW homodyne over L\"{u}ders homodyne; in essence, it squeezes faster, and for homodyne it is squeezing---not thermal cooling---that is the route to a geometric purification, resulting in a rank-$1$ projector.

For the initial conditions
\bea
\omega_0=\rho_0=\zeta_0=\mu_0=\nu_0=z_0=0,
\eea
the solution to the SDEs is
\begin{widetext}
\bea\label{FHomoFilt:rho}
\omega_t &=& \frac{c\tanh\kappa_c t}{1-c+\tanh\kappa_ct} = \frac{c}{2-c}\zeta_t \\ \label{FHomoFilt:zeta}
e^{-\rho_t} &=& \frac{(1-c)\,\mathrm{sech}\,\kappa_c t}{1-c+\tanh\kappa_ct} \\ \label{FHomoFilt:omega}
\zeta_t &=& \frac{(2-c)\tanh\kappa_c t}{1-c+\tanh\kappa_ct}\\ \label{FHomoFilt:nu}
\nu_t &=& \frac{c}{2}e^{-\rho_t}\int_0^t e^{-\kappa_c s}\,\sqrt{\kappa}dW_s \\ \label{FHomoFilt:mu}
\mu_t &=& \frac{2-c}{2}\int_0^t \Bigl(e^{-\rho_s} - \bigl(\omega_t-\omega_s\bigr)e^{-\kappa_c s} \Bigr)\sqrt{\kappa}dW_s \\ \label{FHomoFilt:z}
z_t &=& \frac{2-c}{4}\rho_t + \frac{c}{4}\kappa_c t
+ \frac{2-c}{4}\int_0^t 2\nu_s\,\sqrt{\kappa}dW_s - \nu_s^2\,\kappa ds,
\eea
\end{widetext}
where
\bea
    \kappa_c\equiv\frac{1-c}{2}\kappa.
\eea
Note the normalizer coordinates are deterministic (a Riccati equation for $\omega_t$, with $\rho_t$ and $\zeta^*_t$ slaved to it) while the pre- and post-measurement phase space coordinates take the form of linear stochastic filters and the center coordinate takes the form of a nonlinear stochastic filter.\\

For GGW homodyne ($c=0$) this gives $\rho_t=\kappa t/2$, $\omega_t=\nu_t=0$, $\zeta^*_t=1-e^{-\kappa t}$, $\mu^*_t=\sqrt{\kappa}\int_0^t e^{-\kappa s/2}dW_s$ and $z_t=\kappa t/4$.
Similarly, foor L\"{u}ders homodyne ($c=1$) it gives $e^{\rho_t}=1+\kappa t/2$, $\omega_t=\zeta^*_t=\kappa t/(2+\kappa t)$, and $\mu_t=\nu_t=\tfrac{\sqrt{\kappa}}{2}e^{-\rho_t}W_t$. This is the Harish-Chandra-like statement that the L\"{u}ders homodyne Kraus operator is Hermitian, with $W_t=\int_0^t dW_s$.

The post-measurement unitary, too, behaves differently than for heterodyne. For $c=1$, Eq. \ref{RIDupdeltaHomoF} becomes $\updelta_t^{L-Homo}=-q^2\,\frac{\kappa dt}{2}+q\,\sqrt{\frac{\kappa}{2}}\,dW_t$ with $q=(a+a^\dagger)/\sqrt{2}$, since $H_0+A+A^\dagger=q^2$. Every step is then a function of the single quadrature $q$, so the steps commute, and the accumulated Kraus operator $K_t=e^{-q^2\kappa t/2+q\sqrt{\kappa/2}\,W_t}$, with $W_t=\int_0^t dW_s$, is Hermitian. Thus L\"{u}ders homodyne, unlike L\"{u}ders heterodyne, carries no post-measurement unitary at all: $K^\dagger K$ and $KK^\dagger$ are the same ridge, centered at $q=W_t/(\sqrt{2\kappa}\,t)$ (Fig. \ref{fig:kickforHomodyne}). In this case the SDEs of Eqs. \ref{FHomoSDEs:zeta}-\ref{FHomoSDEs:z} preserve $\nu_t=\mu_t$ and $\omega_t=\zeta_t$.\footnote{That this occurs is somewhat remarkable since these are not coordinates of the POVM unitary and the post-measurement unitary, and is a manifestation of the coordinate transformations presented in Appendix A.} GGW homodyne instead keeps $\nu_t=\omega_t=0$, and its post-measurement unitary both pulls each ridge towards the $p$ axis, In Cartan-Berger coordinates, $\beta_q=e^{-\rho_t}\alpha_q$ with $\rho_t=\kappa t/2$, and squeezes it, widening it by a factor $e^{\kappa t}$ in variance. L\"{u}dersianity therefore survives the integration precisely when the steps commute, as they do when only a single quadrature is measured i.e. as in L\"{u}ders homodyne.

The set of all homodyne Kraus operators, for all homodynes, from initial time $t=0$ to $t=\infty$ exists along the boundary of trace class operators in $\mathrm{IWH}^{\textbf{12}}$, which we call the homosurface. The boundary of the homosurface at $t=\infty$ is where the fully-realized homodyne measurements live on traditional phase space.\footnote{It is worth pausing to remark on this remarkable fact; the boundary of the boundary is not empty here. This is not the usual case in homology, and implies that the set of fully realized homodyne measurements are such a double boundary and thus are their own boundary by the standard point-set theoretic proof.}

That a homodyne measurement's achievement of a non-trace-class projective measurement (i.e. a measurement of a type III von Neumann algebra) is dependent on staying non-trace-class through its whole evolution is worth pausing to ponder; were homodyne to ever cool (that is, were its Cartan-Berger ruler coordinate to become non-zero), a pure measurement process would have no process by which to heat it up again, and it would stay trace-class for all time. This leads us to conjecture that, for any non-trace class final-measurement, the entire measurement trajectory must stay non-trace class. Indeed, for every measurement in $\mathrm{IWH}^{\textbf{12}}$ the Hermitian quadratic part $hH_0+sA+\bar{s}A^\dagger$ of the forward generator obeys $|s|\le -h$: a measurement can contract each quadrature of a POVM element, but never stretch one, so once both quadratures have been measured they stay measured. This has implications for the phase POVMs of Wiseman (and others), which we discuss in more detail in section IIID, and we hope to prove this formally in a future publication.

\begin{figure*}[t]
\centering
\includegraphics[width=\linewidth]{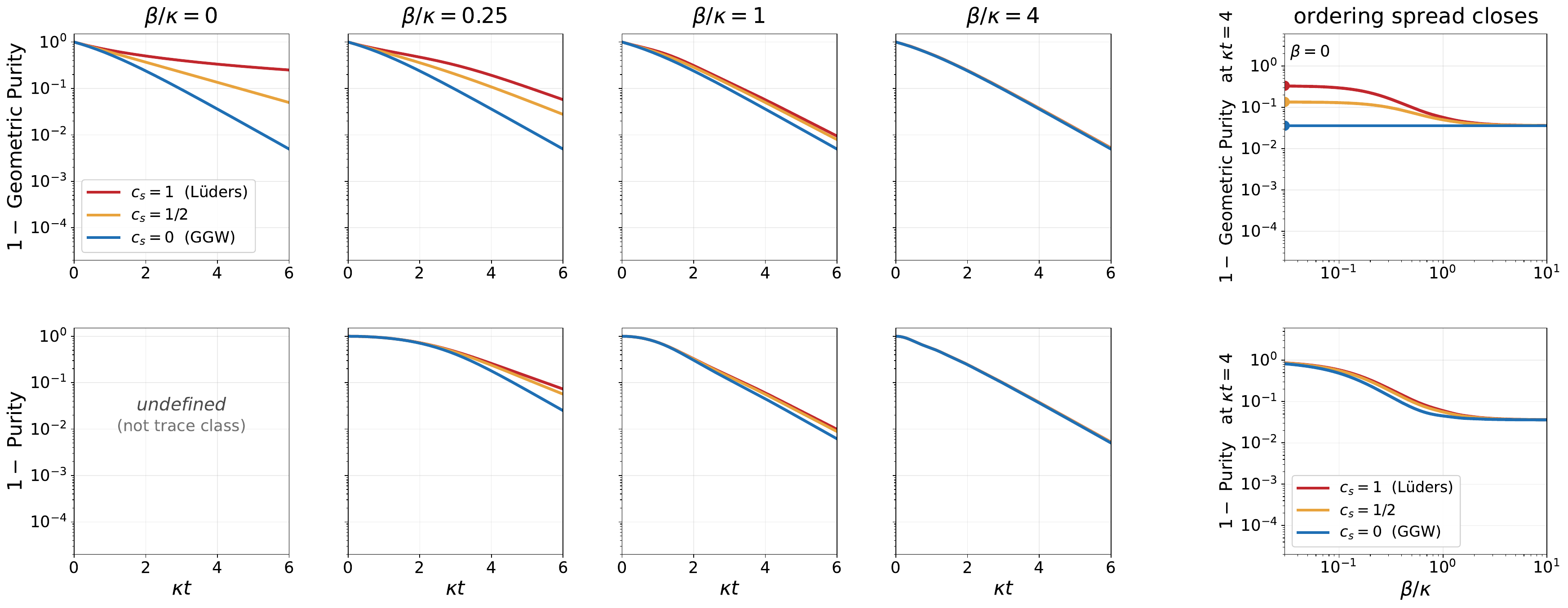}
\caption{\label{fig:slomodynepurity} The purity of slomodyne is captured both through geometric purity and Hilbert space purity; we observe the gap between the different schemes (GGW, L\"{u}ders, and intermediate) closes as $\beta$ increases and the measurement more closely resembles random phase homodyne i.e. heterodyne, where there is no gap in the rate of purification using either metric. At $\beta=0$ the Hilbert space purity is undefined, as homodyne POVM elements are not trace class; at large $\beta$, where the POVM squeeze averages out, the two purities agree and the gap between L\"{u}ders and GGW closes. This is seen for fixed time $\kappa t = 4$ in the rightmost column of panels, though the low-$\beta$ behavior of the bottom right panel requires care; we approach a region where the Hilbert-space POVM purity is simply not defined.}
\end{figure*}

\subsection{Slomodynes}

The realization that squeezing is a non-thermal form of purification that, in principle, can change the rate of purification motivates the modification of the flexible homodyne of Eq. \ref{RIDupdeltaHomoF} by adding a rotation e.g. at a rate $\beta$. This gives a new forward generator

\begin{widetext}
\bea\label{RIDSLomo}
\RInv{\updelta_t}^\text{f-slo}
= \RInv{\updelta_t}^\text{f-homo} - i\beta  \RInv{H_0} dt.
\eea
\end{widetext}

Adding this (non-adaptive) feedforward drastically changes homodyne; for $\beta=0$ we achieve the family of homodynes from the prior section, but as $\beta$ grows large compared to the measurement rate $\kappa$, we approximate random-phase homodyne, which is equivalent to heterodyne \cite{Wiseman1996}; thus slomodyne interpolates between the two families of measurement schemes. This modifies the SDEs of homodyne slightly but significantly, resulting in

\begin{widetext}

\bea
\label{FHomoSDEs:zeta}
d\zeta_t^* &=& \frac{2-c}{2} e^{-2\rho_t}  \kappa dt \\
\label{FHomoSDEs:rho}
d\rho_t &=& \left(\frac{1}{2} + i\frac{\beta}{\kappa}  - \frac{2-c}{2}\omega_t\right)\kappa dt \\
\label{FHomoSDEs:omega}
d\omega_t &=& \left(\frac{c}{2} - \Big(1+2i\frac{\beta}{\kappa}\Big)\omega_t + \frac{2-c}{2}\omega_t^2 \right)\kappa dt \\  
\label{FHomoSDEs:mu}
d\mu_t^* &=& \frac{2-c}{2}e^{-\rho_t} \Big(\sqrt{\kappa} dW_t - \nu_t\kappa dt\Big)\\
\label{FHomoSDEs:nu}
d\nu_t &=& - \nu_t\left( \frac{1}{2} +i\frac{\beta}{\kappa} - \frac{2-c}{2}\omega_t \right)\kappa dt + \Big( \frac{c}{2} - \frac{2-c}{2}\omega_t \Big) \sqrt{\kappa} dW_t \\
\label{FHomoSDEs:z}
dz_t &=& \left(\frac{2-c^2}{8} - \frac{(2-c)^2}{8}\omega_t\right)\kappa dt + \frac{2-c}{2}\nu_t \sqrt{\kappa} dW_t - \frac{2-c}{4} \nu_t^2 \kappa dt.
\eea
\end{widetext}

That this interpolates between homodyne and heterodyne is easily verified by noticing how the difference between GGW and L\"{u}ders purity changes as $\beta$ increases, as illustrated in Fig. \ref{fig:slomodynepurity}.

\begin{figure*}[t]
\centering
\includegraphics[width=\linewidth]{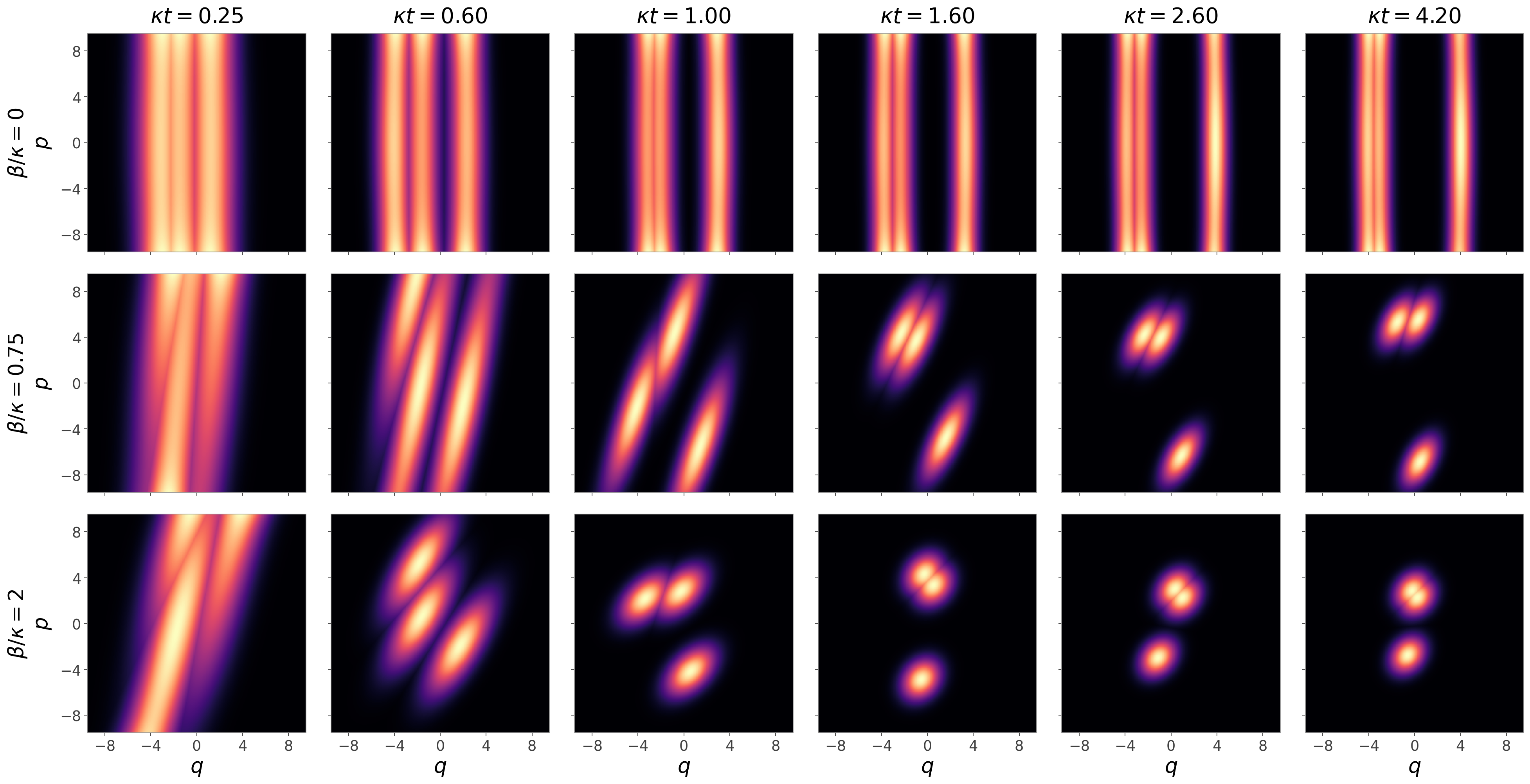}
\caption{\label{fig:slomodyneQfuncs} The Q-functions of GGW-slomodyne are displayed evolving in time for various $\beta$. While for $\beta=0$ we have homodyne, for larger $\beta$ we observe a measurement that increasingly resembles heterodyne, but with an added short-time rotation of the Q-function for Kraus operators further from the origin as well as a contraction towards the origin.}
\end{figure*}

Examining our SDEs, the fixed point structure of $\omega_t$ becomes more interesting, possessing the form of a Riccati equation; $\omega_t$ possesses two fixed points defined

\bea\label{OmegaPM}
\omega^{\pm} =  \frac{\frac{\kappa}{2} - i\beta  \pm \sqrt{(\frac{\kappa}{2} - i\beta )^2 - (1-\frac{c}{2})\frac{c}{2}\kappa^2}}{(1-\frac{c}{2}) \kappa }
\eea

When $\beta=0$ (homodyne) the fixed points are real: $\omega^+=1$ (unstable) and $\omega^-=c/(2-c)$ (stable), with L\"{u}ders the degenerate case $\omega^\pm=1$, approached as a power law, $\omega_t=\kappa t/(2+\kappa t)$. For non-zero $\beta$ the stable fixed point $\omega^-$ becomes complex, tilting the post-measurement squeeze axis, and its modulus falls as $|\omega^-|\simeq c\kappa/4\beta$ for $\beta\gg\kappa$. For the advanced reader: due to the correspondence between the Harish-Chandra-like coordinate $\omega_t$ and the Iwasawa post-measurement unitary squeeze parameter, we can immediately conclude that the post-measurement unitary squeeze vanishes at large $\beta$. Meanwhile the POVM squeeze $\zeta_t^*$, whose source $e^{-2\rho_t}$ carries the phase $e^{2i\beta t}$, rotates and averages out, as illustrated in Fig. \ref{fig:slomodyneQfuncs}, where we observe increasingly circularly Gaussian Q-functions as $\beta$ is increased.

The rotation only complexifies the deterministic sector. With
\bea
\Gamma &\equiv& \tfrac{\kappa}{2}+i\beta,\\
\Delta&\equiv&\bigl[\Gamma^2-\tfrac{c}{2}\bigl(1-\tfrac{c}{2}\bigr)\kappa^2\bigr]^{1/2},
\eea
and
\begin{widetext}
\bea
K(t) &\equiv& 2\cosh\Delta t + \frac{2\Gamma -(2-c)\kappa}{\Delta}\sinh \Delta t\\
&=& \frac{(2-c)\kappa}{2\Delta}\Big((1+\omega^-)e^{-\Delta t} - (1-\omega^+)e^{\Delta t}\Big),
\eea
\end{widetext}
the solution to the SDEs is now
\begin{widetext}
\bea\label{FSlomoFilt:omega}
\omega_t &=& \frac{c\,\kappa \tanh \Delta t}{2\Delta + 2\Gamma\tanh \Delta t} = \frac{c}{2-c}\zeta_t^* \\
\label{FSlomoFilt:rho}
e^{-\rho_t} &=& \frac{\Delta \,\mathrm{sech}\, \Delta t}{2\Delta + 2\Gamma\tanh \Delta t} \\ \label{FSlomoFilt:zeta}
\zeta^*_t &=& \frac{(2-c)\,\kappa \tanh \Delta t}{2\Delta + 2\Gamma\tanh \Delta t}\\
\label{FSlomoFilt:nu}
\nu_t &=& \frac{c}{2}e^{-\rho_t}\int_0^t K(s)\,\sqrt{\kappa}dW_s \\
\label{FSlomoFilt:mu}
\mu_t^* &=& \frac{2-c}{2}\int_0^t \Bigl(e^{-\rho_s} - \bigl(\omega_t-\omega_s\bigr)K(s) \Bigr)\sqrt{\kappa}dW_s \\
\label{FSlomoFilt:z}
z_t &=& \frac{2-c}{4}\rho_t + \frac{(1-c^2)\kappa-(2-c)2\Gamma}{8} t
+ \frac{2-c}{4}\int_0^t 2\nu_s\,\sqrt{\kappa}dW_s - \nu_s^2\,\kappa ds,
\eea
\end{widetext}
which are even in $\Delta$, so the branch of the square root is immaterial. They reduce to Eqs. \ref{FHomoFilt:rho}-\ref{FHomoFilt:zeta} at $\beta=0$. As $t\rightarrow\infty$, $\omega_t\rightarrow\tfrac{c}{2}\kappa/(\Gamma+\Delta)$, which is the stable fixed point $\omega^-$ of Eq. \ref{OmegaPM}. 
Note ${\rm Im}\,\rho_t$ records the accumulated rotation, so the kernels $e^{-(\rho_t-\rho_s)}$ are complex.

While we focus here on slomodyne, it is worth noting that the same approach can be used to study general-dyne \cite{Belavkin1992,WisemanDiosi2001}; here the interpolation is performed by directly imbalancing the coupling to the two meters in the heterodyne forward generator Eq. \ref{RIDupdeltaHeterodyneF}, resulting in measurements that project onto squeezed states.

\section{Adaptidyne}

Having concluded our exploration of standard, non-adaptive 'dyne measurements including our slomodyne measurement, we begin a more exploratory part of this paper, giving the first state-independent description of adaptive measurement for quantum measurements in the literature. 

From the point of view of the IMP, it is straightforward to build adaptive measurements that live within $\mathrm{IWH}^{\textbf{12}}$ using the mathematical technology presented here. This is accomplished by writing forward generators that depend on phase space coordinates e.g. by controlling measurement rates or adaptively applying unitaries. 

 

In the remainder of this section, we consider three such measurements---which we call \emph{adaptidynes}---and comment on their utility. We emphasize that this approach is not unique to the Harish-Chandra-like decomposition, and indeed is better suited to the Iwasawa and Cartan-Berger coordinate systems where the parameters belonging to the POVM element are cleanly distinguished from those belonging to post-measurement unitaries; it is precisely this geometric separation between POVM and unitary that the Harish-Chandra-like decomposition sacrifices for algebraic simplicity. Nonetheless, we can utilize it to generate adaptive measurement schemes, which we now proceed to do.

Before we proceed, we pause to emphasize that, to the authors' knowledge, this is the first general state-independent treatment of adaptive quantum measurement in the literature.\\

\subsection{Adaptidyne 1: Hetero-to-homo-dyne}

One of the measurements we can generate produces Kraus operators shaped like the elements of the well-studied Phase POVM, at least for phase points near the origin. This is generated by making the terms in the forward generator $ \RInv{\updelta}_t$ coordinate dependent i.e. no longer right invariant. Nonetheless, the forward generator can still be expressed in right-invariant basis, and technology of the prior sections still be used (albeit, with increased care when evaluating Stratonovich products). Consider the forward generator

\begin{widetext}
    \bea\label{GaussianPhase}
\RInv{\updelta}_t^\text{A1}
    &=&
    \left(
    -\RInv{a^\dag a}\,2\kappa_hdt
    +\RInv{a}\,2\sqrt{\kappa_h}\,dZ^*_t
    \right)
    +\left(
    -\RInv{a^\dag a}\,\frac{\kappa_p}{2}dt
     \;-\;\RInv{A}\,u(\zeta_t,\mu_t)\,\kappa_p\,dt
     +\;\RInv{a}\,u(\zeta_t,\mu_t)^{\!1/2}\,\sqrt{\kappa_p}\,dW_t
    \right) \\
    &=& -\RInv{H_o}\Bigl(2\kappa_h+\tfrac{\kappa_p}{2}\Bigr)dt
     \;+\;\RInv{1}\Bigl(\kappa_h+\tfrac{\kappa_p}{4}\Bigr)dt
     \;-\;\RInv{A}\,u(\zeta_t,\mu_t)\,\kappa_p\,dt
     +\;\RInv{a}\left(2\sqrt{\kappa_h}\,dZ^*_t\;+\;u(\zeta_t,\mu_t)^{\!1/2}\,\sqrt{\kappa_p}\,dW_t\right)
    \eea
\end{widetext}
where
\bea
    u_t &\equiv& u(\zeta_t,\mu_t)
    \equiv \frac{\zeta^*_t-\gamma\,\mu^*_t/\mu_t}{\bigl|\zeta^*_t-\gamma\,\mu^*_t/\mu_t\bigr|}
\eea
and we have introduced independent parameters $\kappa_{h/p}$ corresponding to a weak heterodyne snf strong homodyne mrsdutrmrny rate, respectively. We also introduce a coefficient $\gamma$ that determines how much squeeze builds up before the homodyne axis locks in: the local oscillator starts perpendicular to the running displacement $\mu_t$ (as in Wiseman's scheme, adaptidyne 3 below) and locks once $|\zeta^*_t|$ exceeds $\gamma$ or reaches its maximum value of $1$.

We now show that this results in a Gaussian approximation of the Phase POVM discussed extensively in the quantum optics literature \cite{Helstrom1976,Holevo1982,SusskindGlogower1964,Leonhardt1995,Wiseman1995,WisemanKillip1998}---at least, near the origin of the phase plane. Whereas the canonical Phase POVM is emphatically not trace class (as it is comprised of projectors onto the non-normalizable phase states \cite{SusskindGlogower1964} $\ket{\varphi}=\sum_n e^{in\varphi}\ket{n}$), our first attempt at a Gaussian approximation of the Phase POVM is trace class and, unlike modern approximations relying on injection of non-Gaussianity via e.g. combination with Fock states \cite{Martin2020}, our measurement is entirely implementable with linear optics, 'dyne detection, and real-time feedback of the local oscillator phase.

Evaluating the forward generator in Eq. \ref{GaussianPhase} yields the SDEs

\begin{widetext}
\bea
\label{AdapativeGaussianSDEs:dzeta}
d\zeta^*_t &=& \,e^{-2\rho_t}\,u_t\,\kappa_p\,dt\\
  d\rho_t    &=& \Bigl(2\kappa_h+\tfrac{\kappa_p}{2}-\omega_t\,u_t\,\kappa_p\Bigr)dt\\
  d\omega_t  &=& -\Bigl(\omega_t\bigl(4\kappa_h+\kappa_p\bigr)-\omega_t^{2}\,u_t\,\kappa_p\Bigr)dt\\
  d\mu^*_t   &=& e^{-\rho_t}\Bigl(2\sqrt{\kappa_h}\,dZ^*_t+u_t^{\!1/2}\sqrt{\kappa_p}\,dW_t
                 -\nu_t\,u_t\,\kappa_p\,dt\Bigr)\\
  d\nu_t     &=& -\nu_t\Bigl(\bigl(2\kappa_h+\tfrac{\kappa_p}{2}\bigr)-\omega_t\,u_t\,\kappa_p\Bigr)dt
                  -\omega_t\Bigl(2\sqrt{\kappa_h}\,dZ^*_t+u_t^{\!1/2}\sqrt{\kappa_p}\,dW_t\Bigr)\\
\label{AdapativeGaussianSDEs:dz}
dz_t       &=& \frac12\Bigl(2\kappa_h+\tfrac{\kappa_p}{2}-\omega_t\,u_t\,\kappa_p+\nu_t^{2}\,u_t\,\kappa_p\Bigr)dt
                  +\;\nu_t\Bigl(2\sqrt{\kappa_h}\,dZ^*_t+u_t^{\!1/2}\sqrt{\kappa_p}\,dW_t\Bigr).
\eea
\end{widetext}
Four of the complex coordinates can be solved explicitly while the remaining two SDEs we only simplify because we suspect the nonlinearity of $u(\zeta_t,\mu_t)$ prevents much further simplification,
\begin{widetext}
\bea
\omega_t &=& 0 \\
\label{A1Filt:rho}
\rho_t &=& \Bigl(2\kappa_h+\tfrac{\kappa_p}{2}\Bigr)t \\
\nu_t &=& 0\\
z_t &=& \tfrac12\rho_t\\
\label{A1Filt:zeta}
d\zeta^*_t &=& e^{-2\rho_t}\,u(\zeta_t,\mu_t)\,\kappa_p\,dt \\
\label{A1Filt:mu}
d\mu^*_t &=& e^{-\rho_t}\Bigl(2\sqrt{\kappa_h}\,dZ^*_t + u(\zeta_t,\mu_t)^{1/2}\,\sqrt{\kappa_p}\,dW_t\Bigr).
\eea
\end{widetext}
Since the feedback $u_t$ is evaluated at the start of each increment, it is a functional of the record on $[0,t)$, and the integral is an It\^o integral. The filter is therefore implicit through $u_t$ alone; without feedback ($u_t$ fixed) it would be explicit.

\begin{figure*}[t]
\centering
\includegraphics[width=\linewidth]{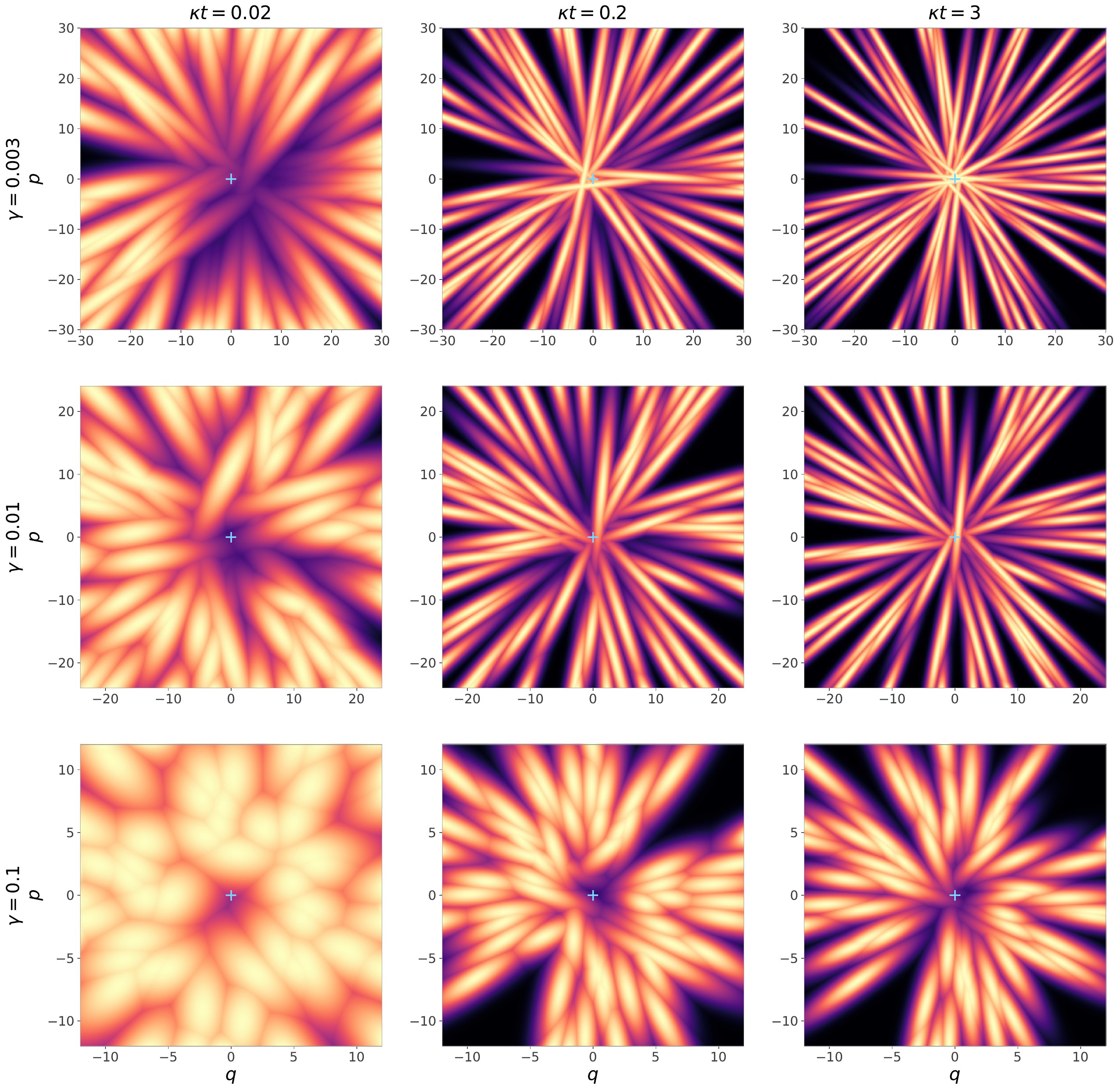}
\caption{\label{fig:gaussianphase} The Q-functions of the POVM elements of adaptidyne 1 (Hetero-to-homo-dyne, Eq. \ref{GaussianPhase}) at $\kappa t=0.02$ (left), $0.2$ (center), and $3$ (right), for $\gamma=0.003$, $0.01$, $0.1$ (rows), with 60 Kraus operators sampled from the KOD, each normalized to unit peak. We choose $\kappa_h = 0.0025\kappa$ and $\kappa_p = 3.995 \kappa$ to yield (to within $0.1\%$) the same purification rate as GGW heterodyne at a rate $\kappa$. Here, the parameter $\gamma$ sets when the homodyne axis locks ($\kappa t\approx\gamma/\kappa_p$), and thus how close the rays come to the origin and how sharply they squeeze. The optimal value of $\gamma$ to achieve rays that usually touch (but not pierce) the origin is approximated $\gamma_{\rm opt} \approx \frac{4\kappa_h}{4\kappa_h + \kappa_p}$; with a lower $\gamma$, they instead pierce the origin. As $\gamma\rightarrow 0$ the axis locks to a direction drawn from the measurement record (i.e. random-phase homodyne) and the elements become lines through the origin with $\sigma_r^2=1+\kappa_p/4\kappa_h$; as $\kappa_h\rightarrow0$ as well, they become projectors onto quadrature eigenstates, each containing both rays $\varphi$ and $\varphi+\pi$, so that the phase is only resolved modulo $\pi$.}
\end{figure*}

Note that the first four coordinates are determinitic essentially because the forward generator contains no raising operators, and only $\zeta^*_t$ and $\mu^*_t$ fluctuate.
Each POVM element then has the exact Husimi function
\begin{widetext}
\bea\label{AdaptiveQ}
Q(\alpha)=\exp\Bigl[\,2\,\mathrm{Re}\bigl(\mu^*_t\alpha\bigr)-\mathrm{Re}\bigl(\zeta^*_t\alpha^2\bigr)
  -\bigl(1-e^{-2\rho_t}\bigr)|\alpha|^2\Bigr],
\eea
\end{widetext}
a Gaussian whose long axis is set by the locked squeeze $\zeta^*_t$. We can, as before, use these to generate POVM Q-functions corresponding to individual Kraus operator trajectories (Fig. \ref{fig:gaussianphase}). At early times they are broad; once the homodyne axis locks, at $\kappa t\approx\gamma/\kappa_p$, they stretch into a sunburst of rays that reach the origin. After locking, only the weak heterodyne measures along the ray, so the radial width $\sigma_r$ of the rays (in units where the vacuum Q-function has unit width) cannot exceed $(1+\kappa_p/4\kappa_h)^{1/2}$. Here, $\gamma$ sets when the axis locks; earlier locking gives longer rays that more often pierce the origin, while later locking shorter rays that stop short of it, with $\gamma_{\rm opt}\approx4\kappa_h/(4\kappa_h+\kappa_p)$ balancing the two.

\begin{figure*}[t]
\centering
\includegraphics[width=\linewidth]{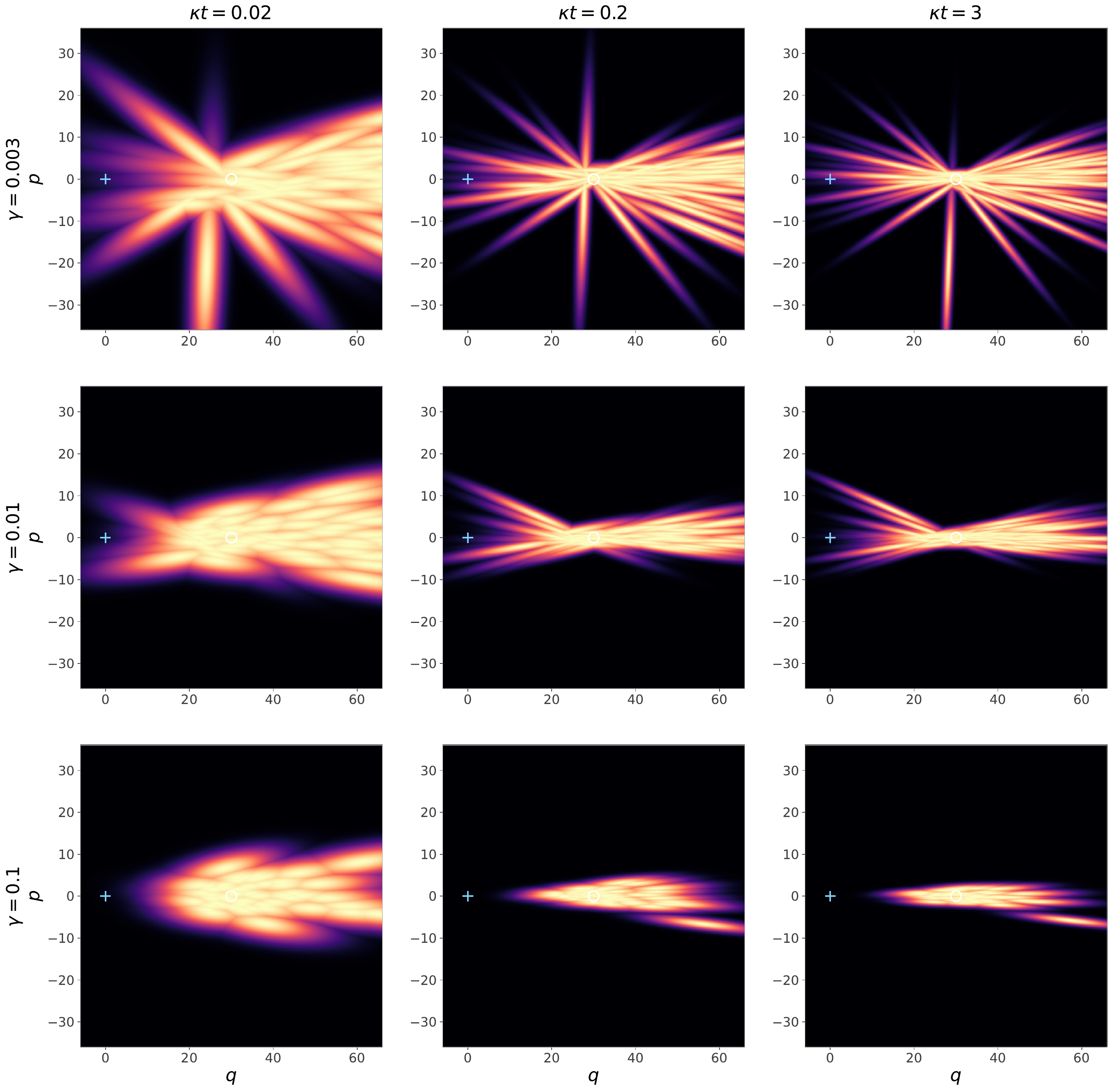}
\caption{\label{fig:gaussianphasefar} Adaptidyne 1 (Hetero-to-homo-dyne) far from the origin: as Fig. \ref{fig:gaussianphase} (same $\gamma$, $\kappa_h$, $\kappa_p$ and times), but for Kraus operators sampled with the ostensible distribution of Sec. \ref{sec:trick}, centered on the phase space point $q_0=30$, $p_0=0$ (white circle); the window is centered on that point, and $+$ marks the origin. For small $\gamma$ the sunburst re-forms around $q_0$: the axis locks before the measurement record has fixed the direction of the POVM element, so the elements are lines through $q_0$ in a spread of directions. At $\kappa t=3$ only $24\%$ ($\gamma=0.003$) and $56\%$ ($\gamma=0.01$) of the sampled axes pass within $2$ of the origin. For $\gamma=0.1$ they all do, but the elements are short radial segments ($\sigma_r\approx7$) that stop well short of the origin. Thus we see a tradeoff; the optimal value of $\gamma$ to ahieve phase-resolution on the region of phase space of interest; this highlights the trickiness of building a measurement resembling the Phase POVM uniformly across phase space.}
\end{figure*}

Notably, however, and very crucially, the sunburst is a near-origin phenomenon. Fig. \ref{fig:gaussianphasefar} shows Kraus operators sampled far from the origin with the trick of Sec. \ref{sec:trick}: we shift the means of the Wiener increments to $\expect{dZ_t}=2\sqrt{\kappa_h}\,\alpha_0e^{-\rho_t}dt$ and $\expect{dW_t}=2\sqrt{\kappa_p}\,{\rm Re}\bigl(u_t^{1/2}\alpha_0e^{-\rho_t}\bigr)dt$, with $\alpha_0=q_0/\sqrt{2}$ and $u_t$ the local-oscillator coefficient of Eq. \ref{GaussianPhase}. This is a time-and-coordinate-dependent $\vec{A}$ in the language of Sec. \ref{sec:trick}, chosen so that every sampled POVM element passes through the phase space point $q_0=30$. For $\gamma\approx\gamma_{\rm opt}$ the sunburst re-forms around that point rather than the origin: the POVM elements are lines through $q_0$ in a spread of directions, and only about a quarter of them point at the origin. The culprit is the timing; the axis locks at $\kappa t\approx\gamma/\kappa_p$, while a measurement record that pulls the POVM element out to amplitude $|\alpha_0|$ needs $\kappa t\approx1/(\kappa_p|\alpha_0|^2)$ to fix its direction, so the elements are aligned only when $\gamma|\alpha_0|^2\gtrsim10$. A larger $\gamma$ restores the alignment (bottom row of Fig. \ref{fig:gaussianphasefar}) but shortens the elements. 


Thus, adaptidyne 1 reproduces the shape of the Phase POVM elements near the origin, but not their statistics far from it, and disappointingly doesn't do a very good job approximating the Phase POVM in general.

\subsection{Adaptidyne 2: Adaptively squeezed heterodyne}

Another measurement with potential to generate a Gaussian approximation of the Phase POVM is generated by a forward generator corresponding to heterodyne with adaptive squeezing and displacement of the form

\begin{widetext}
\bea\label{GaussianPhaseOLD}
  \RInv{\updelta}^{\text{A2}}_t &=& \left(-\RInv{H_o}\,2\kappa\,dt+\RInv{1}\,\kappa\,dt+\RInv{a}\,2\sqrt{\kappa}\,dZ^*_t \right)
   +\bigl(\RInv{A}\mu_t^{*2}-\RInv{A}^\dag\mu_t^{2}\bigr)|\mu_t|^{2n}r_\text{S}dt
     -\bigl(\RInv{a}^\dag\mu_t-\RInv{a}\mu^*_t\bigr)r_\text{D}dt.
\eea
\end{widetext}

Here, as the paramter $\mu$ increases, squeezing and displacement increase as well. This results in SDEs

\begin{widetext}
    \bea
    \label{AdapativeCHECK}
  d\zeta^*_t &=& -\,e^{-2\rho_t}\mu_t^{*2}|\mu_t|^{2n}r_\text{S}\,dt\\
  d\rho_t    &=& \bigl(2\kappa+\omega_t\mu_t^{*2}|\mu_t|^{2n}r_\text{S}\bigr)dt\\
  d\omega_t  &=& \Bigl(-4\omega_t\kappa+\bigl(\mu_t^{2}-\omega_t^{2}\mu_t^{*2}\bigr)|\mu_t|^{2n}r_\text{S}\Bigr)dt\\
  d\mu^*_t   &=& e^{-\rho_t}\Bigl(2\sqrt{\kappa}\,dZ^*_t+\bigl(\mu^*_t r_\text{D}+\nu_t\mu_t^{*2}|\mu_t|^{2n}r_\text{S}\bigr)dt\Bigr)\\
  d\nu_t     &=& -2\omega_t\sqrt{\kappa}\,dZ^*_t
                 -\Bigl(2\nu_t\kappa+\omega_t\nu_t\mu_t^{*2}|\mu_t|^{2n}r_\text{S}+\bigl(\mu_t+\omega_t\mu^*_t\bigr)r_\text{D}\Bigr)dt\\
  dz_t       &=& 2\nu_t\sqrt{\kappa}\,dZ^*_t+\Bigl(\kappa+\nu_t\mu^*_t r_\text{D}+\tfrac12\nu_t^{2}\mu_t^{*2}|\mu_t|^{2n}r_\text{S}\Bigr)dt.
\eea
\end{widetext}

Note that the feedback enters every coordinate, so that the integrated expressions lack explicit solutions and are here omitted. We observe that no It\^o correction appears in $dz_t$ here because $dZ^{*2}_t=0$. Note that $d\omega_t$, $d\nu_t$ and the second term of $d\mu^*_t$ are driven initially entirely by the feedback ($r_\text{S}$, $r_\text{D}$): with the feedback switched off they vanish, recovering the GGW heterodyne filter of Eqs. \ref{FHetFilt:rho}-\ref{FHetFilt:z} at $c_d=c_n=0$.


Both feedback terms are anti-Hermitian; they are a unitary squeeze and a unitary displacement applied after each heterodyne step, locked to the phase of the running displacement $\mu_t$. With $r_\text{S}>0$ the squeeze narrows the POVM elements in the azimuthal direction, so that they are elongated radially (the opposite sign elongates them tangentially). Notably, pairing $\mu^2$ with $A^\dagger$ and $\mu^{*2}$ with $A$ makes the forward generator covariant under phase rotations; the opposite pairing is covariant only under rotations by multiples of $\pi/2$.

\begin{figure*}[t]
\centering
\includegraphics[width=\linewidth]{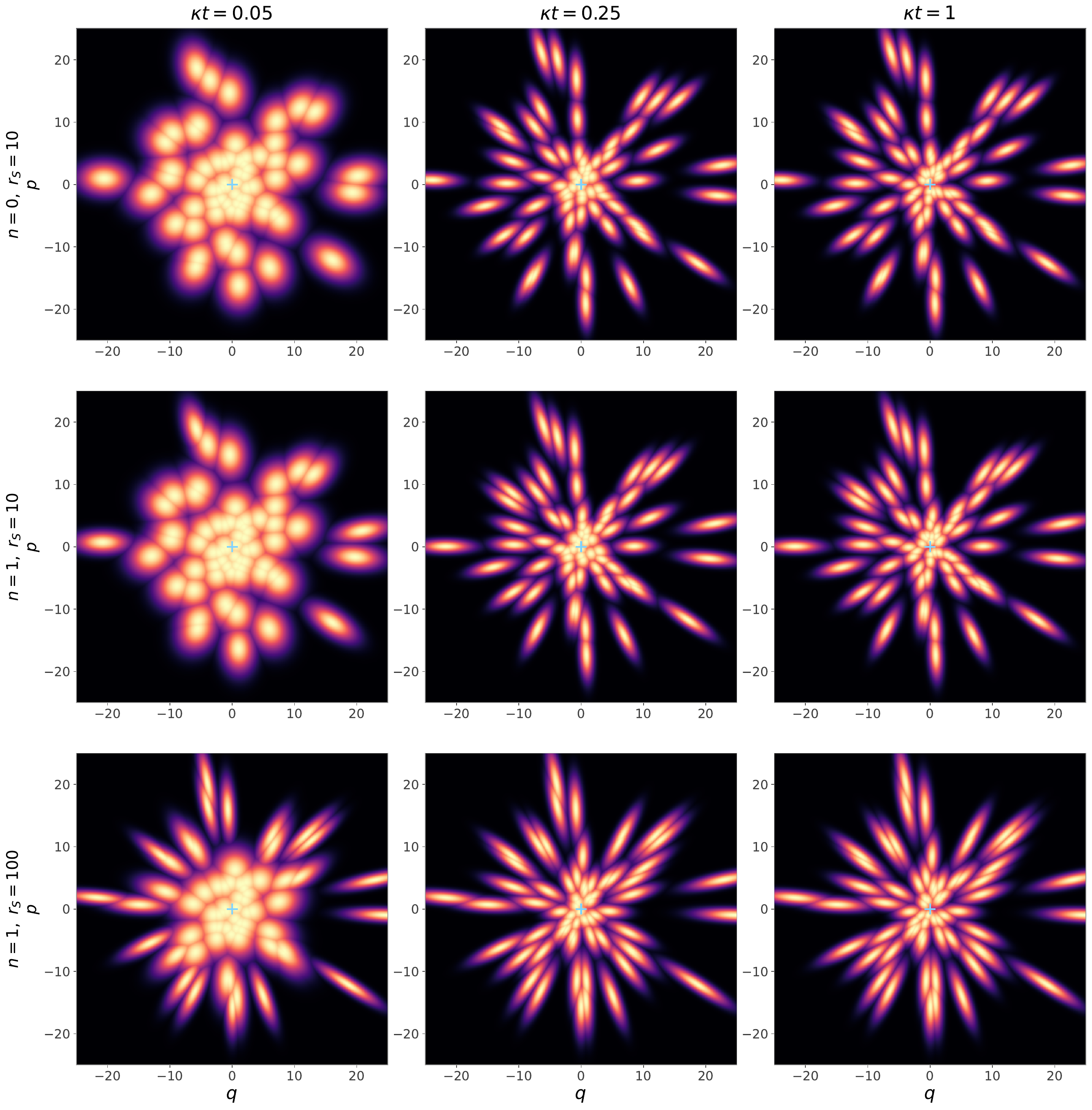}
\caption{\label{fig:DisplaceAndSqueeze} The Q-functions of the POVM elements of adaptidyne 2 (Adaptively squeezed heterodyne, Eq. \ref{GaussianPhaseOLD}) at $\kappa t=0.05$, $0.25$ and $1$, for $(n,r_\text{S})=(0,10)$, $(1,10)$ and $(1,100)$ (rows), with $r_\text{D}=1$ and $\kappa=1$. Each panel overlays 20 Kraus operators sampled from the KOD and 48 sampled with a constant mean shift (Sec. \ref{sec:trick}) from the origin of $|m|=5$, $10$, $15$, $20$, each at 12 angles, with each element normalized to unit peak (since the weights are ostensible). The elements' orientation freeze by $\kappa t\approx0.25$ into radially elongated, detached Gaussians; $n$ barely matters, and a ten times larger $r_\text{S}$ only squeezes sooner.}
\end{figure*}

This measurement seems like the prime candidate for a good Gaussian approximation of the Phase POVM, but also falls short; no amount of squeezing or displacing acts quickly enough to generate the ray-structure seen in Fig. \ref{fig:gaussianphase}. Instead, we observe a collection of azimuthally squeezed Gaussians circling the origin, as seen in Fig. \ref{fig:DisplaceAndSqueeze}; by $\kappa t\approx0.25$ every element has frozen, with an aspect ratio of $\approx3$ far from the origin and nearly round near it, aimed at the origin to within $0.3^\circ$ but detached from it, its radial width only about a fifth of its distance from the origin. A higher power $n$ hardly matters, and a ten times stronger squeeze only acts sooner (far-out aspect ratios $2.9$, $3.0$ and $4.0$ for the three rows of Fig. \ref{fig:DisplaceAndSqueeze}). Making the gains grow with the purification, e.g. $\propto e^{{\rm Re}\,\rho_t}$, makes matters worse: the squeeze feedback itself accelerates the purification, and the coupled equations run away in finite time. This can be understood as follow; unitary feedback never acts on the POVM element directly, only through later measurement steps, and the heterodyne measures the radial quadrature at full rate initially. Since no measurement in $\mathrm{IWH}^{\textbf{12}}$ can stretch a quadrature of a POVM element (the Hermitian quadratic part $hH_0+sA+\bar sA^\dagger$ of every forward generator obeys $|s|\le-h$), the radial information acquired in the first $\kappa t\sim1$ can never be undone; to obtain rays, one must never acquire any amplitude information in the first place e.g. as in homodyne.\\

\subsection{Adaptidyne 3: Wiseman-style adaptive homodyne}

A variant of Wiseman's celebrated approximate phase measurement constructed from adaptive homodyne \cite{Wiseman1995} lives naturally within $\mathrm{IWH}^{\textbf{12}}$ as well. It is the $\gamma\to\infty$ limit of Eq. \ref{GaussianPhase}: the local oscillator is always set at $90^\circ$ to the running displacement $\mu_t$, so that the homodyne measures the quadrature that carries the phase. Thus, the forward generator is
\begin{widetext}
\bea\label{WisemanGenerator}
  \RInv{\updelta}^{\text{A3}}_t &=& -\RInv{H_o}\Bigl(2\kappa_h+\tfrac{\kappa_p}{2}\Bigr)dt
     \;+\;\RInv{1}\Bigl(\kappa_h+\tfrac{\kappa_p}{4}\Bigr)dt
     \;+\;\frac{\mu^{*2}_t}{|\mu_t|^2}\,\RInv{A}\,\kappa_p\,dt
     \;+\;\RInv{a}\left(2\sqrt{\kappa_h}\,dZ^*_t
     \;-\; i\frac{\mu^*_t}{|\mu_t|}\,\RInv{a}\,\sqrt{\kappa_p}\,dW_t\right).
\eea
\end{widetext}

The last term is a homodyne measurement with an adaptive local-oscillator phase $e^{-i\Phi_t}=-i\mu^*_t/|\mu_t|$, and the $\RInv{A}$ term is the back-action from this. The feedback coefficients are evaluated before each increment (i.e. non-anticipating), and the weak heterodyne seeds the direction at $t=0$, where $\mu^*_t=0$. Evaluating Eq. \ref{StochasticEqGenerator} with Eqs. \ref{WierbienHCInv:zeta}-\ref{WierbienHCInv:z} gives the same equations as Eqs. \ref{AdapativeGaussianSDEs:dzeta}-\ref{AdapativeGaussianSDEs:dz}, but with
\bea
u_t^{1/2}\rightarrow (u_t')^{1/2}\equiv -i\mu_t^*/|\mu_t|.
\eea


As for adaptidyne 1, $\omega_t=\nu_t=0$ and $\rho_t$, $z_t$ are deterministic, so that only $d\zeta^*_t$ and $d\mu^*_t$ move, and the POVM elements again have the Husimi function of Eq. \ref{AdaptiveQ}.


The phase of the homodyne kicks is set by the running estimate $\mu^*_s$ itself, which makes this filter nonlinear in the record even though each kick is linear in $dW_s$.

\begin{figure*}[t]
\centering
\includegraphics[width=\linewidth]{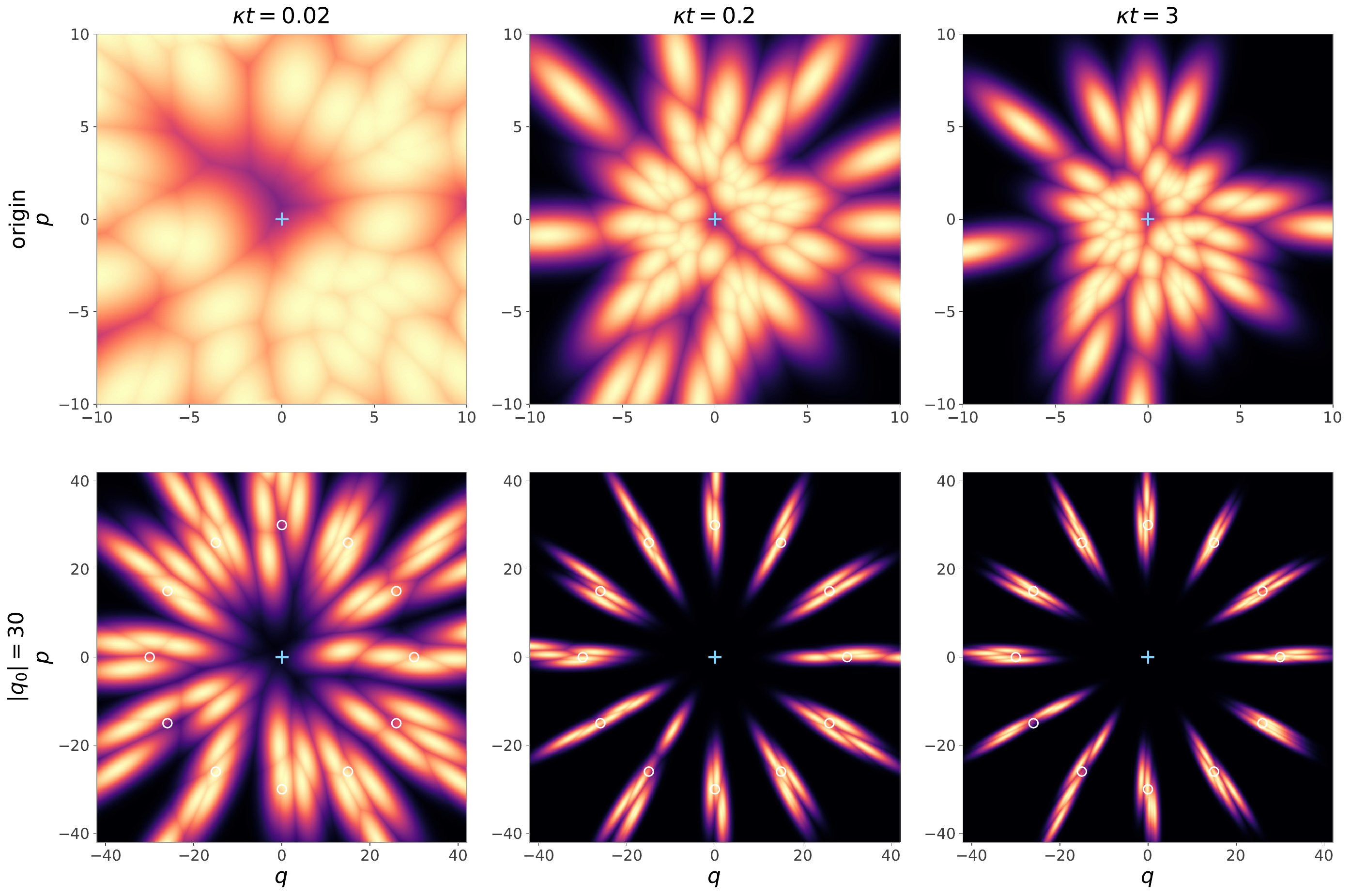}
\caption{\label{fig:wiseman} The Q-functions of the POVM elements of adaptidyne 3 (Wiseman's adaptive homodyne, Eq. \ref{WisemanGenerator}) at $\kappa t=0.02$, $0.2$ and $3$, with $\kappa_h=0.0025\kappa$ and $\kappa_p=3.99\kappa$. Top: 60 Kraus operators sampled from the KOD (note the smaller window). Bottom: five Kraus operators sampled with the ostensible distribution of Sec. \ref{sec:trick} centered on each of 12 points on the ring $|q_0|=30$ (white circles). Every element is radial, but it is a short amplitude-localized segment rather than a ray: far from the origin it stops well short of it.}
\end{figure*}

\begin{figure*}[t]
\centering
\includegraphics[width=\linewidth]{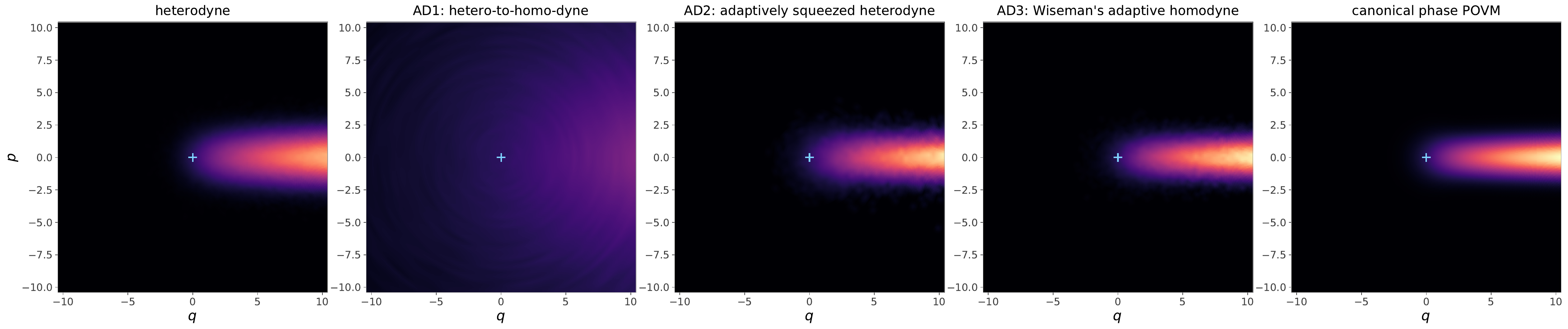}
\caption{\label{fig:coarsegrained} The Q function of a radially coarse-grained phase POVM element $\Pi_{\varphi=0}$, constructed by summing POVM element representations $K_g^\dagger K_g$ over all measurement records whose element is centered at angle $\theta=0$, sampled from the KOD centered on the origin so that they are properly weighted i.e. scaled by the KOD and measure. Shown, on a common square-root color scale, are GGW heterodyne, adaptidyne 1 (hetero-to-homo-dyne), adaptidyne 2 (adaptively squeezed heterodyne), and adaptidyne 3 (Wiseman's adaptive homodyne)---all at $\kappa t=3$ (with $\kappa_h=0.0025\kappa$ and $\kappa_p=3.99\kappa$)---as well as the canonical Phase POVM Q function. Coarse-graining over amplitude turns heterodyne and adaptidynes 2 and 3 into rays from the origin close to the canonical one, whereas adaptidyne 1, whose individual elements are rays, carries almost no phase information.}
\end{figure*}

\begin{figure}[b]
\centering
\includegraphics[width=\linewidth]{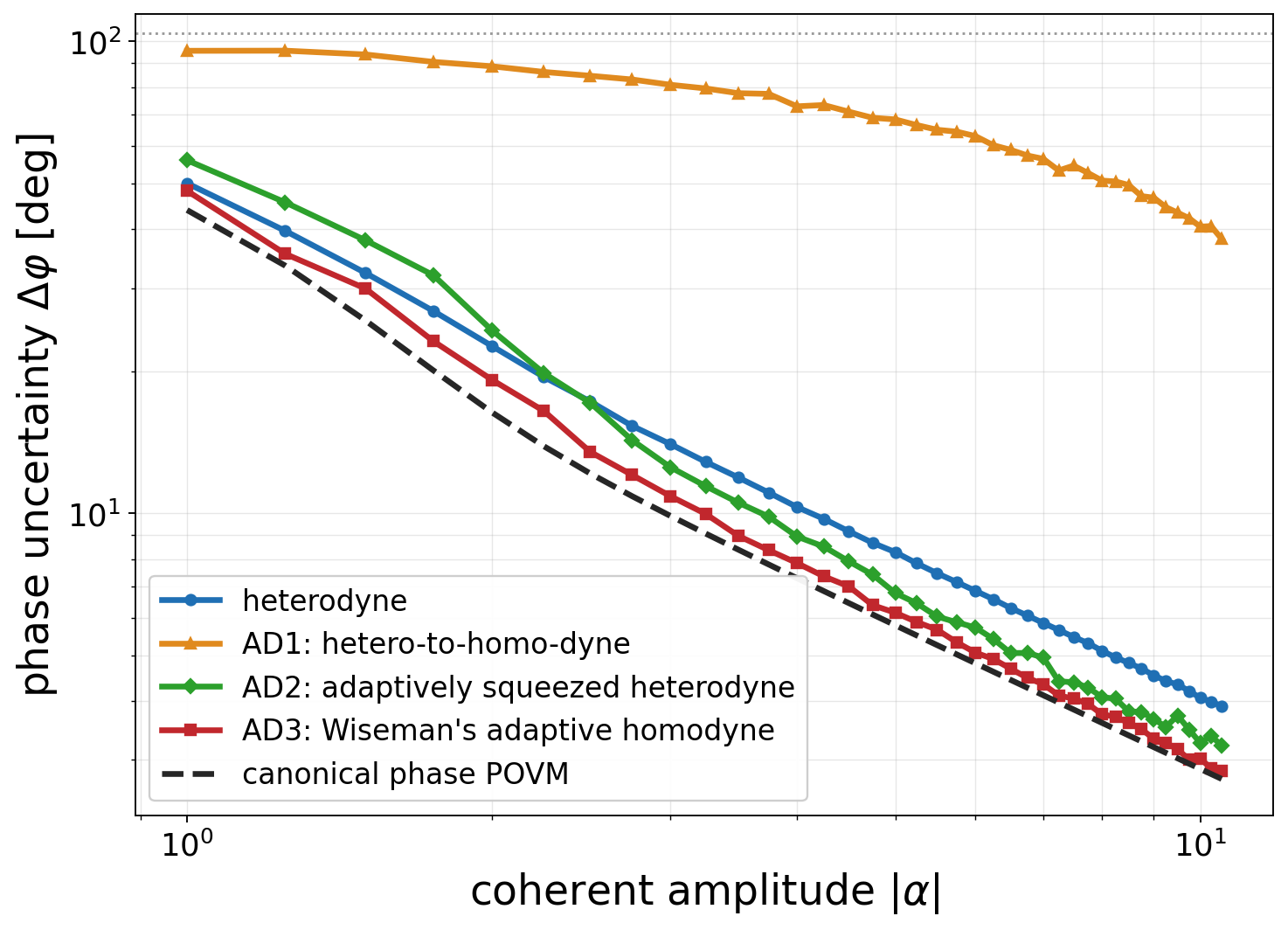}
\caption{\label{fig:phaseuncertainty} Phase uncertainty (the standard deviation of the outcome phase, i.e. the angle of the coarse grained POVM element $\Pi_{\varphi}$) for coherent inputs $\ket{\alpha}$: heterodyne, our three adaptidynes (at $\kappa t=3$, $\kappa_h=0.0025\kappa$, and $\kappa_p=3.99\kappa$), and the canonical Phase POVM (dashed). Wiseman's scheme in adaptidyne 3 closes $76$--$88\%$ of the gap between heterodyne and the canonical Phase POVM for $|\alpha|\ge3$, while adaptidyne 1 stays close to a uniform guess. Adaptidyne 2 outperforms heterodyne and comes close to Wiseman's scheme as $|\alpha|$ increases. }
\end{figure}

Adaptidyne 3 captures the other half of the Phase POVM missed by the two other schemes. Because the homodyne keeps measuring the quadrature perpendicular to the current displacement, every element is radial at every distance, and the phase statistics approach the canonical ones as seen in Fig. \ref{fig:phaseuncertainty}.
Coarse-grained over amplitude, its effective Phase POVM elements $\Pi_\varphi$ are, like the canonical one, rays from the origin (Fig. \ref{fig:coarsegrained}). The individual non-coarse grained POVM elements, however, are not rays (Fig. \ref{fig:wiseman}): they are amplitude-localized segments whose radial width $\sigma_r$ grows only like the square root of the radius. 
Emphatically, each element therefore still measures the amplitude, which a canonical Phase POVM element does not, resulting in suboptimal performance compared to the true Phase POVM. Although built from homodyne, adaptidyne 3 is trace class; as in slomodyne, its rotating local oscillator measures more than one quadrature and removes it instantly from the homosurface of non-trace-class objects at the very first time-step.

\vspace{1em}
Taken together, the three adaptidynes illustrate why a Gaussian approximation of the Phase POVM is so hard to build from continuous measurement. A canonical Phase POVM element has two properties: it is a ray from the origin, blind to the amplitude, and the family resolves the phase as sharply as quantum mechanics allows. Adaptidyne 1 has the first near the origin but not the second; adaptidyne 3 has the second but not the first; adaptidyne 2, whose elements are aimed at the origin but detached from it, does not have the first.

Emphatically, the obstruction is the same in each case: to aim a homodyne at the phase quadrature, the measurement record must first fix the direction of the POVM element it is building, and the measurement that fixes the direction also measures the amplitude. Locking early (adaptidyne 1) avoids measuring the amplitude but aims the elements at random; following continuously (adaptidyne 3) aims them correctly but keeps measuring the amplitude; and unitary feedback (adaptidyne 2) cannot undo amplitude information already acquired. Since every element generated within $\mathrm{IWH}^{\textbf{12}}$ is Gaussian while canonical phase states are not, no adaptidyne can reach the canonical Phase POVM exactly. What remains open is the best compromise, and in particular whether some forward generator gets the elements far from the origin---which matter most for phase estimation---closer to correct than the Wiseman-like scheme.

Here we have given several paths towards achieving a trace-class approximation of the Phase POVM. While it is still possible that a non-trace-class implementation is possible, it is challenging to see how this would work within $\mathrm{IWH}^{\textbf{12}}$; the squeezing must act extremely rapidly, as even a small amount of rotation (as in slomodyne) or measurement of a second quadrature (as in heterodyne) is sufficient to break the non-trace-class property, and---to restate our conjecture---we believe that, while it is possible for a non-trace class measurement to become trace-class, the inverse is not possible; one cannot reenter the homosurface once it is left.

\section{A Trick to Sample Phase Space Far from the Origin}\label{sec:trick}

As discussed above and illustrated in Fig. \ref{fig:samplingregionsforHetero}, the KOD naturally samples points in phase space near the origin by construction; the Wiener increments are mean-zero and since these drive the stochastic displacement of heterodyne, they do not cover phase space evenly. However, the POVM \emph{must} cover phase space uniformly; how is this consistent?

The answer is that the choice of division of the physical Kraus operator $d\Omega$ into a representation of a group element $K$ and a measure is arbitrary i.e. a choice of gauge. This sort of feature has been emphasized in sampling stochastic processes by Wiseman \cite{Wiseman1996}, who calls it ostensibility. Here, our KOD is ostensible; while it appears real and fundamental due to its naturalness, a deeper inspection suggests its total arbitrariness.\footnote{While the choice of gauge is arbitrary, we note another interesting case is to force the positive piece of the group element to be normalized so that the POVM element $\pi(g)$ corresponds to a normalized quantum state; in the limit that $\hbar\rightarrow0$ so that the POVM elements are orthogonal, this allows one to read the measure as a maximum efficiency with which any state can be detected, in analogy with single photon detection \cite{vanEnk2017,ProppNet,ProppPOVM}.}

More technically, the answer is that the KOD carries with it an anti-Gaussian weighting in the center coordinate ($z$ in Harish Chandra coordinates) of the Kraus operator representations; Kraus trajectories far from the origin are weighted higher to compensate for being sampled more rarely. Emphatically, this is a feature, not a bug, and one we may use. For a further in-depth discussion, we refer the reader to Appendix \ref{app:Sampling}; for our purposes, it suffices to observe that the Gaussian measure can be rewritten in terms of a shift 

\bea\label{HaarMeasureWithShift}
d\mu(d\vec{W}_t) &=& \frac{d^n (d\vec{W}_t)}{(2\pi dt)^{n/2}} e^{-\frac{|d\vec{W}_t|^2}{2dt}}   \\
&=& d\mu(d\vec{W}_t - 2q\vec{A} dt) e^{-2q\,d\vec{W}_t \cdot \vec{A} + 2 q^2 |\vec{A}|^2 dt}.\nonumber 
\eea Thus, instead of sampling an operation defined by $\updelta_t$ from the mean-zero distribution $d\mu(d\vec{W}_t)$, we can (by Girsanov's theorem \cite{Girsanov1960}) sample an operation $\updelta_t \rightarrow \updelta_t + \bigl(-q\,d\vec{W}_t \cdot \vec{A} + q^2 |\vec{A}|^2 dt\bigr)\id$ from the distribution with mean $\expect{d\vec{W}_t} = 2q\vec{A}dt$; the weight enters with half the exponent of Eq. \ref{HaarMeasureWithShift} because $d\Omega$ carries with it the measure $\sqrt{d\mu}$. Thus by varying $q\vec{A}$ (which is suggestively written reminiscent of a vector potential), we can state-independently sample Kraus trajectories from regions far from the origin of the phase plane. The cost is that the left-invariant measure is no longer stationary but co-evolves with the trajectories. Here that evolution is deterministic but, in principle, $\vec{A}$ could be any function of the measurement record or coordinates so that the trajectories sampled follow regions of interest as the measurement evolves.

\section{A Preview of Measurement Geometry}

The canonical projection reveals a deeper structure to the geometry of measurement: a positive-operator-valued base manifold, and over each of its points a fiber comprised of post-measurement unitaries. Hence, the $12$ dimensional $\mathrm{IWH}^{\textbf{12}}$ is a fiber bundle, composed of a $6$ dimensional POV Manifold and a $6$ dimensional fiber. These latter dimensions correspond to the post-measurement unitary, now including the rotation from $e^{\RInv{iH_0}}$ and phase from $e^{\RInv{i1}}$ in addition to the two-dimensional squeeze and two-dimensional displacement.\footnote{Technically this is an oversimplification; movement in the direction $\RInv{i1}$ never occurs, so it is not in the adjoint bundle at all; the adjoint $\mathrm{IWH}$ is, effectively, $11$ dimensional.} The post-measurement unitary undergoes transport along with the POVM element as they move through the POV Manifold. This view of measurement as fiber bundle aligns deeply with the original view of measurement put forward by Riemann \cite{Riemann1868}, inspired by his reading of Euclid \cite{Euclid}; measurement was a key part of Riemann's vision of geometry, with the fiber bundle \cite{Steenrod1951} being a modern reframing of parallel transport \cite{LeviCivita1917} i.e. non-Euclidean geometry. It is thus satisfying to see, over a century and a half after Riemann's project, measurement returns to a geometric setting in the realm of non-relativistic quantum theory. Emphatically, this non-Euclidean geometric structure---which is often considered to be limited in scope to general relativity---is also at the heart of how diffusive measurement processes generate POVMs i.e. learning.

Returning to our problem of quantum optics, we emphasize that the Harish-Chandra-like coordinates that we've used here to generate our Kraus trajectories are simply ill-suited to the study of the $6$ dimensional POV Manifold. While derivation of SDEs in Harish-Chandra-like coordinates are algebraically simple, the coordinates themselves obscure the POVM element and the post-measurement unitary i.e. the fiber. Working in Iwasawa or Cartan-Berger coordinates, a POVM element trajectory is naturally accompanied with a parallel transport of the post-measurement unitary. In this work, we have learned that---while the Cartan-Berger coordinates are pathological for non-trace class measurements (i.e. homodyne)---the Iwasawa coordinates maintain a faithful description throughout for all measurements in $\mathrm{IWH}^{\textbf{12}}$. This now sets us up for the project of describing this parallel transport of POVM elements and studying their trajectories in a forthcoming publication, both motivating the introduction of Iwasawa coordinates here to the literature of quantum optics and pointing the way towards the genuine geometry of quantum optical measurements.

\section{The Born Rule Revisited}

Before we conclude, we briefly note another geometric feature of our work, first presented in Ref. \cite{Jackson2023}; given the decomposition in terms of a KOD (or another ostensible distribution) we can re-write the Born rule \cite{Born1926}. The infinitesimal probability to observe a measurement outcome corresponding to $g\in {\mathrm{IWH}^{\textbf{12}}}$ is given by

\bea\label{BornRule2}
dP_t(g|\rho_0) =  \underbrace{d_{\mathrm{L}} g D_t(g)}_{\text{Dynamical}} \;\underbrace{{\rm Tr} \left(K_{g}^\dagger K_{g} \rho_0 \right)}_{\text{Geometric}}
\eea with a clear separation between a time-evolving dynamical component \emph{completely} characterized by the KOD $D_t(g)$, and a time-independent geometric component where the initial state is traced with the group element associated with the coordinates.

We can use our Kraus trajectories to sample the Born rule with the helpful property $\int d_\mathrm{L} g D_t(g) f(g) \approx \frac{1}{N} \sum_{i=1}^N f(g_{t,i})$, with $g_{t,i}$ the sampled group element evolved for a time $t$.

Though we call these positive operators, we emphasize these are pre-Hilbert space, where positivity is defined. These are just as universal as rotations, which are represented by unitaries. Thus, we see that the structure of the Born rule has, at its core, nothing to do with Hilbert space, and can exist independently from it; the group structure of $\mathrm{IWH}^{\textbf{12}}$ is, in a sense, at least equivalently primitive. We also see that the statistical role of the quantum state in the Born rule is to temper the KOD, generating a genuinely sampleable distribution.\\

It is widely accepted and, sometimes, acknowledged that the standard extensions to the Heisenberg picture are persistently state dependent for open quantum systems e.g. commutators aren't preserved in the adjoint master equation \cite{BreuerPetruccione2002}, and it is only expectation values that can be interpreted as quantum mechanical. The decomposition presented in Eq. \ref{BornRule2} points towards state-independent open quantum system dynamics (e.g. to quantify non-Markovianity \cite{Propp2022}, analyze decoherence channels for convexity \cite{Propp2023}, or construct a state-independent, measurement-centric analog of Quantum Darwinism \cite{Zurek2025}); one merely integrates over a region of phase space on both sides to achieve a description of a decoherence channel where the state does not evolve in time, only the distribution on the right hand side and the conditional probability on the left hand side. This is very much in the spirit of QBism's (honest) treatment of quantum probabilities \cite{FuchsSchack2013,FuchsMerminSchack2014,DeBrota2018,DeBrota2023decoherence,DeBrotaList2026,DeBrota2026locality}, but with substantially more mathematical elegance. Additionally, the Heisenberg picture has certain advantages over the Schr\"{o}dinger picture for relativistic quantum theory \cite{Rovelli1991}, and we are curious to see whether Eq. \ref{BornRule2} and the state-independent-yet-time-evolving representation of the channel

\bea\label{mathcalZchannel}
\mathcal{Z}_{t} = \int_{\mathrm{IWH}^{\textbf{12}}} d_{\mathrm{L}} g D_t(g) \; K_{g} \odot K_{g}^\dagger
\eea both inherit these advantages and provide new tools for studying measurement in relativistic quantum theory \cite{Unruh1976,DeWitt1979, UnruhWald1984}.

\section{Conclusions and Outlook}

Learning is a locomotion of information; in continuous measurement, an observer is transformed by their repeated interaction with a system of interest i.e. learning.  Physically, this is the result of repeated interactions between a meter and system. In the spirit of Lie and Klein \cite{LieEngel1888,Klein1872}, who placed transformation groups at the foundation of geometry, the knowledge created during measurement is not revelation (i.e. of a hidden variable), but transformation of something that already exists. In the Kraus trajectories we've studied here, the initial Kraus operator is always the identity, corresponding to pure ignorance, which is dynamically transformed into states of maximum knowledge. More generally, systems---even those without embodiment such as probabilistic automata \cite{Dupont2005}---make use of filtering, which is precisely the process of converting raw measurement outcomes (e.g. $dW_t$'s) into a more usable form of information (e.g. coordinates in the manifold). In this work, we have focused on the principal instrument bundle for quantum optics: $\mathrm{IWH}^{\textbf{12}}$. This is principal in precisely the sense that it is not chaotic; there are a finite number of coordinates, such that a finite memory stack is needed to explore the group. This makes $\mathrm{IWH}^{\textbf{12}}$ an example of what is called an automatic group  \cite{Jackson2025}; for us, this manifests in the easily traversable and interpretable manifold structure we have explored in this project. 

In this paper, we have leveraged this manifold structure  and the Principle of Instrument Universality to explore a variety of quantum optical measurements without specifying a particular quantum state. This has allowed us to study---for the first time---these quantum optical measurements on their own terms, and generate real measurement records without having to specify any input state whatsoever. By placing our analysis on a generalized phase space, we circumvent the problems associated with type-III von Neumann algebras regarding non-trace class measurements. Emphatically, these ``exotic'' mathematical objects already show up in the physics of a single quantum optical mode; one doesn't need to go all the way to quantum gravity to encounter them! We are here able to describe their purification quantitatively using a here-introduced geometric POVM purity, making use of the here-introduced Iwasawa coordinates for $\mathrm{IWH}^{\textbf{12}}$. This innovation sets the stage for further exploration of the POV Manifold e.g. of open sets, and how the point set topology ensures that the manifold of quadrature eigenstates are their own boundary (which we will prove in a subsequent publication). Further connections are pointed towards by the observation that, in deSitter space, trace-class operators are time-like, but non-trace-class operators are light-like.

Our analysis in this paper is limited to $\mathrm{IWH}^{\textbf{12}}$. Here, we have investigated the well-studied homodynes and heterodynes, our newly developed slomodynes, as well as three adaptidynes, each capturing part of the Phase POVM and the last of which reproduces Wiseman's adaptive homodyne scheme \cite{Wiseman1995,WisemanKillip1998,Armen2002}. This last development necessitated the inclusion of adaptivity, which can now be described in a state-independent manner for quantum instruments for the first time to the authors' knowledge; we emphasize that this is the tip of the iceberg, and that there are more bespoke adaptive measurements within $\mathrm{IWH}^{\textbf{12}}$ to discover and engineer.

Beyond $\mathrm{IWH}^{\textbf{12}}$, there are also more quantum optical measurements  to discover. Future work will take us into the chaotic extension e.g. by inclusion of Kerr interactions and other higher-order non-linearities. Now the group will not close, and new technology must be introduced to describe and calculate Kraus trajectories. It will also be interesting to study decoherence concurrent with measurement, corresponding to multiple Lindblad channels with a partial trace. A connection is also ready to be made to thermodynamics; since each Kraus trajectory purifies, it can be associated with a reduction in entropy, and thus a thermodynamic minimum cost.

Historically, quantum processes were divided by von Neumann into two types: measurements (process $\text{\fontfamily{phv}\selectfont\bfseries 1}$) and unitary dynamics (process $\text{\fontfamily{phv}\selectfont\bfseries 2}$) \cite{vonNeumann1955}. Many members of the contemporary quantum community grant ontological status to the latter, but not the former i.e. unitary evolution is real, but measurement is merely epistemic and thus illusory \cite{Gibney2025}. We hope this paper, by exploring the geometric structure of measurement, helps motivate and guide the refinement of a quantum ontology of learning by our philosophically inclined colleagues. 

The door to the geometry of quantum optics is now opened to us; it is up to us to walk on through to the other side and see what we find.

\section{Acknowledgments}

We are grateful to Carlton M. Caves and Steven van Enk for the training we received in these methods and their precursors, as well as for many productive and inspiring conversations. TzBP gratefully acknowledges support from the Quantum Software Consortium Ada Lovelace Fellowship. CSJ acknowledges support from the Perimeter Institute, the Stanford Q-FARM initiative, and the John Templeton Foundation Grant No.~63796, ``Quantum spacetime from the inside: embodied agents and the spacetimes they construct''.  Research at Perimeter Institute is supported in part by the Government of Canada through the Department of Innovation, Science
and Economic Development Canada and by the Province of Ontario through the Ministry of Colleges and Universities.  The opinions expressed in this publication are those of the authors and do not necessarily reflect the views of the John Templeton Foundation. 

\bibliography{MIPQO}

@article{ricci1895dei,
  title={Dei sistemi di congruenze ortogonali in una variet{\`a} qualunque},
  author={Ricci, G},
  journal={Memorie Acc. dei Lincei},
  volume={2},
  year={1895}
}

@misc{jammer1966conceptual,
  title={The conceptual development of quantum mechanics},
  author={Jammer, Max},
  year={1966},
  publisher={McGraw-Hill}
}

@misc{cassidy1992uncertainty,
  title={Uncertainty: The life and science of Werner Heisenberg},
  author={Cassidy, David C and Dresden, Max},
  year={1992},
  publisher={American Institute of Physics}
}

@article{nolte2010tangled,
  title={The tangled tale of phase space},
  author={Nolte, David D},
  journal={Physics today},
  volume={63},
  number={4},
  pages={33--38},
  year={2010},
  publisher={American Institute of Physics}
}

@article{d2002quantum,
  title={A quantum measurement of the spin direction},
  author={D'Ariano, GM and Presti, P Lo and Sacchi, MF},
  journal={Physics Letters A},
  volume={292},
  number={4-5},
  pages={233--237},
  year={2002},
  publisher={Elsevier}
}

@article{chantasri2018simultaneous,
  title={Simultaneous continuous measurement of noncommuting observables: Quantum state correlations},
  author={Chantasri, Areeya and Atalaya, Juan and Hacohen-Gourgy, Shay and Martin, Leigh S and Siddiqi, Irfan and Jordan, Andrew N},
  journal={Physical Review A},
  volume={97},
  number={1},
  pages={012118},
  year={2018},
  publisher={APS}
}

@article{Wiseman1995,
  title = {Adaptive Phase Measurements of Optical Modes: Going Beyond the Marginal {Q} Distribution},
  volume = {75},
  ISSN = {1079-7114},
  url = {http://dx.doi.org/10.1103/PhysRevLett.75.4587},
  DOI = {10.1103/physrevlett.75.4587},
  number = {25},
  journal = {Physical Review Letters},
  publisher = {American Physical Society (APS)},
  author = {Wiseman,  H. M.},
  year = {1995},
  month = Dec,
  pages = {4587–4590}
}

@article{Jackson2023,
  title = {Simultaneous Measurements of Noncommuting Observables: Positive Transformations and Instrumental Lie Groups},
  volume = {25},
  ISSN = {1099-4300},
  url = {http://dx.doi.org/10.3390/e25091254},
  DOI = {10.3390/e25091254},
  number = {9},
  journal = {Entropy},
  publisher = {MDPI AG},
  author = {Jackson,  Christopher S. and Caves,  Carlton M.},
  year = {2023},
  month = Aug,
  pages = {1254}
}

@book{Newton1687,
  author    = {Newton, Isaac},
  title     = {Philosophi{\ae} Naturalis Principia Mathematica},
  publisher = {Royal Society},
  address   = {London},
  year      = {1687}
}

@book{Lagrange1788,
  author    = {Lagrange, Joseph-Louis},
  title     = {M{\'e}chanique analitique},
  publisher = {Veuve Desaint},
  address   = {Paris},
  year      = {1788}
}

@article{Galois1846,
  author  = {Galois, {\'E}variste},
  title   = {{\OE}uvres math{\'e}matiques d'{\'E}variste {G}alois},
  journal = {Journal de Math{\'e}matiques Pures et Appliqu{\'e}es},
  volume  = {11},
  pages   = {381--444},
  year    = {1846},
  note    = {Includes the M{\'e}moire sur les conditions de r{\'e}solubilit{\'e} des {\'e}quations par radicaux (written 1830--1831), ed. J. Liouville}
}

@article{Weyl1927,
  author  = {Weyl, Hermann},
  title   = {Quantenmechanik und {G}ruppentheorie},
  journal = {Zeitschrift f{\"u}r Physik},
  volume  = {46},
  pages   = {1--46},
  year    = {1927}
}

@book{Weyl1928,
  author    = {Weyl, Hermann},
  title     = {Gruppentheorie und {Q}uantenmechanik},
  publisher = {S. Hirzel},
  address   = {Leipzig},
  year      = {1928}
}

@article{Wigner1932,
  author  = {Wigner, Eugene},
  title   = {On the Quantum Correction For Thermodynamic Equilibrium},
  journal = {Physical Review},
  volume  = {40},
  pages   = {749--759},
  year    = {1932}
}

@book{Wigner1931,
  author    = {Wigner, Eugen},
  title     = {Gruppentheorie und ihre {A}nwendung auf die {Q}uantenmechanik der {A}tomspektren},
  publisher = {Vieweg},
  address   = {Braunschweig},
  year      = {1931}
}

@article{Wigner1939,
  author  = {Wigner, Eugene},
  title   = {On Unitary Representations of the Inhomogeneous {L}orentz Group},
  journal = {Annals of Mathematics},
  volume  = {40},
  pages   = {149--204},
  year    = {1939}
}

@article{Moyal1949,
  author  = {Moyal, J. E.},
  title   = {Quantum mechanics as a statistical theory},
  journal = {Mathematical Proceedings of the Cambridge Philosophical Society},
  volume  = {45},
  pages   = {99--124},
  year    = {1949}
}

@article{Glauber1963coherence,
  author  = {Glauber, Roy J.},
  title   = {The Quantum Theory of Optical Coherence},
  journal = {Physical Review},
  volume  = {130},
  pages   = {2529--2539},
  year    = {1963}
}

@article{Warszawski2020,
  author  = {Warszawski, Prahlad and Wiseman, Howard M. and Doherty, Andrew C.},
  title   = {Solving quantum trajectories for systems with linear {H}eisenberg-picture dynamics and {G}aussian measurement noise},
  journal = {Physical Review A},
  volume  = {102},
  pages   = {042210},
  year    = {2020}
}

@article{Unruh1976,
  author  = {Unruh, W. G.},
  title   = {Notes on black-hole evaporation},
  journal = {Physical Review D},
  volume  = {14},
  pages   = {870--892},
  year    = {1976}
}

@article{UnruhWald1984,
  author  = {Unruh, William G. and Wald, Robert M.},
  title   = {What happens when an accelerating observer detects a {R}indler particle},
  journal = {Physical Review D},
  volume  = {29},
  pages   = {1047--1056},
  year    = {1984}
}

@article{Jackson2022photodetector,
  author  = {Jackson, Christopher S.},
  title   = {The photodetector, the heterodyne instrument, and the principle of instrument autonomy},
  journal = {arXiv preprint},
  year    = {2022},
  note    = {arXiv:2210.11100}
}

@incollection{DeWitt1979,
  author    = {DeWitt, Bryce S.},
  title     = {Quantum gravity: the new synthesis},
  booktitle = {General Relativity: An Einstein Centenary Survey},
  editor    = {Hawking, S. W. and Israel, W.},
  publisher = {Cambridge University Press},
  address   = {Cambridge},
  pages     = {680--745},
  year      = {1979}
}

@article{Benoist2019,
  author  = {Benoist, Tristan and Fraas, Martin and Pautrat, Yan and Pellegrini, Cl{\'e}ment},
  title   = {Invariant measure for quantum trajectories},
  journal = {Probability Theory and Related Fields},
  volume  = {174},
  pages   = {307--334},
  year    = {2019}
}

@article{Benoist2021,
  author  = {Benoist, Tristan and Fraas, Martin and Pautrat, Yan and Pellegrini, Cl{\'e}ment},
  title   = {Invariant measure for stochastic {S}chr{\"o}dinger equations},
  journal = {Annales Henri Poincar{\'e}},
  volume  = {22},
  pages   = {347--374},
  year    = {2021},
  note    = {arXiv:1907.08485}
}

@misc{CavesNotes2017,
  author       = {Caves, Carlton M.},
  title        = {Quantum Information Science: Lecture Notes},
  howpublished = {Course lecture notes, University of New Mexico, Spring 2017, \url{http://info.phys.unm.edu/~caves/courses/qinfo-s17/lectures/}},
  year         = {2017},
  note         = {The $\odot$ notation for superoperators appears in Lectures 15--17}
}

@book{Wiener1948,
  author    = {Wiener, Norbert},
  title     = {Cybernetics: Or Control and Communication in the Animal and the Machine},
  publisher = {MIT Press / John Wiley \& Sons},
  address   = {Cambridge, MA / New York},
  year      = {1948}
}

@article{Rosenblueth1943,
  author  = {Rosenblueth, Arturo and Wiener, Norbert and Bigelow, Julian},
  title   = {Behavior, Purpose and Teleology},
  journal = {Philosophy of Science},
  volume  = {10},
  number  = {1},
  pages   = {18--24},
  year    = {1943}
}

@techreport{vonNeumann1945,
  author      = {von Neumann, John},
  title       = {First Draft of a Report on the {EDVAC}},
  institution = {Moore School of Electrical Engineering, University of Pennsylvania},
  address     = {Philadelphia},
  year        = {1945}
}

@book{vonNeumann1966,
  author    = {von Neumann, John},
  title     = {Theory of Self-Reproducing Automata},
  editor    = {Burks, Arthur W.},
  publisher = {University of Illinois Press},
  address   = {Urbana},
  year      = {1966}
}

@article{Turing1950,
  author  = {Turing, A. M.},
  title   = {Computing Machinery and Intelligence},
  journal = {Mind},
  volume  = {59},
  number  = {236},
  pages   = {433--460},
  year    = {1950}
}

@article{Glauber1963,
  author  = {Glauber, Roy J.},
  title   = {Coherent and Incoherent States of the Radiation Field},
  journal = {Physical Review},
  volume  = {131},
  pages   = {2766--2788},
  year    = {1963}
}

@article{Perelomov1972,
  author  = {Perelomov, A. M.},
  title   = {Coherent states for arbitrary {L}ie group},
  journal = {Communications in Mathematical Physics},
  volume  = {26},
  pages   = {222--236},
  year    = {1972}
}

@article{Boltzmann1872,
  author  = {Boltzmann, Ludwig},
  title   = {Weitere {S}tudien {\"u}ber das {W}{\"a}rmegleichgewicht unter {G}asmolek{\"u}len},
  journal = {Sitzungsberichte der Kaiserlichen Akademie der Wissenschaften in Wien, Mathematisch-Naturwissenschaftliche Classe},
  volume  = {66},
  pages   = {275--370},
  year    = {1872}
}

@article{Jacobi1837,
  author  = {Jacobi, C. G. J.},
  title   = {{\"U}ber die {R}eduction der {I}ntegration der partiellen {D}ifferentialgleichungen erster {O}rdnung zwischen irgend einer {Z}ahl {V}ariabeln auf die {I}ntegration eines einzigen {S}ystemes gew{\"o}hnlicher {D}ifferentialgleichungen},
  journal = {Journal f{\"u}r die reine und angewandte Mathematik},
  volume  = {17},
  pages   = {97--162},
  year    = {1837}
}

@book{Jacobi1866,
  author    = {Jacobi, C. G. J.},
  title     = {Vorlesungen {\"u}ber {D}ynamik},
  editor    = {Clebsch, A.},
  publisher = {G. Reimer},
  address   = {Berlin},
  year      = {1866},
  note      = {Lectures delivered 1842--1843}
}

@book{Tacquet1651,
  author    = {Tacquet, Andr{\'e}},
  title     = {Cylindricorum et annularium libri {IV}},
  publisher = {Jacobus Meursius},
  address   = {Antwerp},
  year      = {1651}
}

@book{Alexander2014,
  author    = {Alexander, Amir},
  title     = {Infinitesimal: How a Dangerous Mathematical Theory Shaped the Modern World},
  publisher = {Scientific American / Farrar, Straus and Giroux},
  address   = {New York},
  year      = {2014}
}

@article{Clausius1850,
  author  = {Clausius, Rudolf},
  title   = {{\"U}ber die bewegende {K}raft der {W}{\"a}rme und die {G}esetze, welche sich daraus f{\"u}r die {W}{\"a}rmelehre selbst ableiten lassen},
  journal = {Annalen der Physik},
  volume  = {155},
  pages   = {368--397},
  year    = {1850}
}

@article{Clausius1865,
  author  = {Clausius, Rudolf},
  title   = {{\"U}ber verschiedene f{\"u}r die {A}nwendung bequeme {F}ormen der {H}auptgleichungen der mechanischen {W}{\"a}rmetheorie},
  journal = {Annalen der Physik},
  volume  = {201},
  pages   = {353--400},
  year    = {1865}
}

@incollection{Rovelli1991,
  author    = {Rovelli, Carlo},
  title     = {Is there incompatibility between the ways time is treated in general relativity and in standard quantum mechanics?},
  booktitle = {Conceptual Problems of Quantum Gravity},
  editor    = {Ashtekar, Abhay and Stachel, John},
  publisher = {Birkh{\"a}user},
  address   = {Boston},
  pages     = {126--140},
  year      = {1991}
}

@article{vanEnk2017,
  title = {Photodetector figures of merit in terms of POVMs},
  volume = {1},
  ISSN = {2399-6528},
  url = {http://dx.doi.org/10.1088/2399-6528/aa90ce},
  DOI = {10.1088/2399-6528/aa90ce},
  number = {4},
  journal = {Journal of Physics Communications},
  publisher = {IOP Publishing},
  author = {van Enk,  Steven J.},
  year = {2017},
  month = Nov,
  pages = {045001}
}

@book{Kraus1983,
  author = {Kraus, Karl},
  title = {States, Effects, and Operations: Fundamental Notions of Quantum Theory},
  series = {Lecture Notes in Physics},
  volume = {190},
  ISBN = {9783540387251},
  url = {http://dx.doi.org/10.1007/3-540-12732-1},
  DOI = {10.1007/3-540-12732-1},
  journal = {Lecture Notes in Physics},
  publisher = {Springer Berlin Heidelberg},
  year = {1983}
}

@book{dixmier1977calgebras,
  author    = {Dixmier, Jacques},
  title     = {{$C^*$}-algebras},
  translator= {Jellett, Francis},
  series    = {North-Holland Mathematical Library},
  volume    = {15},
  publisher = {North-Holland Publishing Company},
  address   = {Amsterdam},
  year      = {1977},
  isbn      = {978-0-7204-0762-4}
}

@article{ProppNet,
 title = {Quantum networks for single photon detection},
  author = {Propp, {\relax{Tz}}ula. B. and van Enk, Steven J.},
  journal = {Phys. Rev. A},
  volume = {100},
  issue = {3},
  pages = {033836},
  numpages = {19},
  year = {2019},
  month = {Sep},
  publisher = {American Physical Society},
}

@article{ProppPOVM,
  title = {How to project onto an arbitrary single-photon wave packet},
  author = {Propp, {\relax{Tz}}ula. B. and van Enk, Steven J.},
  journal = {Phys. Rev. A},
  volume = {102},
  issue = {5},
  pages = {053707},
  numpages = {13},
  year = {2020},
  month = {Nov},
  publisher = {American Physical Society},
}

@article{Wiseman1996,
  author = {Wiseman, H. M.}, title = {Quantum trajectories and quantum measurement theory},
  journal = {Quantum and Semiclassical Optics: Journal of the European Optical Society Part B}, volume = {8}, pages = {205--222}, year = {1996}}

@article{Belavkin1992,
  author = {Belavkin, Viacheslav P.}, title = {Quantum stochastic calculus and quantum nonlinear filtering},
  journal = {Journal of Multivariate Analysis}, volume = {42}, pages = {171--201}, year = {1992}}

@article{Dupont2005,
  author = {Dupont, P. and Denis, F. and Esposito, Y.}, title = {Links between probabilistic automata and hidden {M}arkov models: probability distributions, learning models and induction algorithms},
  journal = {Pattern Recognition}, volume = {38}, pages = {1349--1371}, year = {2005}}

@article{JacksonCaves2023Quantum,
  author  = {Jackson, Christopher S. and Caves, Carlton M.},
  title   = {How to perform the coherent measurement of a curved phase space by continuous isotropic measurement. {I}. {S}pin and the {K}raus-operator geometry of {SL}(2,{C})},
  journal = {Quantum},
  volume  = {7},
  pages   = {1085},
  year    = {2023},
  doi     = {10.22331/q-2023-08-08-1085}
}

@article{Shojaee2018,
  author  = {Shojaee, Ezad and Jackson, Christopher S. and Riofr{\'\i}o, Carlos A. and Kalev, Amir and Deutsch, Ivan H.},
  title   = {Optimal Pure-State Qubit Tomography via Sequential Weak Measurements},
  journal = {Physical Review Letters},
  volume  = {121},
  pages   = {130404},
  year    = {2018}
}

@book{Knapp2002,
  author    = {Knapp, Anthony W.},
  title     = {Lie Groups Beyond an Introduction},
  edition   = {2nd},
  series    = {Progress in Mathematics},
  volume    = {140},
  publisher = {Birkh{\"a}user},
  address   = {Boston},
  year      = {2002}
}

@article{Iwasawa1949,
  author  = {Iwasawa, Kenkichi},
  title   = {On some types of topological groups},
  journal = {Annals of Mathematics},
  volume  = {50},
  pages   = {507--558},
  year    = {1949}
}

@article{Poincare83,
  author  = {Poincar{\'e}, Henri},
  title   = {Sur un th{\'e}or{\`e}me de la th{\'e}orie g{\'e}n{\'e}rale des fonctions},
  journal = {Bulletin de la Soci{\'e}t{\'e} Math{\'e}matique de France},
  volume  = {11},
  pages   = {112--125},
  year    = {1883}
}

@article{Poincare99,
  author  = {Poincar{\'e}, Henri},
  title   = {Sur les groupes continus},
  journal = {Transactions of the Cambridge Philosophical Society},
  volume  = {18},
  pages   = {220--255},
  year    = {1899}
}

@book{Weyl13,
  author    = {Weyl, Hermann},
  title     = {Die {I}dee der {R}iemannschen {F}l{\"a}che},
  publisher = {B. G. Teubner},
  address   = {Leipzig},
  year      = {1913},
  note      = {English translation: \emph{The Concept of a Riemann Surface}, 3rd ed. (Dover, 2009)}
}

@book{Edwards1984,
  author    = {Edwards, Harold M.},
  title     = {Galois Theory},
  series    = {Graduate Texts in Mathematics},
  volume    = {101},
  publisher = {Springer-Verlag},
  address   = {New York},
  year      = {1984}
}

@article{Lie1880,
  author  = {Lie, Sophus},
  title   = {Theorie der {T}ransformationsgruppen {I}},
  journal = {Mathematische Annalen},
  volume  = {16},
  pages   = {441--528},
  year    = {1880}
}

@book{vonNeumann1955,
  author    = {von Neumann, John},
  title     = {Mathematical Foundations of Quantum Mechanics},
  note      = {Translated by R. T. Beyer},
  publisher = {Princeton University Press},
  address   = {Princeton, NJ},
  year      = {1955}
}

@article{Gibney2025,
  author  = {Gibney, Elizabeth},
  title   = {Physicists disagree wildly on what quantum mechanics says about reality, {N}ature survey shows},
  journal = {Nature},
  volume  = {643},
  number  = {8074},
  pages   = {1175--1179},
  year    = {2025},
  doi     = {10.1038/d41586-025-02342-y}
}

@book{Zurek2025,
  author    = {Zurek, Wojciech Hubert},
  title     = {Decoherence and Quantum {D}arwinism: From Quantum Foundations to Classical Reality},
  publisher = {Cambridge University Press},
  address   = {Cambridge},
  year      = {2025},
  doi       = {10.1017/9781009552868}
}

@misc{CartanDelphenich1922,
  author       = {Cartan, {\'E}lie},
  title        = {Lessons on Integral Invariants},
  note         = {Translated by D. H. Delphenich from \emph{Le{\c{c}}ons sur les invariants int{\'e}graux} (Hermann, Paris, 1922)},
  howpublished = {\url{http://www.neo-classical-physics.info/uploads/3/4/3/6/34363841/cartan_-_lessons_on_integral_invariants.pdf}},
  year         = {1922}
}

@book{Cartan1922,
  author    = {Cartan, {\'E}lie},
  title     = {Le{\c{c}}ons sur les invariants int{\'e}graux},
  publisher = {Hermann},
  address   = {Paris},
  year      = {1922}
}

@article{Hawkins1991,
  author  = {Hawkins, Thomas},
  title   = {Jacobi and the Birth of {L}ie's Theory of Groups},
  journal = {Archive for History of Exact Sciences},
  volume  = {42},
  number  = {3},
  pages   = {187--278},
  year    = {1991}
}

@article{Cartan1899,
  author  = {Cartan, {\'E}lie},
  title   = {Sur certaines expressions diff{\'e}rentielles et le probl{\`e}me de {P}faff},
  journal = {Annales Scientifiques de l'{\'E}cole Normale Sup{\'e}rieure},
  volume  = {16},
  pages   = {239--332},
  year    = {1899}
}

@article{Fraser2023,
  title = {Canonical transformations from Jacobi to Whittaker},
  volume = {77},
  ISSN = {1432-0657},
  url = {http://dx.doi.org/10.1007/s00407-022-00303-9},
  DOI = {10.1007/s00407-022-00303-9},
  number = {3},
  journal = {Archive for History of Exact Sciences},
  publisher = {Springer Science and Business Media LLC},
  author = {Fraser,  Craig and Nakane,  Michiyo},
  year = {2023},
  month = Jan,
  pages = {241–343}
}

@book{Borel2001,
  author    = {Borel, Armand},
  title     = {Essays in the History of {L}ie Groups and Algebraic Groups},
  series    = {History of Mathematics},
  volume    = {21},
  publisher = {American Mathematical Society},
  address   = {Providence, RI},
  year      = {2001}
}

@book{Chirikjian2009,
  author    = {Chirikjian, Gregory S.},
  title     = {Stochastic Models, Information Theory, and {L}ie Groups, Volume 1: Classical Results and Geometric Methods},
  series    = {Applied and Numerical Harmonic Analysis},
  publisher = {Birkh{\"a}user},
  address   = {Boston},
  year      = {2009}
}

@book{Hawkins2000,
  author    = {Hawkins, Thomas},
  title     = {Emergence of the Theory of {L}ie Groups: An Essay in the History of Mathematics 1869--1926},
  series    = {Sources and Studies in the History of Mathematics and Physical Sciences},
  publisher = {Springer},
  address   = {New York},
  year      = {2000},
  doi       = {10.1007/978-1-4612-1202-7}
}

@article{Jackson2026Notes,
  author  = {Jackson, Christopher S.},
  title   = {Quantum Measurement and Continuous {M}arkov Processes},
  journal = {arXiv preprint},
  year    = {2026},
  note    = {arXiv:2606.15958}
}

@book{Ackerman1975,
  author    = {Lie, Sophus},
  title     = {Sophus {L}ie's 1880 Transformation Group Paper},
  note      = {Translated by M. Ackerman, comments by R. Hermann},
  series    = {Lie Groups: History, Frontiers and Applications},
  volume    = {1},
  publisher = {Math Sci Press},
  address   = {Brookline, MA},
  year      = {1975}
}

@article{Jackson2023SPQM,
  title = {Simultaneous Momentum and Position Measurement and the Instrumental Weyl-Heisenberg Group},
  volume = {25},
  ISSN = {1099-4300},
  url = {http://dx.doi.org/10.3390/e25081221},
  DOI = {10.3390/e25081221},
  number = {8},
  journal = {Entropy},
  publisher = {MDPI AG},
  author = {Jackson,  Christopher S. and Caves,  Carlton M.},
  year = {2023},
  month = Aug,
  pages = {1221}
}

@book{Bourbaki71,
  author    = {Bourbaki, Nicolas},
  title     = {{\'E}l{\'e}ments de math{\'e}matique. {G}roupes et alg{\`e}bres de {L}ie, {C}hapitre 1},
  publisher = {Hermann},
  address   = {Paris},
  year      = {1971},
  note      = {English translation: \emph{Lie Groups and Lie Algebras, Chapters 1--3} (Springer, 1989)}
}

@article{Jackson2025,
  title = {Sequential Quantum Measurements and the Instrumental Group Algebra},
  volume = {7},
  ISSN = {2624-960X},
  url = {http://dx.doi.org/10.3390/quantum7040057},
  DOI = {10.3390/quantum7040057},
  number = {4},
  journal = {Quantum Reports},
  publisher = {MDPI AG},
  author = {Jackson,  Christopher S.},
  year = {2025},
  month = Nov,
  pages = {57}
}

@book{Helgason1978,
  author    = {Helgason, Sigurdur},
  title     = {Differential Geometry, {L}ie Groups, and Symmetric Spaces},
  publisher = {Academic Press},
  address   = {New York},
  year      = {1978}
}

@article{WisemanMilburn1993,
  author  = {Wiseman, H. M. and Milburn, G. J.},
  title   = {Quantum theory of field-quadrature measurements},
  journal = {Physical Review A},
  volume  = {47},
  pages   = {642--662},
  year    = {1993}
}

@article{GoetschGraham1994,
  author  = {Goetsch, Peter and Graham, Robert},
  title   = {Linear stochastic wave equations for continuously measured quantum systems},
  journal = {Physical Review A},
  volume  = {50},
  pages   = {5242--5255},
  year    = {1994}
}

@book{WisemanMilburn2010,
  author    = {Wiseman, Howard M. and Milburn, Gerard J.},
  title     = {Quantum Measurement and Control},
  publisher = {Cambridge University Press},
  address   = {Cambridge},
  year      = {2010}
}

@article{Barchielli1990,
  author  = {Barchielli, Alberto},
  title   = {Direct and heterodyne detection and other applications of quantum stochastic calculus to quantum optics},
  journal = {Quantum Optics: Journal of the European Optical Society Part B},
  volume  = {2},
  pages   = {423--441},
  year    = {1990}
}

@article{ArthursKelly1965,
  author  = {Arthurs, E. and Kelly, J. L., Jr.},
  title   = {On the Simultaneous Measurement of a Pair of Conjugate Observables},
  journal = {Bell System Technical Journal},
  volume  = {44},
  pages   = {725--729},
  year    = {1965}
}

@article{Luders1951,
  author  = {L{\"u}ders, Gerhart},
  title   = {{\"U}ber die {Z}ustands{\"a}nderung durch den {M}e{\ss}proze{\ss}},
  journal = {Annalen der Physik},
  volume  = {443},
  pages   = {322--328},
  year    = {1950},
  note    = {(6) 8, 322 (1951). English translation: Ann. Phys. (Leipzig) 15, 663 (2006)}
}

@article{Husimi1940,
  author  = {Husimi, K{\^o}di},
  title   = {Some Formal Properties of the Density Matrix},
  journal = {Proceedings of the Physico-Mathematical Society of Japan. 3rd Series},
  volume  = {22},
  pages   = {264--314},
  year    = {1940}
}

@article{Lindblad1976,
  author  = {Lindblad, G.},
  title   = {On the generators of quantum dynamical semigroups},
  journal = {Communications in Mathematical Physics},
  volume  = {48},
  pages   = {119--130},
  year    = {1976}
}

@article{GKS1976,
  author  = {Gorini, Vittorio and Kossakowski, Andrzej and Sudarshan, E. C. G.},
  title   = {Completely positive dynamical semigroups of {N}-level systems},
  journal = {Journal of Mathematical Physics},
  volume  = {17},
  pages   = {821--825},
  year    = {1976}
}

@book{Davies1976,
  author    = {Davies, E. B.},
  title     = {Quantum Theory of Open Systems},
  publisher = {Academic Press},
  address   = {London},
  year      = {1976}
}

@article{Dalibard1992,
  author  = {Dalibard, Jean and Castin, Yvan and M{\o}lmer, Klaus},
  title   = {Wave-function approach to dissipative processes in quantum optics},
  journal = {Physical Review Letters},
  volume  = {68},
  pages   = {580--583},
  year    = {1992}
}

@book{Carmichael1993,
  author    = {Carmichael, Howard},
  title     = {An Open Systems Approach to Quantum Optics},
  series    = {Lecture Notes in Physics Monographs},
  volume    = {18},
  publisher = {Springer},
  address   = {Berlin},
  year      = {1993}
}

@book{BreuerPetruccione2002,
  author    = {Breuer, Heinz-Peter and Petruccione, Francesco},
  title     = {The Theory of Open Quantum Systems},
  publisher = {Oxford University Press},
  address   = {Oxford},
  year      = {2002}
}

@article{Witten2018,
  author  = {Witten, Edward},
  title   = {Notes on some entanglement properties of quantum field theory},
  journal = {Reviews of Modern Physics},
  volume  = {90},
  pages   = {045003},
  year    = {2018}
}

@book{Gardiner2009,
  author    = {Gardiner, Crispin},
  title     = {Stochastic Methods: A Handbook for the Natural and Social Sciences},
  edition   = {4th},
  publisher = {Springer},
  address   = {Berlin},
  year      = {2009}
}

@article{Girsanov1960,
  author  = {Girsanov, I. V.},
  title   = {On Transforming a Certain Class of Stochastic Processes by Absolutely Continuous Substitution of Measures},
  journal = {Theory of Probability and Its Applications},
  volume  = {5},
  pages   = {285--301},
  year    = {1960}
}

@book{Helstrom1976,
  author    = {Helstrom, Carl W.},
  title     = {Quantum Detection and Estimation Theory},
  publisher = {Academic Press},
  address   = {New York},
  year      = {1976}
}

@book{Holevo1982,
  author    = {Holevo, A. S.},
  title     = {Probabilistic and Statistical Aspects of Quantum Theory},
  publisher = {North-Holland},
  address   = {Amsterdam},
  year      = {1982}
}

@article{SusskindGlogower1964,
  author  = {Susskind, Leonard and Glogower, Jonathan},
  title   = {Quantum mechanical phase and time operator},
  journal = {Physics Physique Fizika},
  volume  = {1},
  pages   = {49--61},
  year    = {1964}
}

@article{Leonhardt1995,
  author  = {Leonhardt, U. and Vaccaro, J. A. and B{\"o}hmer, B. and Paul, H.},
  title   = {Canonical and measured phase distributions},
  journal = {Physical Review A},
  volume  = {51},
  pages   = {84--95},
  year    = {1995}
}

@article{WisemanKillip1998,
  author  = {Wiseman, H. M. and Killip, R. B.},
  title   = {Adaptive single-shot phase measurements: The full quantum theory},
  journal = {Physical Review A},
  volume  = {57},
  pages   = {2169--2185},
  year    = {1998}
}

@article{Armen2002,
  author  = {Armen, Michael A. and Au, John K. and Stockton, John K. and Doherty, Andrew C. and Mabuchi, Hideo},
  title   = {Adaptive Homodyne Measurement of Optical Phase},
  journal = {Physical Review Letters},
  volume  = {89},
  pages   = {133602},
  year    = {2002}
}

@article{Martin2020,
  author  = {Martin, Leigh S. and Livingston, William P. and Hacohen-Gourgy, Shay and Wiseman, Howard M. and Siddiqi, Irfan},
  title   = {Implementation of a canonical phase measurement with quantum feedback},
  journal = {Nature Physics},
  volume  = {16},
  pages   = {1046--1049},
  year    = {2020},
  note    = {CHECK: is this the non-Gaussian-resource reference intended in Sec. III D?}
}

@article{FuchsSchack2013,
  author  = {Fuchs, Christopher A. and Schack, R{\"u}diger},
  title   = {Quantum-{B}ayesian coherence},
  journal = {Reviews of Modern Physics},
  volume  = {85},
  pages   = {1693--1715},
  year    = {2013}
}

@article{FuchsMerminSchack2014,
  author  = {Fuchs, Christopher A. and Mermin, N. David and Schack, R{\"u}diger},
  title   = {An introduction to {QB}ism with an application to the locality of quantum mechanics},
  journal = {American Journal of Physics},
  volume  = {82},
  pages   = {749--754},
  year    = {2014}
}

@article{Born1926,
  author  = {Born, Max},
  title   = {Zur {Q}uantenmechanik der {S}to{\ss}vorg{\"a}nge},
  journal = {Zeitschrift f{\"u}r Physik},
  volume  = {37},
  pages   = {863--867},
  year    = {1926}
}

@article{Riemann1868,
  author  = {Riemann, Bernhard},
  title   = {{\"U}ber die {H}ypothesen, welche der {G}eometrie zu {G}runde liegen},
  journal = {Abhandlungen der K{\"o}niglichen Gesellschaft der Wissenschaften zu G{\"o}ttingen},
  volume  = {13},
  pages   = {133--150},
  year    = {1868},
  note    = {Habilitation lecture delivered 1854}
}

@book{Euclid,
  author    = {{Euclid}},
  title     = {The Thirteen Books of {E}uclid's {E}lements},
  editor    = {Heath, Thomas L.},
  publisher = {Cambridge University Press},
  address   = {Cambridge},
  year      = {1908}
}

@book{Steenrod1951,
  author    = {Steenrod, Norman},
  title     = {The Topology of Fibre Bundles},
  publisher = {Princeton University Press},
  address   = {Princeton, NJ},
  year      = {1951}
}

@article{LeviCivita1917,
  author  = {Levi-Civita, Tullio},
  title   = {Nozione di parallelismo in una variet{\`a} qualunque e conseguente specificazione geometrica della curvatura {R}iemanniana},
  journal = {Rendiconti del Circolo Matematico di Palermo},
  volume  = {42},
  pages   = {173--204},
  year    = {1917}
}

@article{WisemanDiosi2001,
  author  = {Wiseman, H. M. and Di{\'o}si, L.},
  title   = {Complete parameterization, and invariance, of diffusive quantum trajectories for {M}arkovian open systems},
  journal = {Chemical Physics},
  volume  = {268},
  pages   = {91--104},
  year    = {2001}
}

@article{DeBrota2026locality,
  author  = {DeBrota, John B. and Fuchs, Christopher A. and Schack, R{\"u}diger},
  title   = {{QB}ism on Locality and Nonlocality},
  journal = {arXiv preprint},
  year    = {2026},
  note    = {arXiv:2608.05368}
}

@article{DeBrotaList2026,
  author  = {DeBrota, John B. and List, Christian},
  title   = {A Heptalemma for Quantum Mechanics},
  journal = {Foundations of Physics},
  volume  = {56},
  number  = {2},
  pages   = {24},
  year    = {2026},
  doi     = {10.1007/s10701-026-00919-9}
}

@article{DeBrota2023decoherence,
  author  = {DeBrota, John B. and Fuchs, Christopher A. and Schack, R{\"u}diger},
  title   = {Quantum Dynamics Happens Only on Paper: {QB}ism's Account of Decoherence},
  journal = {arXiv preprint},
  year    = {2023},
  note    = {arXiv:2312.14112}
}

@article{DeBrota2018,
  author  = {DeBrota, John B. and Stacey, Blake C.},
  title   = {{FAQB}ism},
  journal = {arXiv preprint},
  year    = {2018},
  note    = {arXiv:1810.13401}
}

@misc{Preskill1998,
  author       = {Preskill, John},
  title        = {Lecture Notes for {P}hysics 229: {Q}uantum Information and Computation},
  howpublished = {California Institute of Technology, \url{http://theory.caltech.edu/~preskill/ph229/}},
  year         = {1998}
}

@article{vonNeumann1927,
  author  = {von Neumann, J.},
  title   = {Wahrscheinlichkeitstheoretischer {A}ufbau der {Q}uantenmechanik},
  journal = {Nachrichten von der Gesellschaft der Wissenschaften zu G{\"o}ttingen, Mathematisch-Physikalische Klasse},
  volume  = {1927},
  pages   = {245--272},
  year    = {1927}
}

@book{vonNeumann1932,
  author    = {von Neumann, John},
  title     = {Mathematische {G}rundlagen der {Q}uantenmechanik},
  publisher = {Springer},
  address   = {Berlin},
  year      = {1932}
}

@article{Propp2023,
  title = {Decoherence limiting the cost to simulate an anharmonic oscillator},
  volume = {108},
  ISSN = {2469-9934},
  url = {http://dx.doi.org/10.1103/PhysRevA.108.062219},
  DOI = {10.1103/physreva.108.062219},
  number = {6},
  journal = {Physical Review A},
  publisher = {American Physical Society (APS)},
  author = {Propp,  Tzula B. and Ray,  Sayonee and DeBrota,  John B. and Albash,  Tameem and Deutsch,  Ivan H.},
  year = {2023},
  month = Dec 
}

@article{Propp2022,
  title = {How to describe collective decay of uncoupled modes in the input–output formalism},
  volume = {39},
  ISSN = {1520-8540},
  url = {http://dx.doi.org/10.1364/JOSAB.468251},
  DOI = {10.1364/josab.468251},
  number = {12},
  journal = {Journal of the Optical Society of America B},
  publisher = {Optica Publishing Group},
  author = {Propp,  Tzula B.},
  year = {2022},
  month = Nov,
  pages = {3128}
}

@article{Propp2019,
  title = {On nonlinear amplification: improved quantum limits for photon counting},
  volume = {27},
  ISSN = {1094-4087},
  url = {http://dx.doi.org/10.1364/OE.27.023454},
  DOI = {10.1364/oe.27.023454},
  number = {16},
  journal = {Optics Express},
  publisher = {Optica Publishing Group},
  author = {Propp,  Tzula B. and van Enk,  S. J.},
  year = {2019},
  month = July,
  pages = {23454}
}

@book{Klein1872,
  author    = {Klein, Felix},
  title     = {Vergleichende {B}etrachtungen {\"u}ber neuere geometrische {F}orschungen},
  publisher = {A. Deichert},
  address   = {Erlangen},
  year      = {1872},
  note      = {Reprinted in Math. Ann. \textbf{43}, 63--100 (1893); English translation by M. W. Haskell, Bull. New York Math. Soc. \textbf{2}, 215--249 (1893)}
}

@book{LieEngel1888,
  author    = {Lie, Sophus and Engel, Friedrich},
  title     = {Theorie der {T}ransformationsgruppen},
  publisher = {B. G. Teubner},
  address   = {Leipzig},
  year      = {1888--1893},
  note      = {3 vols.}
}

\appendix

\section{Coordinate Transformations}\label{app:coordtrans}

There are three important coordinate systems for the instrumental group manifold $\mathrm{IWH}^{\textbf{12}}$, corresponding to the following decompositions of an aribraty group element, $g \in \mathrm{IWH}^{\textbf{12}}$. This has discussed elsewhere for Harish-Chandra-like and Cartan-Bergers coordinates \cite{Jackson2023}, but the transformations to and from Iwasawa coordinates are here introduced to the literature for the first time. 

There is the Harish-Chandra-like decomposition:
\begin{align}\label{HClin}
	g = e^{z_\text{HC}}e^{a^\dag \nu} Q e^{a\mu^*}.
\end{align}
where
\begin{align}\label{HCquad}
	Q = e^{-A^\dag \omega} e^{-iK_0 \varphi} e^{-K_0 c} e^{-A\zeta^*}.
\end{align}

There is the Iwasawa-like decomposition:
\begin{align}\label{Ilin}
	g = e^{z_\text{I}}D_\gamma Q e^{a\lambda^*},
\end{align}
where
\begin{align}\label{Iquad}
	Q = S_\omega e^{-iK_0 \varphi} e^{-K_0 b} e^{-A\chi^*}.
\end{align}

There is the Cartan-Bergers-like decomposition:
\begin{align}\label{CBlin}
	g = e^{z_\text{CB}}D_\beta Q D_\alpha^\inv,
\end{align}
where
\begin{align}\label{CBquad}
	Q = S_\eta e^{-iK_0\psi} S_\xi e^{-K_0 a} S_\xi^\inv.
\end{align}

\subsection{From Harish-Chandra-like to Iwasawa}

To transform from Harish-Chandra-like coordinates to Iwasawa coordinates, first transform the quadratic coordinates with (\ref{HC2Iquad})
\begin{align}\label{HC2Iquad}
	\begin{cases}
		e^{-b} = \frac{e^{-c}}{1-|\omega|^2}\\
		\chi = \zeta + e^{-c +i\varphi}\frac{\omega}{1-|\omega|^2}
	\end{cases}
\end{align}
and then transform the remaining linear and central coordinates with (\ref{HC2Ilinear})
\begin{align}\label{HC2Ilinear}
	\begin{cases}
		\lambda = \mu + \frac{\nu -\omega \nu^*}{1-|\omega|^2}e^{-\upsilon^*/2}\\
		\gamma = \frac{\nu -\omega \nu^*}{1-|\omega|^2}\\
		z_\text{I} = z_\text{HC} + \frac12(\gamma+\gamma^*\omega)\gamma^*
	\end{cases}
\end{align}
where
\begin{align}
    \upsilon = c + i \varphi.
\end{align}\\

\subsection{From Iwasawa to Cartan-Berger}

To transform from Iwasawa coordinates to Cartan-Bergers coordinates, again first transform the quadratic coordinates with (\ref{I2CBquadbase}, \ref{I2CBquadfiber}, and \ref{Itrans})
\begin{align}\label{I2CBquadbase}
	\begin{cases}
		\cosh a = \cosh b -\frac12 e^b |\chi|^2\\
		e^{i\phi}\tanh\eta = \frac{\chi}{1-e^{-2b}+|\chi|^2}
	\end{cases}
\end{align}
and
\begin{align}\label{I2CBquadfiber}
	\begin{cases}
		e^{-i\psi} = e^{-i\varphi}\frac{1-\omega e^{i\varphi}\uptau^*}{1-\omega^* e^{-i\varphi}\uptau}\\
		\eta = \frac{\omega - e^{-i\varphi}\uptau}{1-\omega^* e^{-i\varphi}\uptau}
	\end{cases}
\end{align}
where the transfer parameter in Iwasawa coordinates is
\begin{align}
	\uptau
	&= \frac{\chi}{e^{-b}+1}\label{Itrans}.
\end{align}
Then transform the remaining linear and central coordinates with (\ref{I2CBlinear}):
\begin{widetext}
\begin{align}\label{I2CBlinear}
	\begin{cases}
		\alpha = \frac{(1-e^{-b})\lambda-\chi\lambda^*}{(1-e^{-b})^2-|\chi|^2}\\
		\beta = \gamma + \frac{1}{(1-e^{-b})^2-|\chi|^2}\left(\big((1-e^{-b})e^{-u/2}+\omega\chi^*e^{-u^*/2}\big)\lambda-\big(\chi e^{-u/2}+\omega(1-e^{-b})e^{-u^*/2}\big)\lambda^*\right)\\
		z_\text{CB} = z_\text{I} - \frac12(\gamma+\gamma^*\omega)\gamma^* + \frac12\alpha(\alpha^* + \zeta^* \alpha) + \frac12(\beta+\beta^*\omega)\beta^*
		- \beta^* e^{-u/2} \alpha\sqrt{1-|\omega|^2}
	\end{cases}
\end{align}
\end{widetext}
where
\begin{equation}
	u \equiv b+i\varphi.
\end{equation}

\subsection{Inverse transformations}

To transformation from Cartan-Bergers coordinates back to Iwasawa coordinates between the quadratic coordinates is (\ref{CB2Iquadbase}, \ref{CB2Iquadfiber}, and \ref{CBtrans})
\begin{align}\label{CB2Iquadbase}
	\begin{cases}
		e^{-b} = e^{-a}\frac{1-|\xi|^2}{1-|\xi|^2e^{-2a}}\\
		\chi = \xi \frac{1-e^{-2a}}{1-|\xi|^2e^{-2a}}
	\end{cases}
\end{align}
and
\begin{align}\label{CB2Iquadfiber}
	\begin{cases}
		e^{-i\varphi} = e^{-i\psi}\frac{1+\eta e^{i\psi}\uptau^*}{1+\eta^* e^{-i\psi}\uptau}\\
		\omega = \frac{\eta + e^{-i\psi}\uptau}{1+\eta^* e^{-i\psi}\uptau}.
	\end{cases}
\end{align}
where the transfer parameter in Cartan-Bergers coordinates is
\begin{align}
	\uptau
    &= \xi \frac{1-e^{-a}}{1-|\xi|^2e^{-a}}\label{CBtrans}.
\end{align}
Then, the transformation between the linear and central coordinates is (\ref{CB2HClinear})
\begin{align}\label{CB2HClinear}
	\begin{cases}
		\mu = \alpha + \alpha^* \zeta - \beta e^{-\upsilon^*/2}\\
		\nu = \beta + \beta^*\omega -\alpha e^{-\upsilon/2}\\
		z_\text{HC} = z_\text{CB} - \frac12\alpha(\alpha^* + \zeta^* \alpha) - \frac12(\beta+\beta^*\omega)\beta^* + \beta^* e^{-\upsilon/2} \alpha
	\end{cases}
\end{align}
composed with equations \ref{HC2Ilinear}.\\

To transformation from Iwasawa coordinates back to Harish-Chandra-like coordinates between the quadratic coordinates is (\ref{I2HCquad})
\begin{align}\label{I2HCquad}
	\begin{cases}
		e^{-c} = e^{-b}(1-|\omega|^2)\\
		\zeta = \chi - e^{-b+i\varphi}\omega
	\end{cases}
\end{align}
and between the linear and central coordinates is (\ref{I2HClinear})
\begin{align}\label{I2HClinear}
	\begin{cases}
		\mu = \lambda - \gamma e^{-\upsilon^*/2} \\
		\nu = \gamma + \gamma^*\omega\\
		z_\text{HC} = z_\text{I} - \frac12(\gamma+\gamma^*\omega)\gamma^*.
	\end{cases}
\end{align}\\

\section{Sampling Other Parts of Phase Space}\label{app:Sampling}

As emphasized in the maintext, the Kraus Operator Distribution (K.O.D.) only samples near the origin. To sample other regions of phase space, we shift our Wiener increments to have non-zero mean.

\bea\label{SamplingPhaseSpace}
d\vec{W}_t \rightarrow d\vec{W}_t - 2q\vec{A} dt.
\eea 

For all coordinates, this only modifies the center SDE
\bea \label{SamplingPhaseSpace}
dz_t \rightarrow dz_t - q \vec{A} \dot\vec{d}W_t + q^2 |\vec{A}|^2 dt.
\eea 

Note that while this suppresses individual trajectory weights, the correct way to calculate the average weight is by the average of the exponential i.e. $\expect{e^{2{\rm Re}[z]}}$ and not $e^{2\expect{{\rm Re}[z]}}$. For an individual trajectory, we expect $e^{2{\rm Re}[z]}$ to become very small, corresponding to the sampling of each phase space point with (non-orthogonal) Kraus operators.

This is straightforward to verify from the Haar measure of the Wiener increments

\bea \label{SamplingPhaseSpace}
d\mu (d\vec{W}_t) = d^n \frac{(d\vec{W}_t)}{(2\pi)^{n/2}} e^{-\frac{|d\vec{W}_t|^2}{2dt}}.
\eea 

Now our new distribution, which is also ostensible is the sense Wiseman, will be centered on a drifting point i.e. one that ballistically increases in mean. We use this to state-independently sample different regions of phase space.



\end{document}